\documentclass[a4paper,11pt]{article}
\usepackage{jheppub} 
\pdfoutput=1
\usepackage{graphicx,array}
\usepackage{changepage}
\usepackage{comment}
\usepackage{multirow}
\definecolor{orange}{rgb}{1,0.5,0}
\definecolor{brown}{rgb}{0.59, 0.29, 0.0}

\allowdisplaybreaks

\newcommand{\beq}{\begin{equation}}
\newcommand{\eeq}{\end{equation}}
\newcommand{\TeV}{\mathrm{\;TeV}}
\newcommand{\GeV}{\mathrm{\;GeV}}

\def\beq{\begin{equation}}
\def\eeq{\end{equation}}
\def\bea{\begin{eqnarray}}
\def\eea{\end{eqnarray}}
\def\bit{\begin{itemize}}
\def\eit{\end{itemize}}

\def\baa{\begin{array}}
\def\eaa{\end{array}}

\newcommand{\met}{{/\!\!\!\! E_T}} 
\newcommand{\mpt}{{/\!\!\!\! \vec{P}_T}} 
\newcommand{\kk}{\textrm{KK}}

\newcommand{\dk}[1]{\textcolor{brown}{#1}}

\newcolumntype{C}[1]{>{\centering}m{#1}}
\newcolumntype{x}[1]{>{\raggedright\arraybackslash}p{#1}}

\def\beq{\begin{equation}}
\def\eeq{\end{equation}}
\def\bea{\begin{eqnarray}}
\def\eea{\end{eqnarray}}
\def\bal{\begin{align}}
\def\eal{\end{align}}

\title{LHC Signals from Cascade Decays of 
Warped Vector Resonances
}

\author[a]{Kaustubh S.~Agashe,}
\author[a,b]{Jack Collins,}
\author[a]{Peizhi Du,}
\author[a]{Sungwoo Hong,}
\author[c,d]{Doojin Kim,}
\author[a]{Rashmish K.~Mishra}

\affiliation[a]{Maryland Center for Fundamental Physics,
     Department of Physics,
     University of Maryland,
     College Park, MD 20742, USA}
\affiliation[b]{Department of Physics and Astronomy, Johns Hopkins University, Baltimore, MD 21218, USA}
\affiliation[c]{Department of Physics, University of Florida, Gainesville, FL 32611 USA}
\affiliation[d]{Theory Division, CERN, CH-1211 Geneva 23, Switzerland}

\emailAdd{kagashe@umd.edu}
\emailAdd{jhc296@umd.edu}
\emailAdd{pdu@umd.edu}
\emailAdd{sungwoo83hong@gmail.com}
\emailAdd{doojin.kim@cern.ch}
\emailAdd{rashmish@umd.edu}

\abstract{
Recently (arXiv:1608.00526), a new framework for warped higher-dimensional compactifications with ``bulk" standard model (SM) was proposed: in addition to the UV (Planck scale) and IR (a couple of TeV) branes, there is an intermediate brane, taken to be around 10 TeV. The SM matter and Higgs fields propagate from the UV brane down to this intermediate brane only, while gauge and gravity fields propagate in the entire bulk. Such a configuration renders the lightest gauge Kaluza-Klein (KK) states within LHC reach, simultaneously satisfying  flavor and CP constraints. In addition, the usual leading decay modes of the lightest KK gauge bosons into top and Higgs bosons are suppressed. This effect permits erstwhile subdominant channels to become significant. These include flavor-universal decays to SM fermions and Higgs bosons, and a novel channel -- decay to a radion and a SM gauge boson, followed by radion decay to a pair of SM gauge bosons. In this work, we first delineate the parameter space where the above mentioned cascade decay of gauge KK particles dominates, and thereby can be the discovery mode at the LHC. We then perform a detailed analysis of the LHC signals from this model, finding that 300/fb suffices for evidence of KK-gluon in tri-jet, jet + di-photon and jet + di-boson channels. However, KK photon in photon + di-jet, and KK-W in leptonic W + di-jet require 3000/fb. The crucial feature of this decay chain is a ``double" resonance, i.e. 3-particle and 2-particle invariant mass peaks, corresponding to the KK gauge boson and the radion respectively.
}

\begin{document} 

\begin{flushright}
CERN-TH-2016-247
\end{flushright}

\maketitle
\flushbottom


\section{Introduction}
\label{sec:introduction}

Addressing the Planck-weak hierarchy and dark matter problems of the Standard Model (SM) motivates new physics around the TeV scale. 
Thus far, there is no sign of it in a wide range of collider experiments performed both at low and high energies. However, many of these searches are done in obvious channels. This situation motivates further theoretical and phenomenological thinking about hitherto hidden, unexpected, and subtler ways in which new physics might be revealed: often a slight non-minimal extension of an established framework suffices to cause such surprises.
This exercise must of course be followed by conducting {\em experimental} searches in the new suggested channels.

\begin{figure}
\center
\includegraphics[width=0.8\linewidth]{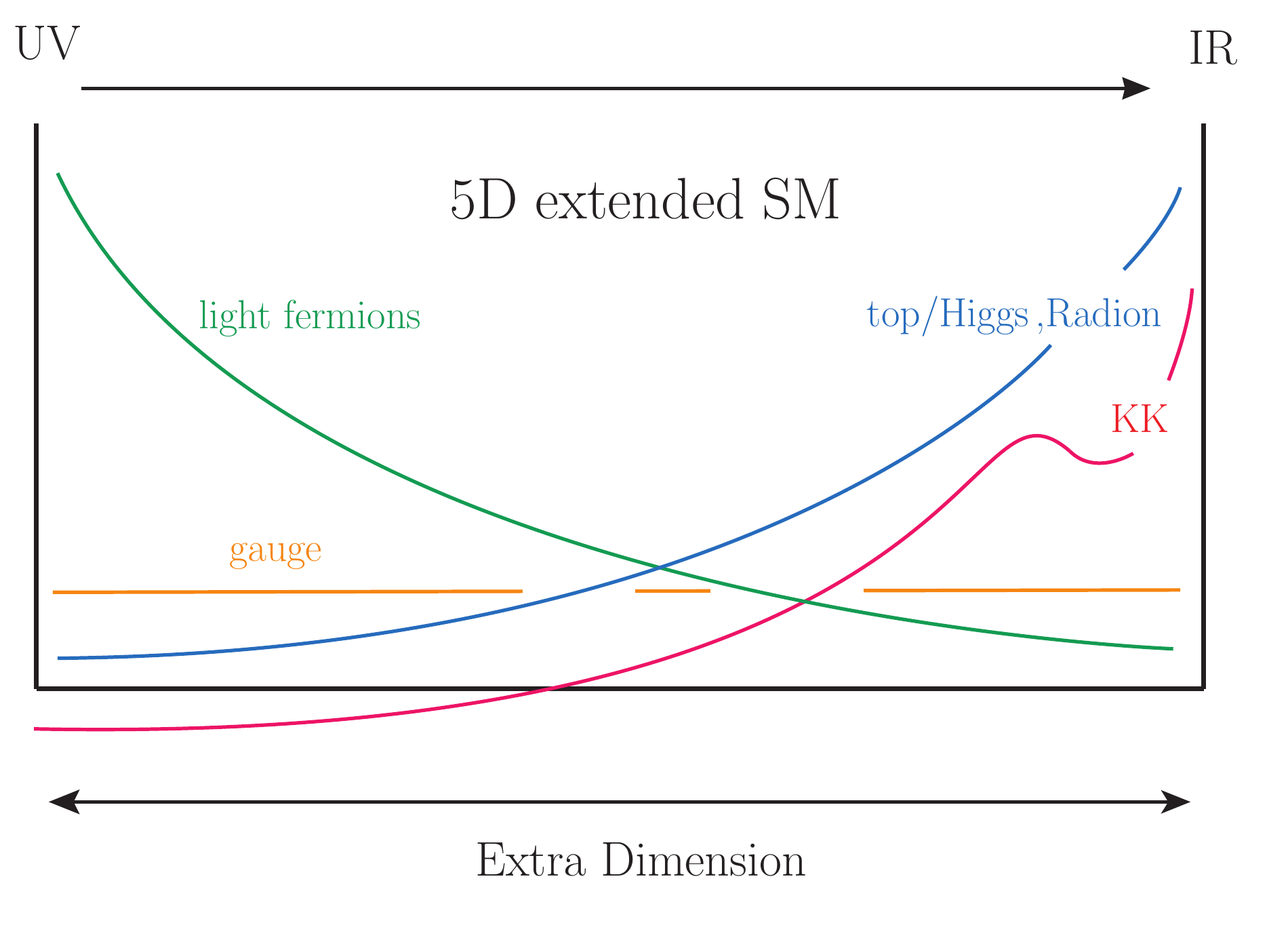}
\caption{Warped extra dimensional model with SM fields in bulk (standard framework). Schematic shapes of extra-dimensional wavefunctions for various particles (zero modes and a generic KK mode) are shown. }
\label{fig:standard}
\end{figure}

Some of the authors of this paper have recently proposed one such possibility, 
in the context of warped higher-dimensional compactifications~\cite{Agashe:2016rle}. 
We provide a brief review here for the sake of completeness. 
To begin with, the usual framework of warped higher-dimensional compactifications 
involves fields corresponding to \textit{all} SM particles (including graviton) propagating in the bulk of a warped extra dimension, which is terminated on the two ends by the UV and IR branes (see Fig.~\ref{fig:standard}). 
The metric is taken to be a slice of anti de-Sitter (AdS) space. This is a very attractive extension of the SM since it addresses the Planck-weak and flavor hierarchy problems of the SM.

The four-dimensional (4D) SM particles arise from a Kaulza-Klein (KK) decomposition of the 5D fields and correspond to the lightest modes of this expansion. All the modes get a profile in the extra dimension which indicates where the mode is localized. Specifically, the 4D graviton is localized near the UV brane, where the characteristic 4D energy scale is Planckian, whereas the Higgs boson occupies the region near the IR brane, with the associated scale being warped-down to $\mathcal{O}(\text{TeV})$. 
This aspect of the framework of warped geometry is what underlies the solution to the Planck-weak hierarchy problem.

On the other hand, the SM fermions can have disparate profiles in the extra dimension. 
When evaluated near the IR brane where the Higgs boson is localized, these profiles have hierarchical values.
This translates into a similar feature in their overlap -- and hence into the Yukawa couplings with Higgs boson. In particular, the top quark resides near the IR brane, reflecting its large mass. The massive excitations of all the SM fields (called KK modes) -- whether fermions, gauge bosons, or graviton -- are localized near the IR brane. Their masses are essentially given by the scale corresponding to the IR brane. Finally, we have the radion, 
%
%
which is the modulus corresponding to fluctuations in the size of the extra dimension. The radion is also localized near the IR brane, just like KK modes, thus its mass 
is generically of the order of the IR brane scale.
However, with mild tuning, the radion mass can be arranged to be a factor of few below the mass of the first KK 
mode~\cite{Konstandin:2010cd, Chacko:2012sy, Chacko:2013dra}.

However, moving on to finer details of the KK mass scale, it is well known that stringent constraints from flavor and CP tests on effects of (lightest) gauge {\it and} fermion KK modes 
%
%
require their masses 
%
to be $\gtrsim \mathcal{O}(10)$ TeV~\cite{Csaki:2008zd,Blanke:2008zb,Bauer:2009cf,KerenZur:2012fr}, unless some additional flavor structure is imposed (see Refs.~\cite{Redi:2011zi,Barbieri:2012tu,Straub:2013zca,Konig:2014iqa} for recent work in the context of a ``simplified'' version of the 5D model). This equivalently means that the IR brane scale should be $\gtrsim \mathcal{O}(10)\TeV$.~\footnote{Electroweak Precision Measurements also impose strong constraints on the IR brane scale. For example, with only SM gauge group in the bulk, the consistency with the electroweak precision measurements requires KK scale $\gtrsim \mathcal{O}(10)$ TeV~\cite{Huber:2000fh,Huber:2001gw,Csaki:2002gy,Hewett:2002fe}. However, it was shown in Ref.~\cite{Agashe:2003zs} that extension of the bulk gauge group to $SU(2)_{\rm L} \times SU(2)_{\rm R} \times U(1)_{\rm B-L}$, which contains the built-in custodial symmetry of the electroweak sector, can relax this bound and KK scale $\gtrsim 3$ TeV is still allowed.} We will refer to this setup as ``standard" from here on. This creates a ``meso"-tuning to be imposed on the theory (see for example~\cite{Agashe:2016rle}), since a fully natural solution would require the IR brane scale to be $\sim \mathcal{O}(1)$ TeV. Of more concern to us is, however, the possible lack of LHC signals resulting from direct production of the associated new physics, namely, the KK modes, simply based on the kinematic reach of the LHC.


With the above situation in mind, the new idea
%
%
in~\cite{Agashe:2016rle} involves,
%
%
%
%
broadly speaking, the introduction of extra branes in-between the UV and IR ones. 
Various bulk fields are allowed to propagate different amounts in the bulk, consistent with general principles and symmetries. In particular, gravity must propagate in the entire spacetime due to its dynamical nature, while the gauge fields must propagate at least equal or more than the matter fields in the extra dimension. This is because the matter currents need a gauge field to couple to, while the gauge fields can exist on their own.

\begin{figure}
\centering
\includegraphics[width=0.8\linewidth]{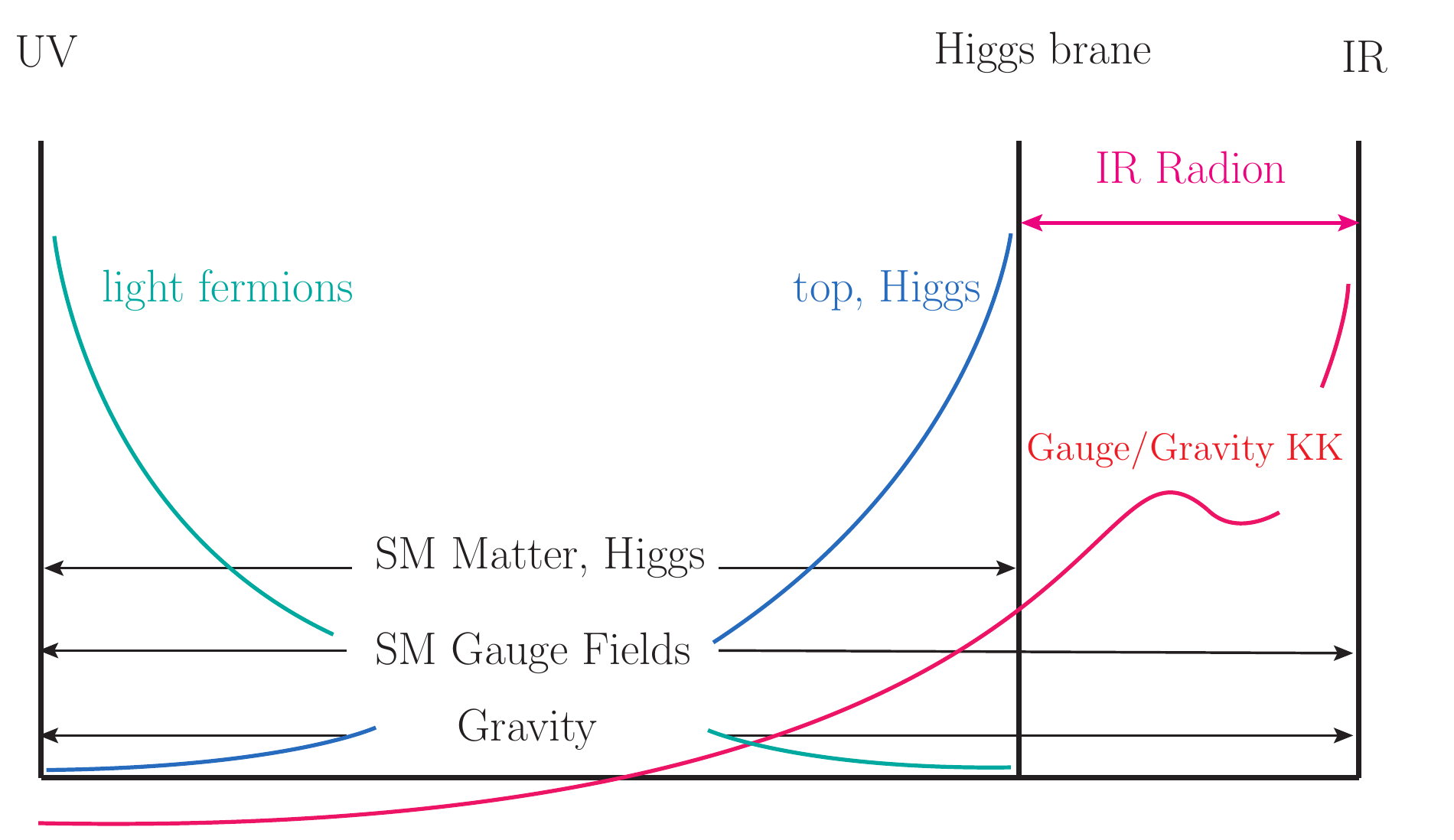}
\caption{Warped extra dimensional model with SM fields in the bulk (extended framework). Schematic shapes of extra-dimensional wave functions for various particles (zero modes and generic KK modes) are shown.  }
\label{fig:extended}
\end{figure}

In the simplest incarnation of this proposal the basic setup is modified by the inclusion of one such extra brane, chosen to be located very close to the IR brane.
The SM matter and Higgs fields are allowed to propagate only in the subspace from UV to this ``intermediate" brane, whereas gauge and gravity occupy the entire bulk (see Fig.~\ref{fig:extended}). 
We will henceforth refer to this framework as the ``extended'' framework, 
and the intermediate brane as the ``Higgs'' brane. 
We choose the Higgs brane scale to be $\gtrsim \mathcal{O}(10)\TeV$, i.e., same as the IR brane scale of the standard scenario.
We then see that in this extended setup, we retain solutions to both the Planck-weak and flavor hierarchy problems. This is of course modulo the meso-tuning mentioned earlier. 
It is useful to keep in mind that the standard framework described above is a special case of this extended framework, if the Higgs brane and the IR brane are identified as one. 

In order to determine how the bound from flavor and CP tests on the lightest {\em gauge} KK mass scale is modified,\footnote{
By the above
construction,
%
%
the KK fermions satisfy these bounds even in the extended framework.} we need to, in turn, figure out
the couplings of gauge KK modes to the light SM fermions.
To this end, we make use of the usual conceptual approach that couplings between 4D particles are dictated by the overlap of their respective profiles in the extra dimension. 
The point is that flavor \textit{dependence} of these couplings of the gauge 
%
%
KK modes arises \textit{primarily} from the part of the overlap in the infra-red region,
where the KK modes are localized.
%
%
%
%
Because of the splitting of
IR brane (where gauge KK are peaked) from the brane where matter fields end, we see that 
the flavor {\em non}-universal component of gauge KK couplings to SM fermions is \textit{reduced}.
Thus, bounds on gauge KK mass from flavor and CP violation are relaxed in the extended case.
It is noteworthy that gauge KK couplings to SM fermions/Higgs also have a contribution from 
overlap near the UV brane: this is, however, universal, given the constant profile of the gauge KK in that region.
To summarize then, this setup has an important feature: the lightest gauge KK particle mass of \textit{a few} TeV (related to the location of the IR brane) can be consistent with the flavor and CP bounds. 
%
%
This makes the gauge KK modes lie within the kinematic reach of the LHC. 
But in order to complete this story, we need to check the fate of the \textit{couplings} involved in
their \textit{production}. 
In the standard scenario, the gauge KK production at the LHC occurs dominantly via the coupling to the light quarks (inside protons). This coupling is the flavor-universal UV-region-dominated coupling, as mentioned above. It is therefore clear that the size of this coupling is not modified in the extended setup. Combining the above couplings and masses, the stated goal of the gauge KK particles being within the LHC reach is thus achieved. 

Having ensured significant production at the LHC, we next move onto the {\em decays} of the gauge KK modes. As already indicated above, the coupling between modes near the IR brane is the largest. 
In the standard scenario, examples of such couplings would be those between gauge KK modes and top quark/Higgs bosons (including longitudinal $W/Z$, among the SM particles). 
Concomitant to what happens to flavor-violation, these 
%
%
top/Higgs-philic couplings of gauge KK modes -- hence their decays to top/Higgs (which are usually the dominant channels) -- 
are then also suppressed. This is because top/Higgs are localized on the intermediate brane in the new framework, while the KK gauge bosons are localized on the IR brane.
%
Such a twist then opens the door for {\em other} couplings (i.e., involving profiles not necessarily peaked near IR brane) to become relevant for the gauge KK boson decays.
For example, there is a coupling among KK gauge boson, radion and SM gauge boson, which involves two profiles which are IR-localized and one flat profile (of the SM gauge boson). Due to the suppression of the gauge KK modes coupling to the top/Higgs, this coupling becomes important.
As already mentioned, the radion can be lighter than the gauge KK modes by a factor of a few so that the above coupling can mediate the decay of a KK gauge boson into a radion and the corresponding SM gauge boson.
Note that in the standard setup, radion subsequently decays dominantly into top/Higgs, because its profile is peaked near IR brane, where the top/Higgs are localized. Remarkably, in the extended framework, radion instead decays mostly into a pair of SM gauge bosons. This is because the other dominant channels are suppressed for the same reason as for gauge KK -- top/Higgs profiles have now moved away from the radion.\footnote{Note that decays of spin-1 gauge KK into a pair of SM/massless gauge bosons are not allowed by the Landau-Yang theorem.}

Similarly, we have a flavor-universal decay of KK gauge boson into two SM fermions (again, from overlap near UV brane)
which might come into play here. Note that this is the same coupling which is involved in the production of gauge KK modes at the LHC, as mentioned earlier. We would like to emphasize here that both of these couplings are present, with similar strength, in the standard framwork as well, but it is just that the associated decays are swamped by top/Higgs final states.
After this motivation, we summarize the important aspects of this extended setup in Table~\ref{summary_table}, contrasting them with those in the standard setup.

\begin{table}[t]
\centering
\begin{tabular}{c C{3cm} C{4cm} C{5cm} }
\hline
 & & Standard & Extended \tabularnewline
 \hline \hline
 %
%
KK fermion
& Mass & $\gtrsim \mathcal{O}(10)$ TeV& $\gtrsim \mathcal{O}(10)$ TeV \tabularnewline
\hline
\multirow{3}{*}{
%
%
KK gauge
} & Mass & $\gtrsim \mathcal{O}(10)$ TeV & a few TeV  \tabularnewline
 & Production & $q\bar{q}$ & $q\bar{q}$ \tabularnewline
 & Decay & $t\bar{t},\, HH$ & $f\bar{f}$ (universally), radion
 %
%
$
+\gamma/W/Z/g$ \tabularnewline
\hline 
\multirow{3}{*}{
Radion
%
%
} & Mass & $\gtrsim \mathcal{O}(10)$ TeV$/(\rm a \; few)$ & $\mathcal{O}(1)$ TeV \tabularnewline
 & Production & $gg$ & $gg$ \tabularnewline
 & Decay & $t\bar{t},\, HH$ & $gg \gg WW/ZZ \gg \gamma\gamma$ \tabularnewline
 \hline
\end{tabular}
\caption{Summary of properties (masses, dominant production and decay channels) of  
relevant new particles in the extended warped model (fourth column). The corresponding properties in the standard warped model (third column) are listed for comparison.
\label{summary_table} }
\end{table}


Motivated by these characteristics of the production and decay of gauge KK modes, in this paper, we perform a detailed study of the potential LHC signals resulting from the above-mentioned new, cascade decay process into a SM gauge boson and a radion.
As indicated above, this interesting mode competes mainly with decays to a pair of SM fermions (via universal coupling). 
%
%
As the first step, we therefore determine 
the region of parameter space where the decay channel of a KK gauge boson into a radion and a corresponding SM gauge boson
(with the radion decaying into two SM gauge bosons) dominates.
%
%
We also map out the parameter region which respects bounds on gauge KK modes, from dilepton, dijet, and ditop (i.e., the other competing channels) {\em and} direct (or via above gauge KK decay) production of the radion, where the dominant bound arises from the decay into a photon pair.
We then analyse production of KK photon, KK gluon and KK $W/Z$  and their decay into the corresponding SM gauge boson and the radion in this viable and relevant part of parameter space, with all allowed subsequent radion decays.
Among all these possible final states, we focus on a few which can make discovery feasible at the LHC.
Overall, we show that the prospects are quite promising. In particular, 
%
%
an integrated luminosity of $\mathcal{O}(100)$ fb$^{-1}$ suffices for discovery via the new channel of KK gluon,
whereas $\mathcal{O}(1000)$ fb$^{-1}$ is required for KK $W/Z$ and KK photon due to their small production cross sections.  
%
%
 
A word on the big picture is worth putting in here. In general, the IR region of warped higher-dimensional compactification can have non-trivial structure, including presence of localized terms on the IR brane or deformation of metric from pure AdS metric (cf. hard/bare wall assumed -- mostly for simplicity -- in the standard setup).
One of the ideas behind the proposed framework is to provide a way to model this generic possibility, simply by adding intermediate brane(s), taken to be hard/bare themselves. In other words, the presence of extra branes need not necessarily be taken literally because it can merely be a stand-in for modified IR dynamics. We view this extension as being plausible. 

Following the AdS/CFT correspondence, the above classes ({\em both} standard and extended) of warped compactifications are dual to the purely 4D scenario of Higgs compositeness, with rest of the SM (including gauge bosons and fermions) being partially composite and the associated strong dynamics being an (approximate) conformal field theory (CFT).
In particular, the gauge KK modes in 5D models are dual to {\em composite} heavy partners of the SM gauge bosons, whereas the radion is dual to the dilaton, i.e., Nambu-Goldstone boson arising from spontaneous breaking of scale invariance in the IR.
In turn, the framework of SM compositeness (fully for Higgs boson, vs. partially for the rest) can be thought of as an appropriate generalization of strong dynamics in the {\em  real} world, i.e., QCD coupled to QED. The KK gauge boson is then the analogue of the $\rho$ meson in usual QCD, coupled to QED. Let us elaborate on this point.

To begin with, partial compositeness of SM {\em gauge} bosons is similar to $\gamma-\rho$ mixing in 
$\textrm{QCD}+\textrm{QED}$, and this idea can be suitably extended to fermions. Moreover, it is noteworthy that decays corresponding to all three channels for gauge KK particles outlined above are actually present in 
$\textrm{QCD}+\textrm{QED}$, namely, $\rho \rightarrow \pi \pi$, $\pi \gamma$, and $e^+ e^-$. These three modes are arranged here in decreasing order of branching fractions (or equivalently, strength of couplings). The coupling among all three composites is the largest; 
%
%
as we go down the list, a 
%
%
reduction in strength arises,
due to replacement of a composite particle by an elementary one. 
In more detail, in this analogy, Higgs/top is ``like" $\pi$ (i.e., composites which are massless -- in some limit). Indeed, usually the leading decay channel in warped models for KK gauge is precisely to top/Higgs, which maps onto $\rho \rightarrow \pi \pi$. However, as already mentioned above, in the extended 5D framework presented here, this mode is suppressed. The introduction of intermediate brane(s) correspond to a {\em sequence} of confinement scales in the strong dynamics picture (see Ref.~\cite{Agashe:2016rle} for details). In other words, the above coupling becomes
small by virtue of Higgs/top quark being composite at higher scale than the lightest spin-1 states: once again, we see that this is not such a radically different possibility.
Moving onto the next-in-line channel, 
%
$\rho \rightarrow \pi \gamma$, followed by $\pi \rightarrow \gamma \gamma$,
can be understood as analogous to the cascade decay of gauge KK modes, 
%
%
where $\pi$ ``mimics" radion/dilaton (only in the sense that both are composite and neutral under external gauging), and elementary/external gauge boson, i.e., $\gamma$, in $\textrm{QCD}+\textrm{QED}$ 
case stands for {\em all} SM gauge bosons in the warped/composite Higgs model.
Finally, the $\rho\:e^+\:e^-$ coupling drives the decay with the smallest branching ratio (BR) in the real world, and is also relevant for production of $\rho$-meson at $e^+ e^-$ colliders.
With the (elementary) electron being like all (light) SM fermions in the warped/composite Higgs model, this matches the flavor-universal part of the coupling of the gauge KK to SM fermions; again, this mediates a subdominant decay for the gauge KK, but is crucial, both in the standard and the extended framework, for production at {\em both} hadron and lepton colliders.
These latter two couplings are {\em not} directly related to Higgs/top compositeness, thus are similar in the standard and extended setups (as already mentioned in the 5D picture).

We also would like to emphasize here that, although our study is rather specific to the warped/composite Higgs model, the event topology of interest might actually arise in other situations as well. In fact, we would like to mention that our modelling of this decay channel has enough number of independent parameters (for example, roughly one per coupling) so that it can be readily adapted to a more general case.
More importantly, 
%
%
%
we think that 
%
%
such a channel (i.e., of a heavy particle decaying into SM gauge boson plus another -- possibly different - pair of SM gauge bosons from the decay of an intermediary, {\em on}-shell particle)
has not 
received much attention 
%
%
%
(phenomenologically or experimentally) in the past.\footnote{See, however, Ref.~\cite{Agashe:2014wba} for an analysis of a photo-cascade decay of KK {\em graviton} and Ref.~\cite{Bini:2011zb} for KK gluon in
the standard warped model.}
%
%

Nevertheless, some related analysis of experimental data has been performed, which is worth mentioning here.
First one is the resonant channel search such as a single jet plus a photon (from an excited quark, for example, Ref.~\cite{CMS:2016qtb}): this does apply to our case, but only when the radion is very light, thus highly boosted so that the two jets from its decay merge.
On the other hand, searches for dijet resonances produced in association with photon/jet (mostly originating from ISR) have been performed~\cite{ATLAS:2016bvn}, ISR jet/photon here being used for the purpose of tagging to reduce background, especially in the context of looking for {\em low} mass dijet resonances. In this case, there was clearly no reason to simultaneously study the three-particle invariant mass (i.e., dijet $+$ photon/jet). However in our case, it is crucial in reducing background.
Finally, there is a ``general" search performed by the ATLAS Collaboration~\cite{ATLAS:2014sxa}, where invariant mass distributions of various final states (involving combinations of SM objects such as photons, jets and leptons) were studied for possible excesses relative to the SM predictions. The channels studied by the ATLAS Collaboration include some of the three-particle ones found in the new decay channel in our extended warped/composite Higgs model such as dijet $+$ photon. 
However, the invariant masses of a subset of two particles therein were not considered at the same time, presumably for simplicity.
Crucially enough, 
%
%
the striking feature about the new channel that we study here is that the final state features {\em both} three-particle (i.e., KK/composite gauge boson) and two-particle (i.e., radion/dilaton) resonances.
%
%

In order to serve the motivation for this work which was elaborated above, we organize the rest of this paper as follows.
We begin with a rather detailed review on the new framework in Sec.~\ref{sec:review}, including the mass spectrum of relevant particles and their couplings in terms of model parameters.
In Sec.~\ref{Sec:overview}, we take the simplified model for our phenomenological study and provide the allowed parameter space consistent with the existing bounds. 
An overview of the various signal channels that we shall study follows, especially in the sense of their production and decay rates, guiding us to establishing our benchmark points. In Sec.~\ref{EventSimulation}, we then discuss general details of our event simulation and key mass variables for our collider study. Sec.~\ref{results} is reserved for presenting our main results from the data analyses in various signal channels. 
Finally, we summarize and conclude in Sec.~\ref{sec:conclusion}, together with a brief discussion on some potential future work.


\section{Review on the Model \label{sec:review} }

In this section, we review a natural extension of the ``standard'' Randall-Sundrum framework introduced in Ref.~\cite{Agashe:2016rle}. 
We begin
a brief discussion on the motivation for such an extension 
and 4D dual picture in the next subsection. 
We then move our focus onto the mass spectrum of relevant particles and their interactions in detail, providing the corresponding explicit formulae.

\subsection{Motivation for a natural extension: 5D and 4D-dual pictures \label{subsec:natural_extension_5D} }

As discussed in the introductory section, the stringent constraints from 
%
%
flavor/CP
experiments 
%
%
push the IR-brane scale of the ``standard'' RS framework to $\gtrsim \mathcal{O}(10)$ TeV. 
%
%
This bound
implies that the new particles in this framework, i.e., the KK excitations of the SM, might be beyond LHC reach.
%
%
%
%
This situation suggests
%
%
we should 
speculate about other logical possibilities
%
%
within this broad framework 
and study its phenomenological consequences
thoroughly, in particular, in order to see if LHC signals are possible therein.
Indeed, Ref.~\cite{Agashe:2016rle} has pointed out a simple but robust observation along this line: different fields in 5D can propagate different amounts into the IR along the extra dimensions.
\begin{itemize}
\item The {\bf gravity} itself is the dynamics of the spacetime, and therefore, the 5D gravity field should be present in the entire 5D spacetime in the form of 5D Einstein gravity. 
\item The {\bf gauge} fields, however, can propagate into a less extent than the gravity simply because pure gravity theory may stand alone without gauge fields, but not in the opposite fashion. Therefore, the ordering between the gravity and the gauge fields is not random but fixed as described here.
\item Analogously, the {\bf matter} fields can exist in an even smaller amount of 5D than the gauge fields. The reason is that any matter field charged under a certain gauge field can emit the associated gauge field, enforcing the presence of the gauge field wherever matter fields exist.\footnote{
%
%
Of course, for any gauge fields under which the SM matter fields are not charged (if existed), this argument does not directly apply and the fraction in the extra dimensions that they occupy is rather free of constraint.}
\end{itemize}
Based on the above-listed observation, the possibility of letting different fields propagate modestly different degrees into the IR of the warped dimension is not only {\it robust} but {\it natural}.
%
%
%
A concrete realization of the idea is to introduce extra branes relative to the set-up in Fig.~\ref{fig:standard}.
%
%
As an example of minimal extensions, Fig.~\ref{fig:extended} schematically displays the configuration in which 
gravity and gauge fields propagate the same amount along the fifth direction while matter fields are present in a smaller amount. 
%
%
It is straightforward to see that within this generalized framework, the ``standard'' RS setup is merely a special case with the last two branes (i.e., the Higgs and IR branes) in Fig.~\ref{fig:extended} identified.
From now on, we shall focus on this setup for concreteness of our discussion.



In the language of the 4D-dual picture, the above extension can be understood as follows.
In the far UV, the physics is strongly coupled dynamics of preons with conformal invariance. 
This conformally invariant ``UV strong dynamics'' is deformed by some explicit breaking term(s), and as a result the theory runs until it undergoes a confinement at, say, $\Lambda_{\rm Higgs}$. 
Composite hadrons and mesons are ``born'' at this stage and SM top quark and Higgs are part of such massless 
composite states, whereas the massive states correspond to KK fermions of the 5D model. 
This confinement scale is dual to the position of brane in the warped fifth dimension where top and Higgs are localized (i.e., the Higgs brane in Fig.~\ref{fig:extended}). 
Unlike QCD-like strong dynamics, however, this confinement can also produce composite preons;  the resulting theory flows to a new fixed point in the farther IR. 
%
%
%
In addition, 
the physics at $\Lambda_{\rm Higgs }$
may also produce deformation terms to the CFT of the composite preons, including couplings between composite preons and composite hadrons.
Thus, this ``IR strong dynamics'' runs as before until it confronts the second confinement at $\Lambda_{\rm IR}$ which is dual to the position of the IR brane. 
This second confinement then creates
its own composite mesons and glueballs. 
However, these composite states do not
possess the quantum numbers of the SM matter, although they might have SM gauge charges. 
Composite vector mesons 
resulting from
this second confinement are dual to KK-excited gauge bosons, and the dilaton, a pseudo Nambu-Goldstone boson of the spontaneously broken scale invariance, is dual to the radion. 
%
Due to this duality, we refer to these particles as dilaton/radion and composite mesons/KK gauge bosons interchangeably throughout this section.

\subsection{Mass spectrum and couplings}
\label{subsec:spectrum_couplings}

With the model setup delineated in the preceding section in mind, we now consider the mass spectrum of the radion and the lightest KK gauge bosons in terms of model parameters.\footnote{See also Ref.~\cite{Agashe:2016rle} for more detailed derivations and dedicated discussions.}
The discussion on the couplings relevant to our study follows. In particular, we shall demonstrate that light states below $\Lambda_{\rm Higgs}$, e.g., spin-0 glueball (dual to radion) and spin-1 mesons (dual to KK gauge bosons), interact with SM matter fields dominantly via flavor-blind couplings, from which we find interesting and important phenomenology.

\subsubsection{Radion} 
\label{subsubsec:radion}

First of all, the mass of the dilaton $m_{\varphi}$~\cite{Konstandin:2010cd,Eshel:2011wz,Chacko:2012sy,Chacko:2013dra} is given by
\bea
m^2_{ \varphi } & \sim & \epsilon \; \lambda \; \Lambda_{ \rm IR }^2\,, 
\label{radion_mass}
\eea
where $\lambda$ is dual to the amount of detuning of the IR brane tension in 5D and $\epsilon$ denotes the parameter encoding the ratio between the first and the second confinement scales~\cite{Agashe:2016rle}.
Their typical sizes are
\bea
\lambda < 1\,, \,\,\, \epsilon \sim  \frac{1}{\log(\Lambda_{\rm Higgs} / \Lambda_{\rm IR})} < 1\,,
\eea
from which we find
that the mass of the dilaton is generally lighter than that of spin-1 resonances ($\sim \Lambda_{ \rm IR }$)
which opens up the decay mode of a spin-1 resonance into a dilaton along with an associated SM gauge boson.

\paragraph*{Coupling to SM gauge bosons (flavor-blind):}

One can derive the coupling of the dilaton to a pair of SM gauge bosons, considering the running of the SM gauge coupling and using the fact that dilaton is the Goldstone boson that parameterizes the fluctuation of $\Lambda_{\rm IR}$. 
The final form of the coupling~\cite{Chacko:2012sy,Csaki:2007ns,Chacko:2014pqa} is 
\bea
\delta {\cal L} & \sim & \left( \frac{  g_{ \rm SM } } { g^{ \rm gauge }_{ \star \; \rm IR } } \right)^2 \frac{ g^{ \rm grav }_{ \star } }{ \Lambda_{ \rm IR }}  \varphi A_{ \mu \nu} A^{ \mu \nu } 
\label{coupling_dilaton_SM_gauge}
\eea
where $g_{\rm SM}$ is the usual SM gauge coupling associated with the gauge field strength tensor $A_{\mu\nu}$ for which the gauge indices are suppressed for notational brevity.
The stared quantities $g^{ \rm grav }_{ \star }$  and $g^{ \rm gauge }_{ \star \; \rm IR }$ parameterize the cubic couplings of the IR strong dynamics with at least one composite state being of spin-2 and spin-1, correspondingly. 
Denoting $N_{\rm strong}$ as the number of ``color'' charges of strong dynamics, we remark that
in the large-$N_{\rm strong}$ limit,
these composite cubic couplings generically have the size of
\bea
g_{\star} \sim \frac{4\pi}{\sqrt{N_{\rm strong}}}. \\ \nonumber
\label{eq:g_star_large_N}
\eea

\paragraph*{Coupling to top/Higgs (flavor non-universal):} 
Since the radion is localized near the IR brane in the minimal RS setup, it predominantly decays into the pairs of top quark, Higgs, and longitudinal modes of $W/Z$ gauge bosons (through the Goldstone equivalence theorem).
In particular, the decay rate of the radion in a pair of SM gauge bosons via the coupling in~\eqref{coupling_dilaton_SM_gauge} is negligible.
However, in the extended framework, the Higgs brane is {\it de}localized from the IR brane, 
and as a consequence, the radion has a small overlap with top quark or Higgs in their 5D profiles, 
yielding a reduced coupling to top quark or Higgs as follows~\cite{Agashe:2016rle}:
\bea
\delta {\cal L} \left( \Lambda_{ \rm IR } \right) \sim 
%
%
\left( \frac{ \Lambda_{ \rm IR } }{ \Lambda_{ \rm Higgs } } \right)^{ 4 - \epsilon }
\frac{  g^{ \rm grav}_{ \star \; \rm IR }}{ \Lambda_{ \rm IR } } \varphi \Big[  m_t \bar{t} t + \left( \partial_{ \mu } H \right)^{\dagger} \partial^{ \mu } H  \Big]\,.
\label{radion_topHiggs}
\eea
%
%
%
As we will discuss in more detail later, we will (roughly) choose $\Lambda_{ \rm IR }$ a couple of TeV,
whereas $\Lambda_{ \rm Higgs } \gtrsim \mathcal{O}(10)$ TeV and $g^{ \rm gauge }_{ \star \; \rm IR }$ of a few.
With these parameters, we see that the couplings of the radion to top quark (first term) and Higgs (second term) in~\eqref{radion_topHiggs}
are (highly) suppressed as compared to
the coupling of radion to SM gauge bosons in~\eqref{coupling_dilaton_SM_gauge}.
%
%
Thus, an interesting phenomenological implication is that the branching fractions of the radion into SM gauge boson pairs become sizable, playing an important role in our collider study.


\subsubsection{KK gauge boson}
\label{subsubsec:KK_gauge_boson}

As mentioned before, the mass scale of the spin-1 resonance (henceforth represented by $\rho$), which is dual to the KK gauge boson, is simply given by
\bea
m_{\kk} \sim \Lambda_{{\rm IR}}.
\eea

\begin{figure}
\center

\includegraphics[height=40mm]{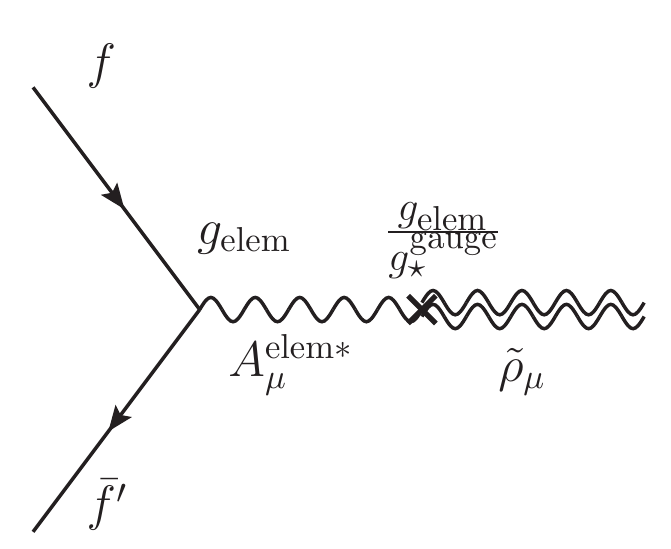}

\caption{Flavor-universal coupling of spin-1 composite states to (light) SM fermions via an elementary-composite mixing. $A_{\mu}^{{\rm elem} \star}$ and $\tilde{\rho}_{\mu}$ denote the (virtual) elementary and composite states before the mixing. $f$ and $f'$ denote SM fermions.}
\label{fig:photon_rho_mixing}
\end{figure}

\paragraph*{Coupling to SM matter (flavor-universal):} 
The flavor universal couplings of $\rho$ to SM fermions and Higgs are given by the famous $\gamma - \rho$ mixing mechanism 
observed in $\textrm{QCD} + \textrm{QED}$ system, which we summarize as follows.
When the strong sector (QCD) is confined and produces hadrons, there exists a vector meson which has the same quantum number as the elementary gauge boson in QED due to the fact that the external or elementary gauge symmetry gauges subgroups of its global symmetries.
Therefore, there arises a mixing between the vector meson $\rho$ and the corresponding elementary gauge boson $\gamma$.
This mixing induces the breakdown of the elementary gauge symmetry in such a way that a certain linear combination between $\rho$ and $\gamma$ remains massless and the associated unbroken symmetry is interpreted as the SM gauge symmetry. 
Physical mass eigenstates are admixture of composite and elementary states and their mixing angle $\theta$~\cite{Agashe:2016rle} is simply given by 
\bea
\sin \theta  =  \frac{ g_{ \rm elem } }{ \sqrt{ g_{ \rm elem }^2 + \left(g^{ \rm gauge }_{ \star }\right)^2 } } \sim \frac{g_{ \rm elem }}{g^{ \rm gauge }_{ \star }}\,,
\label{eq:rho-photon_mixing_angle}
\eea
where $g_{ \rm elem }$ and $g^{ \rm gauge}_{ \star }$ are gauge couplings of elementary and strong sectors, respectively. 
The interaction between the composite state and 
%
%
{\em all} SM fermions via the mixing is shown in Fig.~\ref{fig:photon_rho_mixing}, wherein $A_{\mu}^{{\rm elem}}$ and $\tilde{\rho}_{\mu}$ denote the elementary and composite states before the mixing. 
Using the mixing angle given above, we write the coupling between them~\cite{Agashe:2016rle} as
%
\bea
\delta \mathcal{L} \sim \frac{g_{\rm SM}^2}{g^{ \rm gauge }_{ \star }}   \rho^{\mu} \bar{f} \gamma_{\mu} f
\label{eq:rho_SM_fermion_coupling}
\eea
where 
%
%
%
$f$
%
%
represents the 
%
%
SM fermions and we used the relation
\bea
g_{ \rm SM } = \frac{ g_{ \star }^{ \rm gauge } g_{ \rm elem } }{ \sqrt{ g_{ \rm elem }^2 + g^{ \rm gauge \; 2 }_{ \star } } } \approx g_{ \rm elem }.
\eea
%
%
In addition to the above flavor-{\em universal} coupling, there is a non-universal part which 
is significant only for top/Higgs: we discuss this effect next.
%

\paragraph*{Coupling to top/Higgs (flavor-non-universal):}
The profile of KK gauge bosons is localized near the IR brane, implying that its value at the Higgs brane is suppressed, accordingly.
An explicit calculation shows (either a 5D or 4D analysis) that the flavor {\em non}-universal part of the coupling of gauge KK to top/Higgs can be expressed as follows~\cite{Agashe:2016rle}:
\bea
\delta {\cal L} \left( \Lambda_{ \rm IR } \right) & \sim & \frac{ \left( g^{ \rm gauge }_{ \star \; \rm UV } \right)^2 }{ g^{ \rm gauge}_{ \star \; \rm IR} } 
\left( \frac{ \Lambda_{ \rm IR } }{ \Lambda_{ \rm Higgs } } \right)^2 \rho^{ \mu }
  \left( \bar{t} \gamma_{ \mu } t + H^{ \dagger } D_{ \mu } H \right)\,, 
\label{gauge_topHiggs}
\eea
where $g^{ \rm gauge}_{ \star \; \rm UV}$ and $g^{ \rm gauge}_{ \star \; \rm IR}$ are the composite gauge couplings of UV and IR strong dynamics, correspondingly. 
Again $\rho^{\mu}$ represents a composite state obtained by the confinement of IR strong dynamics.
The size of this coupling 
%
%
depends on the position of the Higgs brane relative to the IR brane as encoded in the factor $\left(\Lambda_{ \rm IR } /\Lambda_{ \rm Higgs }  \right)^2$.
 
Ref.~\cite{Agashe:2016rle} has extensively discussed the significance of this coupling and the resultant, (potentially) striking phenomenology.
%
%
An interesting possibility is that this flavor-non-universal coupling is
%
%
comparable to the flavor-universal in~(\ref{eq:rho_SM_fermion_coupling}).
This happens 
in the case of KK gluon (KK $Z$) for $\Lambda_{ \rm Higgs } \sim 10 \; (15)$ TeV;
%
%
%
remarkably this value of the top/Higgs compositeness scale is (roughly) the flavor/CP bound on the KK scale!
If KK gauge bosons (e.g., KK gluon and KK $Z$) are discovered at the LHC,
%
%
their decay branching fractions would show $\mathcal{O}(1)$ deviation from those in 
%
%
the flavor-blind limit ($\Lambda_{\rm Higgs} \to \infty$), i.e., when we only have the couplings in~(\ref{eq:rho_SM_fermion_coupling}).
At the same time, these are significantly different than the standard warped model, which corresponds to the limit
$\Lambda_{\rm Higgs} \to \Lambda_{\rm IR}$, i.e.,~\eqref{gauge_topHiggs} dominates over~\eqref{eq:rho_SM_fermion_coupling}, so that gauge KK modes decay mostly into $t\bar{t}/HH$ final state.
In other words, the LHC may be sensitive to the top/Higgs compositeness scale, a striking signature for composite physics as a solution to the gauge hierarchy problem. 
%
%
In our current study, we shall demonstrate another possibility, namely, a
cascade decay of KK gauge bosons: while this will not {\it per se} be a probe of top/Higgs compositeness (cf.~above idea), 
it nevertheless is very exciting since it is quite 
different from the ``vanilla'' decay of gauge KK modes into pairs of SM fermions/Higgs.
Furthermore, we will see that these two 
%
%
signals 
are interestingly
independent, i.e., this new channel exists
no matter $\Lambda_{ \rm Higgs} \sim \mathcal{O}(10)$ TeV or much higher (in the latter case, the above probe of top/Higgs compositeness
obviously
fades away).
%
%

%

\paragraph*{Coupling to radion and SM gauge bosons (flavor-blind):}

The interaction among KK gauge boson-radion-SM gauge boson arising as a consequence of radius stabilization was discussed in Ref.~\cite{Agashe:2016rle}. 
%
%
The relevant coupling is given by
\bea
\delta {\cal L } \left( \Lambda_{ \rm IR } \right) & \sim &  \epsilon\,\lambda\, g^{ \rm grav }_{ \star \; \rm IR} \left( \frac{ \Lambda_{ \rm IR } }{ \Lambda_{ \rm Higgs } } \right)^{ - \epsilon }
\frac{ g_{ \rm elem } } { g^{ \rm gauge }_{ \star \; \rm IR } } \frac{\varphi}{\Lambda_{\rm IR}} \rho_{ \mu \nu } A^{ \mu \nu }\,,
\label{dilaton_rho_photon}
\eea
where $\rho^{\mu\nu}$ is the field strength tensor for the spin-1 composite field $\rho^{\mu}$. 
As mentioned earlier, $\epsilon \sim 1 / \log \left( \Lambda_{ \rm Higgs } / \Lambda_{ \rm IR } \right) \sim 1 / ({\rm a\, few})$, thus we find that
$\left( \Lambda_{ \rm IR } / \Lambda_{ \rm Higgs } \right)^{ - \epsilon }$ is an $\mathcal{O}(1)$ factor. 
%
%
This implies that the KK gauge boson-radion-SM gauge boson coupling 
can be (roughly) comparable to the flavor-universal coupling of the KK gauge boson to SM fermions 
in~\eqref{eq:rho_SM_fermion_coupling}
(in turn, the latter
%
%
is comparable to/larger than the non-universal one 
for $\Lambda_{ \rm Higgs } \gtrsim \mathcal{O}(10)$ TeV).
In Ref.~\cite{Agashe:2016rle}, as mentioned above, the focus was on probing top/Higgs compositeness so that,
for simplicity, in the analysis there 
it was assumed that we live in 
the part of parameter space where the new decay channel is smaller (and hence was neglected in the BR's shown),
for example, small $g^{ \rm grav }_{ \star \; \rm IR }$
and/or $\epsilon$ in eq.~(\ref{dilaton_rho_photon}).
While
here, we 
%
%
choose the 
{\em another} part of
parameter space where 
%
%
the branching ratio of the KK gauge boson decay into a radion and the corresponding SM gauge boson can be substantial, 
even dominating over pair of SM fermions.
%
%
Furthermore, as we discussed earlier, the radion, in turn, decays predominantly into a pair of SM gauge bosons. 
We emphasize that in the standard warped model,
%
%
although both the interaction vertices involved in the above new decay channel are present,
%
%
both KK gauge bosons and radion have overwhelming decay rates into top/Higgs final states,
leaving a little chance for the above novel channels to be probable at the LHC. 


\section{Overview of LHC Signals}

\label{Sec:overview}


As we reviewed in Sec.~\ref{sec:review}, the extended warped extra-dimensional framework proposed in Ref.~\cite{Agashe:2016rle} renders significant branching ratios for (i) the decay of KK gauge bosons to radion and the corresponding SM gauge boson and (ii) the decay of radion to a pair of SM gauge bosons. The {\em combination} of these two features creates a very novel search channel for KK gauge bosons and radion. 
Namely, the LHC can produce on-shell KK gauge bosons 
via the same, i.e., flavor-universal, coupling to light quarks as in the standard model.
These
heavy particles
subsequently decay into a radion and a corresponding SM gauge boson,
followed by the radion decay into a pair of SM gauge boson. 
This 
offers final states containing various combinations of three SM gauge bosons from decays of \emph{two resonances}: KK gauge boson and radion. 
Fig.~\ref{fig:KKgauge-radion-SMgauge} displays the decay topology associated with various signal channels.
When it comes to the study on collider signatures, instead of working with a full 5D warped extra-dimensional model or its 4D dual theory, it is much more convenient to conduct the study with a simplified model containing only relevant particles and parameters. Therefore, we first construct the simplified model for our phenomenological study in the next section,
and then discuss the production and decays of all types of KK gauge bosons and radion together with current bounds.
We finally close this section by identifying relevant parameter space for our study and choosing the benchmark points for various channels.

\subsection{Simplified model and allowed parameter space}
\label{SubSec:Simplifedmodel}


We now describe a simplified model on which our collider analyses in Sec.~\ref{results} are based, presenting the relevant particles and their interactions.
%
%
%
%
The notation for the particles (and their masses and couplings) that we will set-up in this section (and which is to be used for rest of the paper) 
is somewhat 
{\em different} than in the earlier section.
However, (as much as is possible) we will try to provide a correspondence between the two sets: 
the one we develop in this section is more convenient for phenomenological studies, whereas the previous might be better suited for a more theoretical discussion.
%
The simplified-model approach 
%
%
also
allows enough generality to encompass a broad class of models 
which could accommodate the same signatures.

\begin{table}[t]
\centering
\begin{tabular}{c c c}
\hline
 & Name & Mass \\
\hline
KK gauge bosons & $A_\kk$ & $m_\kk$ \\
KK photon & $\gamma_\kk$ & $m_{\gamma_\kk}$ \\
KK $W$ gauge boson & $W_\kk$ & $m_{W_\kk}$ \\
KK $Z$ gauge boson & $Z_\kk$ & $m_{Z_\kk}$ \\
KK gluon & $g_\kk$ & $m_{g_\kk}$ \\
radion & $\varphi$ & $m_{\varphi}$ \\
\hline 
\end{tabular}
\caption{\label{tab:particlenotation} Notation of names and mass parameters for new physics particles.}
\end{table}

Relevant particles in our study include four types of (lightest) KK gauge bosons $A_{\kk}=\lbrace \gamma_\kk, W_\kk, Z_\kk, g_\kk \rbrace$,\footnote{Here we assume the masses of electroweak KK gauge bosons are degenerate.} their zero-mode SM gauge bosons $A=\lbrace \gamma, W, Z, g \rbrace$, radion $\varphi$, and (light) SM fermions $\psi$. 
For convenience, we tabulate the symbols for new physics particles together with their respective mass parameters in Table~\ref{tab:particlenotation}. 
%
%
%
%

%
%
We now comment on the choice of $\Lambda_{ \rm Higgs }$ that we will make in our subsequent analysis.
As mentioned just above, the motivation in our paper 
%
%
%
is
different from that in Ref.~\cite{Agashe:2016rle}, where the idea was to obtain signals for top/Higgs compositeness, 
thus the cascade decay channel was 
%
%
neglected.
%
%
Namely, we are {\em now} precisely interested in the new decay channel.
So, for simplicity, here we will instead neglect the top/Higgs {\em non}-universal coupling (which drove sensitivity to 
top/Higgs compositeness) by (formally) setting $\Lambda_{\rm Higgs} \to \infty$; we are then left with only the flavor-universal coupling
of gauge KK modes to pair of SM fermions/Higgs.
Note that, as discussed above, the non-universal can at most be as large as universal, as long as $\gtrsim \mathcal{O}(10)$ TeV (flavor bound) so that, in reality (i.e., if we assume $\Lambda_{ \rm Higgs }$ finite), it will be at most $\mathcal{O}(1)$ effect on
our signal.
We re-iterate that 
the decay rates of KK gauge bosons into top/Higgs pairs are much smaller than those in the standard 
%
%
warped 
model (where the non-universal coupling to top/Higgs dominates over all others).

\begin{figure}
\center

\includegraphics[width=0.5\linewidth]{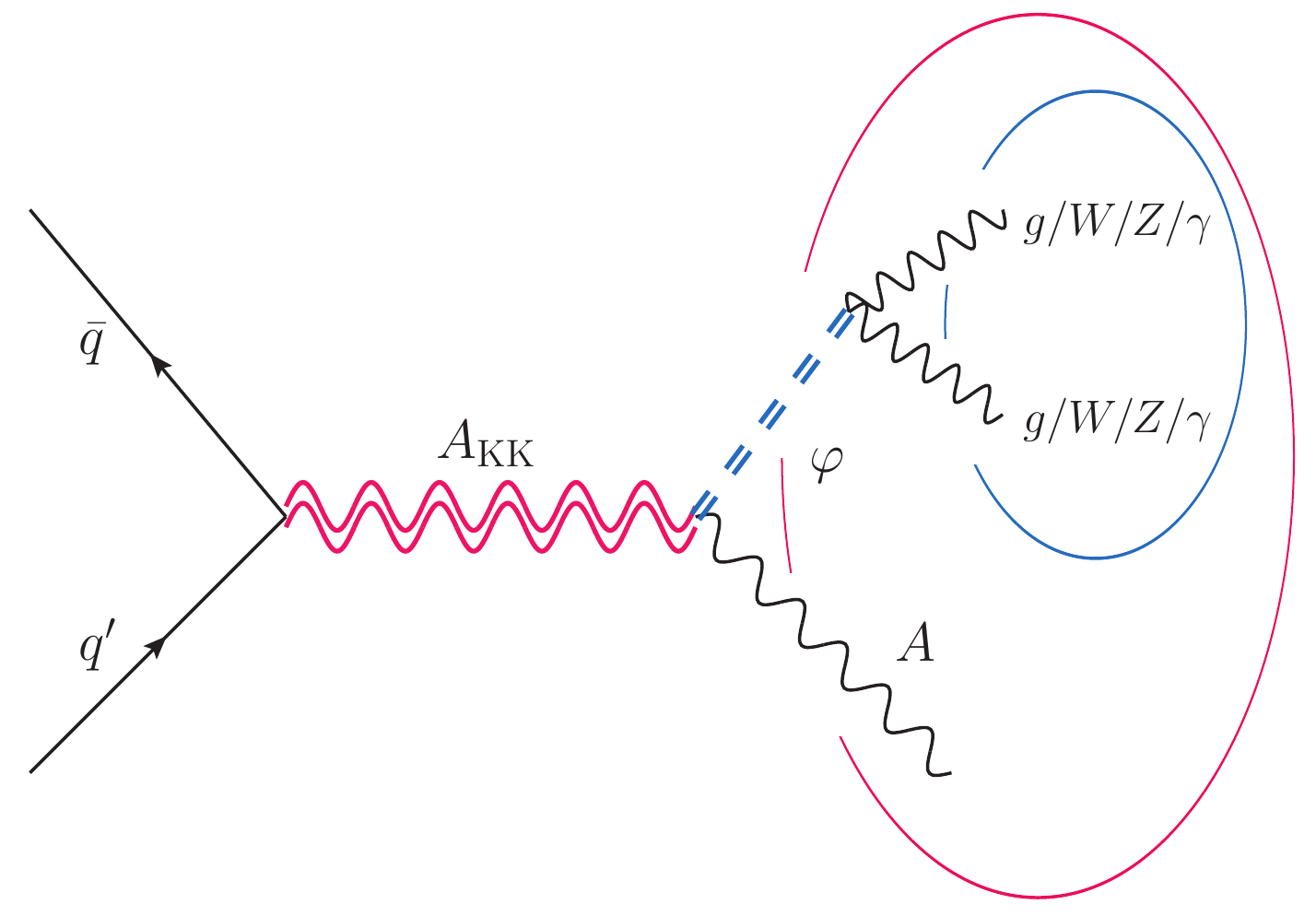}

\caption{ Feynman diagram for the signal process. Double (single) lines represent composite (SM elementary) particles and $q / q'$ denote light quarks inside the proton.
The signal process is characterized by two resonance bumps illustrated by blue and red circles.
}
\label{fig:KKgauge-radion-SMgauge}
\end{figure}

Three types of (new) couplings are relevant in the signal processes as clear from Fig.~\ref{fig:KKgauge-radion-SMgauge}:
(1) KK gauge bosons coupling to SM quarks, (2) KK gauge boson-radion-SM gauge boson coupling, and (3) radion coupling to a pair of SM gauge bosons. 
First of all, KK gauge boson coupling to SM quarks 
has the form:
\bea
\delta \mathcal{L}_{(1)} = Q^2_A \frac{g^2_{A}}{g_{A_{\rm KK}}}A_{\kk}^{\mu} \bar{\psi} \gamma_\mu \psi
 \label{Eqn:KKgauge-fermion-coupling}
\eea
where $g_A$ and $g_{A_{\rm KK}}$ are SM and KK gauge couplings for respective gauge bosons $A$ and $A_{\rm KK}$.
Here $Q_A$ denotes SM $A$-gauge charge of the SM fermion $\psi$. 
One can easily notice that this coupling is nothing but the expression in~\eqref{eq:rho_SM_fermion_coupling}, but with change of notation from $g_\star^{\rm gauge}$ to $g_{A_{\rm KK}}$. 
Second, KK gauge boson-radion-SM gauge boson coupling is of the form:
\bea
\delta \mathcal{L}_{(2)}=\epsilon g_{\rm grav}\frac{g_{A}}{g_{A_ {\rm KK}}}\frac{\varphi}{m_{\rm KK}}A_{\mu\nu}A_{\textrm{ KK}}^{\mu\nu} ,
\label{Eqn:KKgauge-radion-SMgauge-coupling}
\eea
where $g_{\rm grav}$ is the KK gravity coupling and $m_{\rm KK}$ is the mass of KK gauge boson (or equivalently, KK scale). 
$A_{\textrm{ KK}}^{ \mu\nu}$ is the field strength tensor for the KK gauge boson $A_{\textrm{ KK}}$. 
This coupling is just a rewriting of~\eqref{dilaton_rho_photon}, but with $\Lambda_{\rm IR}$ identified as $m_\kk$ and $\mathcal{O}(1)$ factors like $\lambda$ and $\left( \Lambda_{ \rm IR } / \Lambda_{ \rm Higgs } \right)^{ - \epsilon }$ dropped from it.
One can interpret that other parameters like $\epsilon$ and $g_{\rm grav}$ absorb those $\mathcal{O}(1)$ factors and get redefined. 
Finally, the radion coupling to a pair of SM gauge bosons has the structure of
\bea
\delta \mathcal{L}_{(3)}=-\frac{1}{4}\left( \frac{g_A}{g_{A_{\rm KK}}} \right )^2 \frac{g_{\rm grav}}{m_{\rm KK}} \varphi A_{\mu \nu}A^{ \mu \nu}\,,
\label{Eqn:Radion-SMgauge-coupling}
\eea
where again $m_\kk$ corresponds to $\Lambda_{\rm IR}$.
This coupling structure obviously originates from~\eqref{coupling_dilaton_SM_gauge}, while the prefactor $-1/4$ comes from the normalization of gauge kinetic terms.
We will simply neglect the coupling of radion to top/Higgs in~\eqref{radion_topHiggs}, 
just like we did above for gauge KK couplings. We are now about to detail the scheme of scanning the above parameter space in order to obtain the allowed region therein.
%
%



\paragraph*{KK gauge and KK gravity couplings:}
Although there are four KK gauge couplings ($g_{\gamma_{\rm KK}}$, $g_{W_{\rm KK}}$, $g_{Z_{\rm KK}}$, and $g_{g_{\rm KK}}$) under consideration,
just like in the SM, only three of them are independent, which are $g_{g_{\rm KK}}$, $g_{W_{\rm KK}}$, and $g_{B_{\rm KK}}$. The KK gauge couplings of $\gamma_{\rm KK}$ and $Z_{\rm KK}$ are obtained via well-known relations 
\bea
g_{\gamma_{\rm KK}} = \frac{g_{W_{\rm KK}}g_{B_{\rm KK}}}{\sqrt{g_{W_{\rm KK}}^2+g_{B_{\rm KK}}^2}}\,,~~~~ g_{Z_{\rm KK}} = \sqrt{g_{W_{\rm KK}}^2+g_{B_{\rm KK}}^2}.
\label{eq:g_gamma_g_Z_from_others}
\eea
Although perturbativity in 5D warped models demands $g_{g/W/B_{\rm KK}} \lesssim 3$ \cite{Agashe:2016rle}, in this simplified model approach, we allow those KK couplings to be larger. 
This way, we can explore broader parameter space, even covering the possibility that some strongly-coupled 4D theories might be realized in some parameter space without obvious 5D dual. 
%
%
%
However, 
reasonably requiring $N \gtrsim {\rm (a\, few)}$ in the relation  
$g_{A_{\rm KK}}\sim 4\pi/\sqrt{N_\textrm{strong}}$
%
%
does 
set a rough upper limit on $g_{g/W/B_{\rm KK}} $ to be around 6.\footnote{Note that this is also roughly the size of
$\rho\:\pi\:\pi$ coupling in QCD.}
On the other hand, a
lower limit for gauge KK coupling arises from requiring that the Landau pole scale is higher than GUT scale and comes out to be 3. 
Therefore, the allowed ranges for KK gauge couplings are
\bea
3 \lesssim g_{g_{\rm KK}}\,,g_{W_{\rm KK}}\,,g_{B_{\rm KK}} \lesssim 6, \label{eq:paramranges}
\eea
from which we deduce the constraints for $g_{\gamma_{\rm KK}}$ and $g_{Z_{\rm KK}}$ in conjunction with the relation~\eqref{eq:g_gamma_g_Z_from_others}.

Similarly to the case of KK gauge couplings, the KK gravity coupling has the upper limit around 6. However, since there is no Landau pole issue in gravity sector, KK gravity coupling is unbounded below although too small $g_{\rm grav}$, which implies too large $N_{\rm strong}$, may not be reasonable. Hence, the allowed KK gravity coupling is given by $\mathcal{O}(1) \lesssim g_{\rm grav} \lesssim 6$. \\

\paragraph*{KK gauge boson and radion masses:}


Ongoing experimental effort on various resonance searches constrain the masses for KK gauge bosons. 
We shall discuss the associated bounds in Sec.~\ref{Sec:KKgauge_prod_decay} in detail.
%
%
We choose $m_{\rm KK}$ to be somewhat heavier than the current bound: in most channels $m_\kk = 3$ TeV.
When it comes to the radion mass, the diphoton resonance search mainly constrains it:
we consider both $m_\varphi = 1$ TeV and $1.5$ TeV. \\

\paragraph*{Parameter $\epsilon$:}


%
%
The $\epsilon$ parameter appears in the radion mass, 
where its effect can be ``compensated" by the detuning parameter $\lambda$.
Its only other appearance is in the KK gauge-radion-SM gauge coupling (see~\eqref{Eqn:KKgauge-radion-SMgauge-coupling}), i.e., our signal channel;
%
in particular, 
%
%
this 
means that this parameter is not constrained by experimental bounds. 
Generically, $\epsilon$ needs to be $\mathcal{O}$(1/ a few) in order for the hierarchy $\Lambda_{\rm Higgs} / \Lambda_{\rm IR}$ to be stabilized. 
%
%
As is evident from eq.~\eqref{Eqn:KKgauge-radion-SMgauge-coupling}, taking larger value of $\epsilon$ enhances the signal cross section, so for our benchmark points, we set $\epsilon$ to be 0.5 in this study.

\subsection{Radion direct production, decay, and current bounds}
\label{SubSec:Radion_prod_decay}


Radion is produced at the LHC via gluon fusion using flavor-universal coupling in~\eqref{Eqn:Radion-SMgauge-coupling}. The same interaction vertices are responsible for its dominant decays to a pair of SM gauge bosons $gg$, $WW$, $ZZ$, and $\gamma \gamma$. To leading order, the radion decay width is given by
\bea
\Gamma(\varphi \to A A )=N_A g^2_{\rm grav} \left( \frac{g_{A}}{ g_{A_{\rm KK}}} \right )^4\left(\frac{m_{\varphi}}{m_{\rm KK}}\right )^2 \frac{m_\varphi}{64 \pi} 
\label{eq:Radion_BR}
\eea
where $N_A$ is the degrees of freedom of SM gauge boson: 8 for gluon, 2 for $W$, and 1 for $\gamma$ and $Z$.
\begin{figure}
\center

\includegraphics[width=0.41\linewidth]{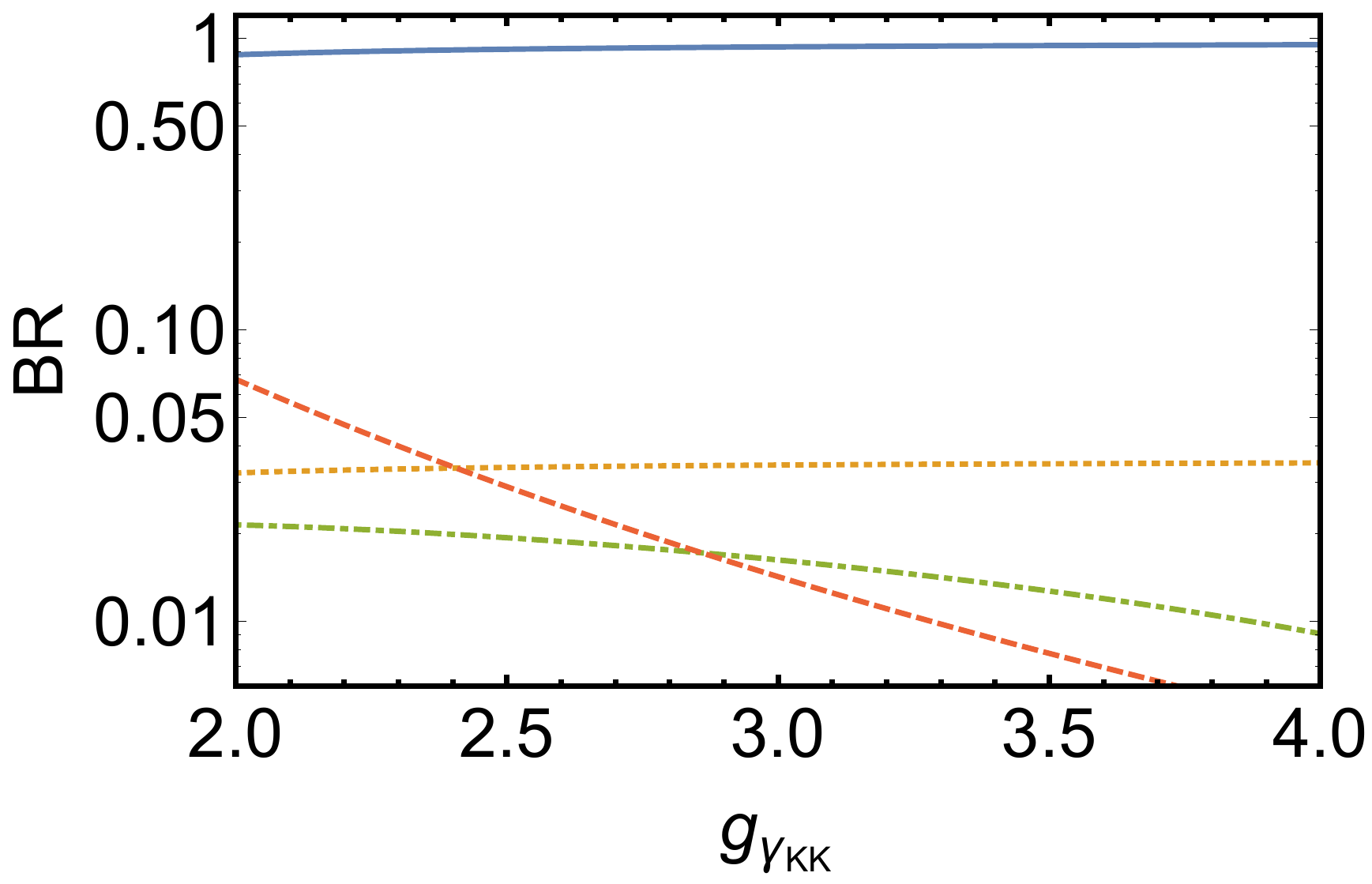}
\includegraphics[width=0.5\linewidth]{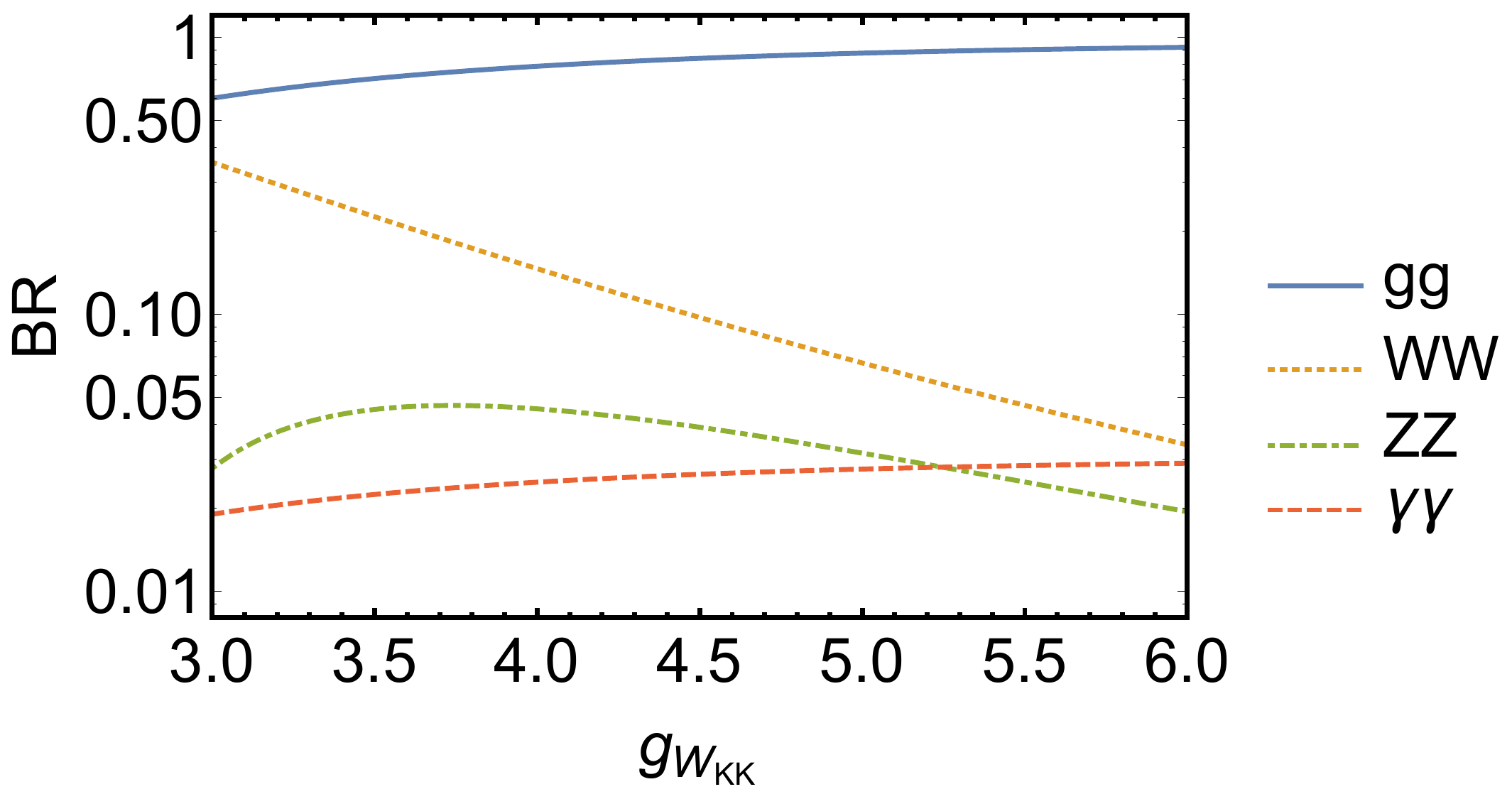}

\caption{
The left panel shows BR of radion as a function of $g_{\gamma_\kk}$, keeping $g_{W_\kk}=6$. The right panel shows BR as a function of $g_{W_\kk}$, keeping $g_{\gamma_\kk}=2.5$. In both cases we choose $g_{g_\kk}=6$. 
}
\label{fig:Radion_BR}
\end{figure}
From this we see that radion decay branching ratios are determined by the relative size of KK gauge couplings. Numerically, we find that BRs to $\gamma\gamma$, $ZZ$, $W W$, and $gg$ are roughly $\mathcal{O}(0.1) \%$, $\mathcal{O}(1) \%$, $\mathcal{O}(1) \%$, and $\mathcal{O}(95) \%$, respectively: 
here, we have used the numerical values $g_{ \gamma } \approx 0.3$, $g_W \approx 0.65$, $g_Z \approx 0.74$, and
$g_g \approx 1$.
%
%
We display
%
%
the branching ratios of various radion decay modes as a function of $g_{\gamma_{\rm KK}}$ ($g_{W_{\rm KK}}$) with $g_{g_{\rm KK}}=6$ and $g_{W_\kk}=6$ ($g_{\gamma_{\rm KK}}=2.5$) in the left (right) panel of Fig.~\ref{fig:Radion_BR}.




Although the diphoton channel has the smallest branching ratio in most of parameter space of interest, 
the cleaner nature of photonic final states than diboson or dijet ones leads to
the most stringent bound for radion. 
The current diphoton searches performed by the ATLAS and CMS Collaborations~\cite{CMS:2016crm,Aaboud:2016tru} suggest
0.7 (0.4) fb for 1 (1.5) TeV radion. Since all our signal channels contain a radion as an intermediary on-shell state, this bound is relevant so that 
we take this into account in our study. 
%
%
As stated before, we choose 1 TeV and 1.5 TeV as benchmark values for the radion mass. 
Even though heavier radions could be safe from the bounds, they would result in smaller signal cross sections  because of the phase space suppression in decay width $\Gamma(A_\kk \to \varphi A)$. On the other hand, lower radion masses would be more constrained by the current diphoton bounds and also develop narrower possible parameter space. 
We shall discuss the bounds again more explicitly in the context of benchmark points for our collider study (see Fig.~\ref{fig:allBP}).





\subsection{Gauge KK production, decay, and current bounds}
\label{Sec:KKgauge_prod_decay}

KK gauge bosons are produced via pair-annihilation of light quarks inside the proton, whose coupling structures are encoded in~\eqref{Eqn:KKgauge-fermion-coupling}.
They can then decay directly into a pair of SM fermions via the same interaction vertices. 
Another decay mode of them is to a radion and a corresponding SM gauge boson, whose coupling is governed by~\eqref{Eqn:KKgauge-radion-SMgauge-coupling}.
Let us call this ``radion channel'' for short. 
As we explained earlier, decays to tops/Higgs via flavor-non-universal couplings are usually very suppressed, and hence neglected.
However, we remark that decays to top/Higgs would
still occur via the flavor-universal coupling in
%
%
\eqref{Eqn:KKgauge-fermion-coupling}. 
We first summarize decay widths for all KK gauge bosons and move onto their current bounds.

\subsubsection{Decay widths of KK gauge bosons}
\paragraph*{KK photon:}

%
%

\begin{figure}
\center

\includegraphics[width=0.37\linewidth]{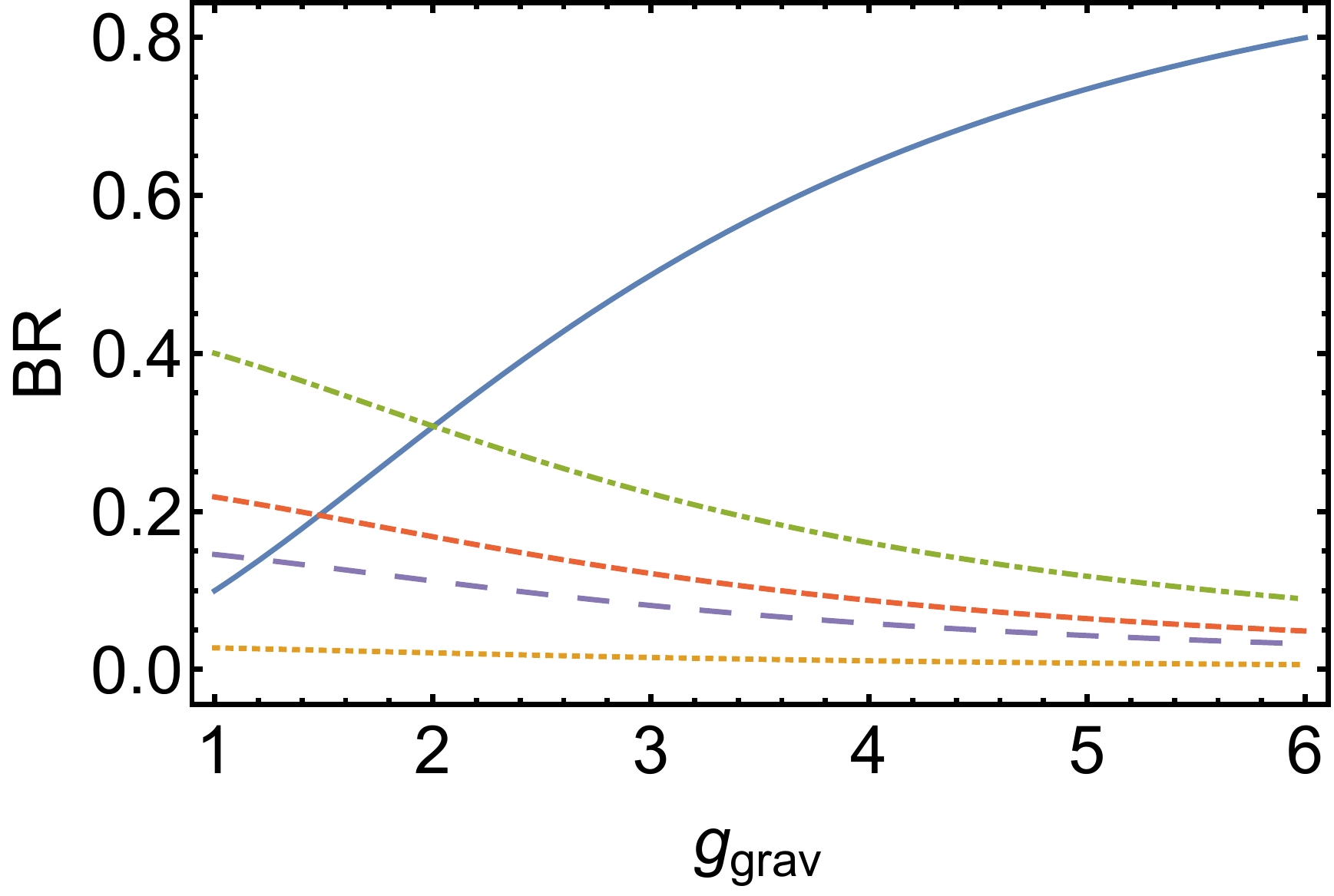}
\includegraphics[width=0.55\linewidth]{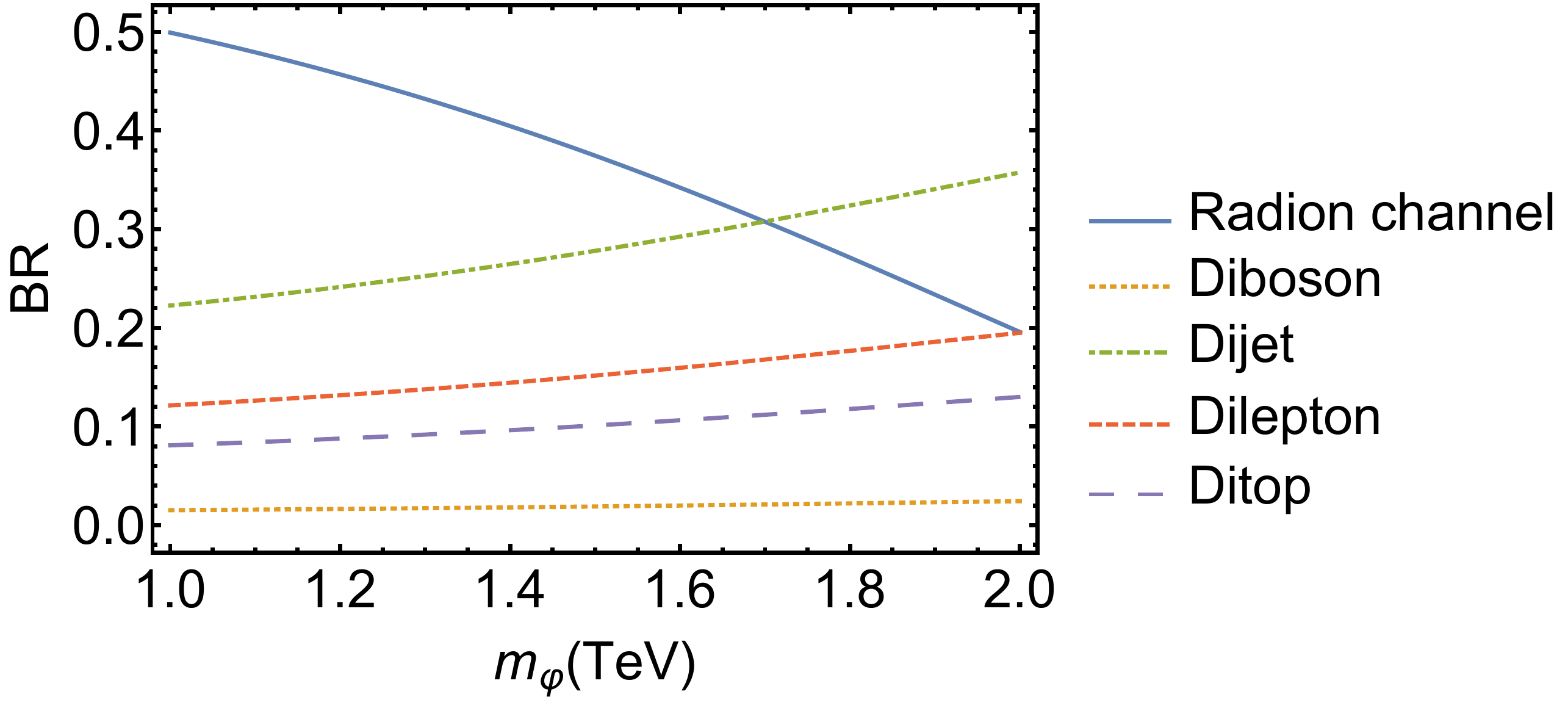}

\caption{The left panel 
shows BR of KK photon as a function of $g_{\rm grav}$, keeping $m_\varphi=1$ TeV. The right panel 
shows BR as a function of $m_\varphi$, keeping $g_{\rm grav}=3$. In both cases we choose $m_{\rm KK}=3$ TeV and $\epsilon=0.5$. 
%
%
}
\label{fig:AKK_BRs}
\end{figure}


Decay channels for the KK photon are radion channel,  $WW$, dilepton, dijet, and ditop channels:
\bea
& \Gamma & (\gamma_{KK} \to \varphi \gamma) = \left(  \epsilon g_{\rm grav} \frac{g_{ \gamma }}{ g_{\gamma_{\rm KK}}} \right )^2\left(1-\left(\frac{m_{\varphi}}{m_{\rm KK}}\right)^2\right)^3 \frac{m_{\rm KK}}{24 \pi}\,, \\
& \Gamma & (\gamma_{KK} \to W W ) \approx \left(\frac{g_{ \gamma }^2}{ g_{\gamma_{\rm KK}}} \right )^2\frac{m_{\rm KK}}{48 \pi}\,, \\
& \Gamma & (\gamma_{KK} \to \psi \psi ) \approx N_\psi Q_\gamma^2 \left(\frac{g_{ \gamma }^2}{ g_{\gamma_{\rm KK}}} \right )^2\frac{m_{\rm KK}}{12 \pi}\,, 
\eea
where $\gamma_{KK} \to \psi \psi $ represents the KK photon decay into a pair of SM fermions.
$N_\psi$ denotes the degrees of freedom of SM fermions (e.g., 3 for quarks and 1 for leptons), while $Q_\gamma$ denotes the electric charge of the associated fermions. The approximation signs in some of the partial decay width formulae in this section originate from taking the massless limit of SM particles.
Based on the formulae listed above, we exhibit branching ratios of KK photon as a function of $g_{\rm grav}$ (left panel) and $m_\varphi$ (right panel) in Fig.~\ref{fig:AKK_BRs}.
For both panels, $m_\kk$ and $\epsilon$ are set to be $3$ TeV and 0.5, respectively, whereas $m_\varphi$ ($g_{\rm grav}$) is fixed
%
%
to 1 TeV (3) for the left (right) panel.
We clearly observe that the radion channel can be the dominant decay mode of KK photon in a wide range of the parameter region of interest. 
The BR of other KK gauge bosons will be roughly similar to that of KK photon, so we only show plots for BR of KK photon as a representative example.
%
%




%
%
%
\paragraph*{KK gluon:}

Decay channels for the KK gluon are radion channel, dijet, and ditop channels:
\bea
& \Gamma & (g_{KK} \to \varphi\  g) = \left( \epsilon  g_{\rm grav} \frac{g_{g}}{ g_{g_{\rm KK}}} \right )^2\left(1-\left(\frac{m_{\varphi}}{m_{\rm KK}}\right)^2\right)^3 \frac{m_{\rm KK}}{24 \pi}\,, \\
& \Gamma & (g_{KK} \to qq ) \approx \left(\frac{g^2_{g}}{ g_{g_{\rm KK}}} \right )^2\frac{m_{\rm KK}}{24 \pi}\,. 
\eea

\paragraph*{KK $W$:}

Decay channels for the KK $W$ boson are radion channel, diboson, dijet, and dilepton channels:
\bea
& \Gamma & (W_{KK} \to \varphi\  W) = \left( \epsilon  g_{\rm grav} \frac{g_{W}}{ g_{W_{\rm KK}}} \right )^2\left(1-\left(\frac{m_{\varphi}}{m_{\rm KK}}\right)^2\right)^3 \frac{m_{\rm KK}}{24 \pi}\,, \\
& \Gamma & (W_{KK} \to WZ/Wh) \approx \left(\frac{g^2_{W}}{ g_{W_{\rm KK}}} \right )^2\frac{m_{\rm KK}}{96 \pi}\,, \\
& \Gamma & (W_{KK} \to \psi \psi') \approx N_\psi \left(\frac{g^2_{W}}{ g_{W_{\rm KK}}} \right )^2\frac{m_{\rm KK}}{48 \pi}\,, 
\eea
where $W_{KK} \to \psi \psi'$ represents KK $W$ decay into a pair of (different-flavored) SM fermions. 

\paragraph*{KK $Z$:}

Decay channels for the KK $Z$ boson are radion channel, diboson, dijet, ditop, and dilepton channels:
\bea
& \Gamma & (Z_{KK} \to \varphi\  Z) = \left( \epsilon  g_{\rm grav} \frac{g_{Z}}{ g_{Z_{\rm KK}}} \right )^2\left(1-\left(\frac{m_{\varphi}}{m_{\rm KK}}\right)^2\right)^3 \frac{m_{\rm KK}}{24 \pi}\,, \\
& \Gamma & (Z_{KK} \to WW/Zh) \approx Q_Z^2 \left(\frac{g^2_{Z}}{ g_{Z_{\rm KK}}} \right )^2\frac{m_{\rm KK}}{24 \pi}\,, \\
& \Gamma & (Z_{KK} \to \psi \psi) \approx  N_\psi Q_Z^2\left(\frac{g^2_{Z}}{ g_{Z_{\rm KK}}} \right )^2\frac{m_{\rm KK}}{24 \pi}\,,
\eea
where $Q_Z$ denotes the SM $Z$ charge of the associated fermion $\psi$.

\subsubsection{Current bounds of KK gauge bosons}
\label{subsubsec:current bounds for KK gauge}

\paragraph*{KK $Z$:}

As mentioned in Ref.~\cite{Agashe:2016rle}, the strongest bound for KK $Z$ comes from the dilepton resonance search. 
%
%
%
We can obtain it by simply 
using the experimental searches for sequential SM $Z^{ \prime }$~\cite{CMS:2016abv}, but taking into account
the coupling to light quarks, which is involved
%
%
in 
the dominant production mechanism, being {\em reduced} by $\sim g_{  Z } / g_{Z_{\rm KK} }$.
We expect that our cascade decay signal channel further relaxes the bounds since
the original dilepton branching ratio is reduced by half for 50\% branching ratio for the radion channel:\footnote{Note that 
in Ref.~\cite{Agashe:2016rle}, the new decay channel for KK $Z$ was neglected so that the bounds quoted there
are slightly stronger than here.} based on the discussion in the previous section,
we see that such a suppression of BR for decay to pair of SM fermions/Higgs 
can be easily achieved. 
We find that the predicted cross section of sequential SM $Z^{ \prime }$ exceeds the bound~\cite{CMS:2016abv} by $\sim 70 \; (25)$ for $m_{ Z^{ \prime } } \sim 2  \; (2.5)$ TeV. Translating this bound for our case, including radion channel, we obtain
%
\bea
m_{Z_{\rm KK}} & \gtrsim &  2.5 \; \hbox{TeV}  \; \hbox{for} \;  g_{Z_{\rm KK} } \sim 5\,, \\
 & \gtrsim &  3 \; \hbox{TeV} \; \hbox{for} \;  g_{Z_{\rm KK} } \sim 3 \,,
\eea
with $g_Z$ set to be around 0.75. 

\paragraph*{KK photon:}


Similarly to the KK $Z$ boson, the mass of the KK photon is most severely constrained by the dilepton resonance search. 
Indeed, $\textrm{BR}(\gamma_\kk \to \ell^+ \ell^-) = (8/3)\cdot \textrm{BR}(Z_\kk \to \ell^+ \ell^-)$ with the assumption of 
the same braching ratio for the radion channel in both cases. 
However, $\sigma(p p\to \gamma_\kk)\cdot \textrm{BR}(\gamma_\kk \to \ell^+ \ell^-)$ is smaller than $\sigma(p p\to Z_\kk)\cdot \textrm{BR}(Z_\kk \to \ell^+ \ell^-)$, 
given that $\gamma_\kk$ and $Z_\kk$ have the same mass. This is because their production rates are proportional to $g^4_{A}/g_{A_{\rm KK}}^2$, and therefore, 
with $g_{Z_\kk}\sim g_{\gamma_\kk}$, $\sigma(p p\to \gamma_\kk)/\sigma(p p\to Z_\kk)$ is roughly $g^4_{\gamma}/g^4_Z < 1$.\footnote{This is just a rough estimate. For a more accurate analysis one needs to take into account the difference
between electric charge and SM $Z$ charge of quarks (i.e., $Q_\gamma$ vs. $Q_Z$). }
So, we expect that KK photon is less constrained than KK $Z$ from dilepton bounds. Considering 50\% branching ratio for the radion channel again, we find that the bound is roughly
\bea
m_{ \gamma_{\rm KK}} & \gtrsim &  2 \; \hbox{TeV}  \; \hbox{for} \;  g_{ \gamma_{\rm KK} } \sim 3\,.
\eea
%

\paragraph*{KK $W$:}


The dominant bound comes from the leptonic decay of KK $W$, i.e., $W_{KK} \to \ell v$~\cite{Agashe:2016rle}. In our model, assuming that the radion channel comprises 50\% of the branching ratio for KK $W$ decays, we see that the leptonic decay of one generation (either $e\nu_e$ or $\mu\nu_\mu$) has the branching ratio of 4\%. 
From the new resonance search in $\ell \nu$ channels conducted by the ATLAS Collaboration~\cite{ATLAS:2016ecs}, we find that the bound therein can be interpreted as 
\bea
m_{W_{\rm KK}} & \gtrsim &  2.5 \; \hbox{TeV}  \; \hbox{for} \;  g_{W_{\rm KK}} \sim 3\,.
\eea

\paragraph*{KK gluon:}

The constraints for the KK gluon come from both ditop and dijet searches. 
The ditop bound can be obtained by rescaling the KK gluon bound given in Ref.~\cite{Khachatryan:2015sma}.
The predicted cross-section (all for $g_{g_{\rm KK} } \sim 5$, as assumed in Ref.~\cite{Agashe:2006hk}, which is quoted in Ref.~\cite{Khachatryan:2015sma}) is larger than the bound by $\sim$ 6 (2) for mass of KK gluon of 2 (2.5) TeV. The above bounds are assuming BR to top quarks $\approx 1$ (as in the standard scenario) so that for our case, with the radion channel having 50\% branching ratio and BR to top quarks is $\approx 1/12$ , we get
\bea
m_{g_{\rm KK}} &   \gtrsim  &  2.0 \; \hbox{TeV} \; \hbox{for} \; g_{g_{\rm KK} } \sim 3.5\,,  \\
 & \gtrsim & 2.5 \; \hbox{TeV} \; \hbox{for} \;  g_{g_{\rm KK} } \sim  2\,.
\eea
For the dijet bound, we may rescale from axigluon bounds in Ref.~\cite{CMS:2016wpz},
i.e., coupling to our composite gluon  is smaller by a factor of $\sim g_{G } / \left(  \sqrt{2} g_{g_{\rm KK} }   \right)$, since coupling of axigluon is larger than QCD by $\sqrt{2}$ (see also the discussion in Ref.~\cite{Chivukula:2013xla} referred to by Ref.~\cite{CMS:2016wpz}). The cross-section is constrained to be smaller than the prediction for axigluon by $\sim 50  \; (30)$ for axigluon mass of 2 (2.5) TeV. So, using the above couplings, and taking radion channel BR to be 50\%, we get for our case:
\bea
m_{g_{\rm KK}} & \gtrsim &  2.0 \; \hbox{TeV}  \; \hbox{for} \; g_{g_{\rm KK} }  \sim  3.5 \,,\\
 & \gtrsim & 2.5 \; \hbox{TeV} \; \hbox{for} \;  g_{g_{\rm KK} }  \sim  3\,.
\eea

\subsection{Benchmark points}
\label{SubSec:Benchmark-points}


In this section, we list the benchmark points (BPs) for all channels that we examine in Sec.~\ref{results}. 
We carefully choose 
them 
%
to
satisfy all experimental/theoretical bounds that we 
discussed in previous sections. 
We tabulate parameter values for each benchmark point in Table~\ref{tab:BPtable}.
The name of each BP obeys the following pattern.
\bea
\hbox{{\it the name of the KK gauge boson - final states - BP1 or BP2}}  \nonumber
\eea
For example, $\gamma$-$\gamma gg$-BP1 means the first benchmark point (BP1) for KK photon ($\gamma$ in the first placeholder) with final states photon $+$ dijet ($\gamma gg$).


\begin{table}[t]
\centering
\hspace*{-0.9cm}
\begin{tabular}{c|l|c|c|c|c|c|c|c}
\hline
 & Process & Name & $m_{\rm KK}$ & $m_\varphi$ & $g_{\gamma_{\rm KK}}$ & $g_{W_{\rm KK}}$ & $g_{g_{\rm KK}}$ & $g_{\rm grav}$ \\
\hline \hline
\multirow{2}{*}{$\gamma_{\rm KK}$}
 & \multirow{2}{*}{$\gamma_{\rm KK} \rightarrow \gamma \varphi \rightarrow \gamma g g$ (\ref{sec:akk})}
  & $\gamma$-$\gamma gg$-BP1 & 3 & 1 &3 &6 &3 &3  \\
 & & $\gamma$-$\gamma gg$-BP2 & 3 & 1.5 & 2.7&6 &3 &4.1  \\
\hline
\multirow{6}{*}{$g_{\rm KK}$}
& \multirow{2}{*}{$g_{\rm KK} \rightarrow g \varphi \rightarrow g \gamma\gamma$ (\ref{sec:jetdiphoton})}
  & $g$-$g\gamma \gamma$-BP1 & 3 & 1 & 2.7&6 &6 &2.25  \\
 & & $g$-$g\gamma \gamma$-BP2 & 3 & 1.5 &2.7 &6 &6 &3 \\
 \cline{2-9}
 & \multirow{2}{*}{$g_{\rm KK} \rightarrow g \varphi \rightarrow g g g$ (\ref{sec:trijet})}
  & $g$-$ggg$-BP1 & 3 & 1 &2.7 &6 &3 &2.45  \\
 & & $g$-$ggg$-BP2 & 3 & 1.5 &2.7&6 &3 & 4\\ 
 \cline{2-9}
 & \multirow{2}{*}{$g_{\rm KK} \rightarrow g \varphi \rightarrow g V_h V_h$ (\ref{sec:jetdiboson})}
  & $g$-$gVV$-BP1 & 3 & 1 & 2.65&3 &6 &3  \\
 & & $g$-$gVV$-BP2 & 3 & 1.5 &2.65 &3 &6 &5  \\
   \hline
 \multirow{2}{*}{$W/Z_{\rm KK}$}
 & \multirow{2}{*}{$W_{\rm KK} \rightarrow W_l \varphi \rightarrow W_l g g$ (\ref{sec:kkwlepton})}
  & $W$-$Wgg$-BP1 & 2.5 & 1 & 3.5 &4.4 &3 &3.5 \\
 & & $W$-$Wgg$-BP2  & 3 & 1.5 &3 &3.5 &3 &5.1  \\
 \hline
\end{tabular}
\caption{\label{tab:BPtable} A list of benchmark points defined by their associated process and chosen parameter values. 
For all of them, the $\epsilon$ parameter is set to be 0.5. 
We assign the name of the channels in the following pattern: {\it the name of the KK gauge boson - final states - BP1 or BP2}. 
The numbers in the parentheses of the second column refer to the section discussing the corresponding collider analysis. 
$V$ refers to either $W$ or $Z$ and the subscript $h$ ($l$) stands for hadronic (leptonic) decay. All mass quantities are in TeV. }
\end{table}

\begin{figure}
\center

\includegraphics[width=0.25\linewidth]{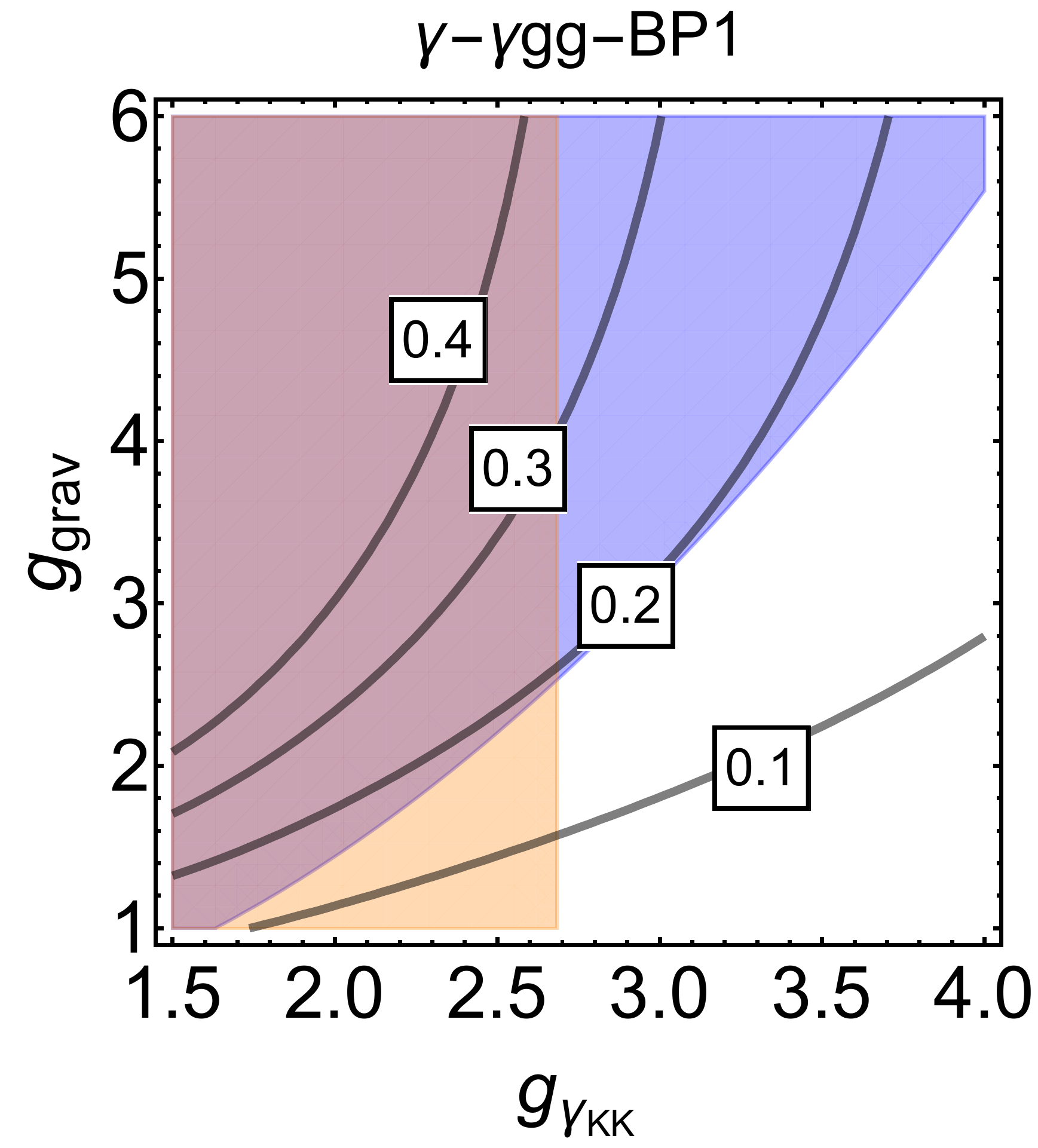}
\includegraphics[width=0.25\linewidth]{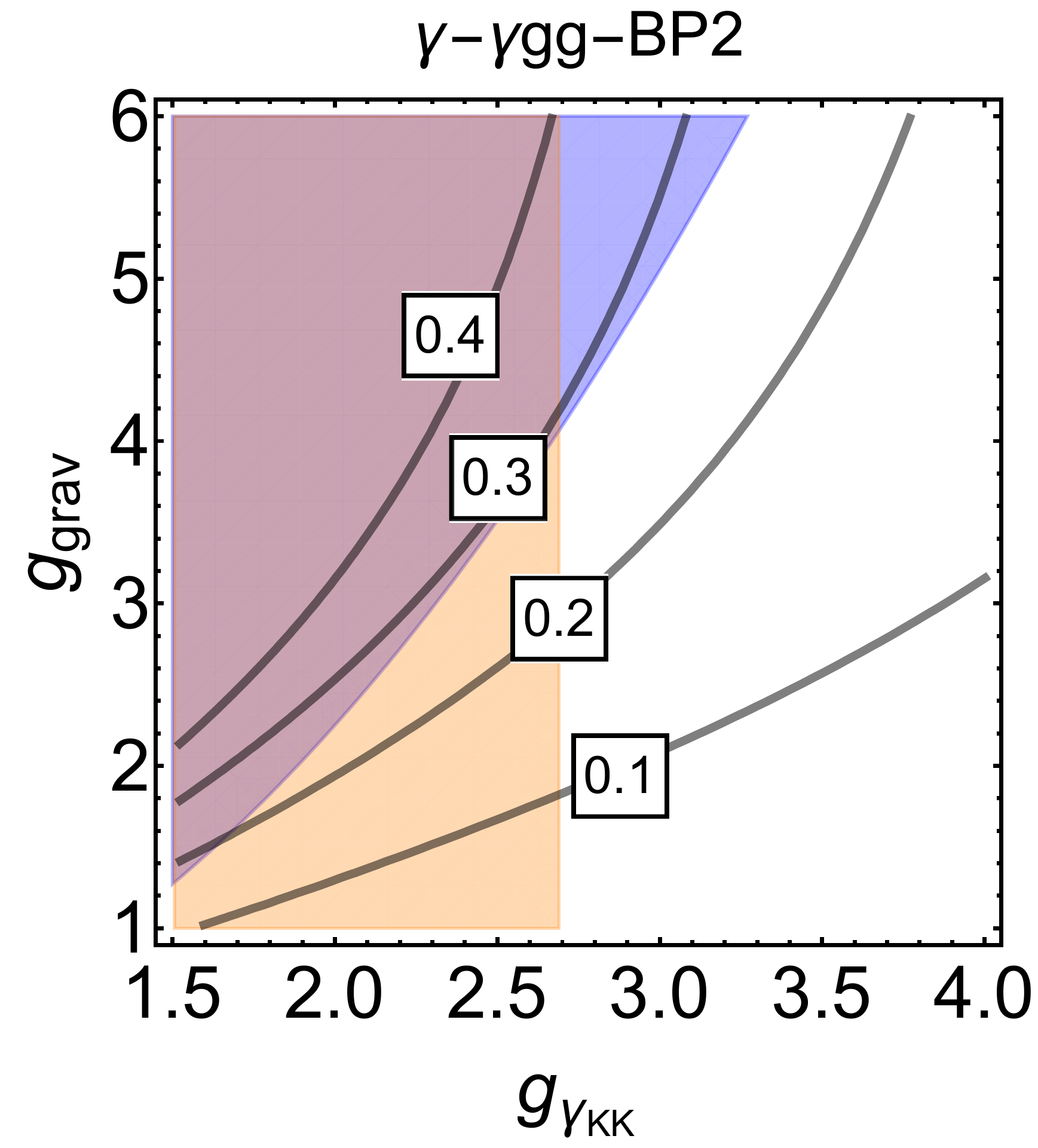}
\includegraphics[width=0.25\linewidth]{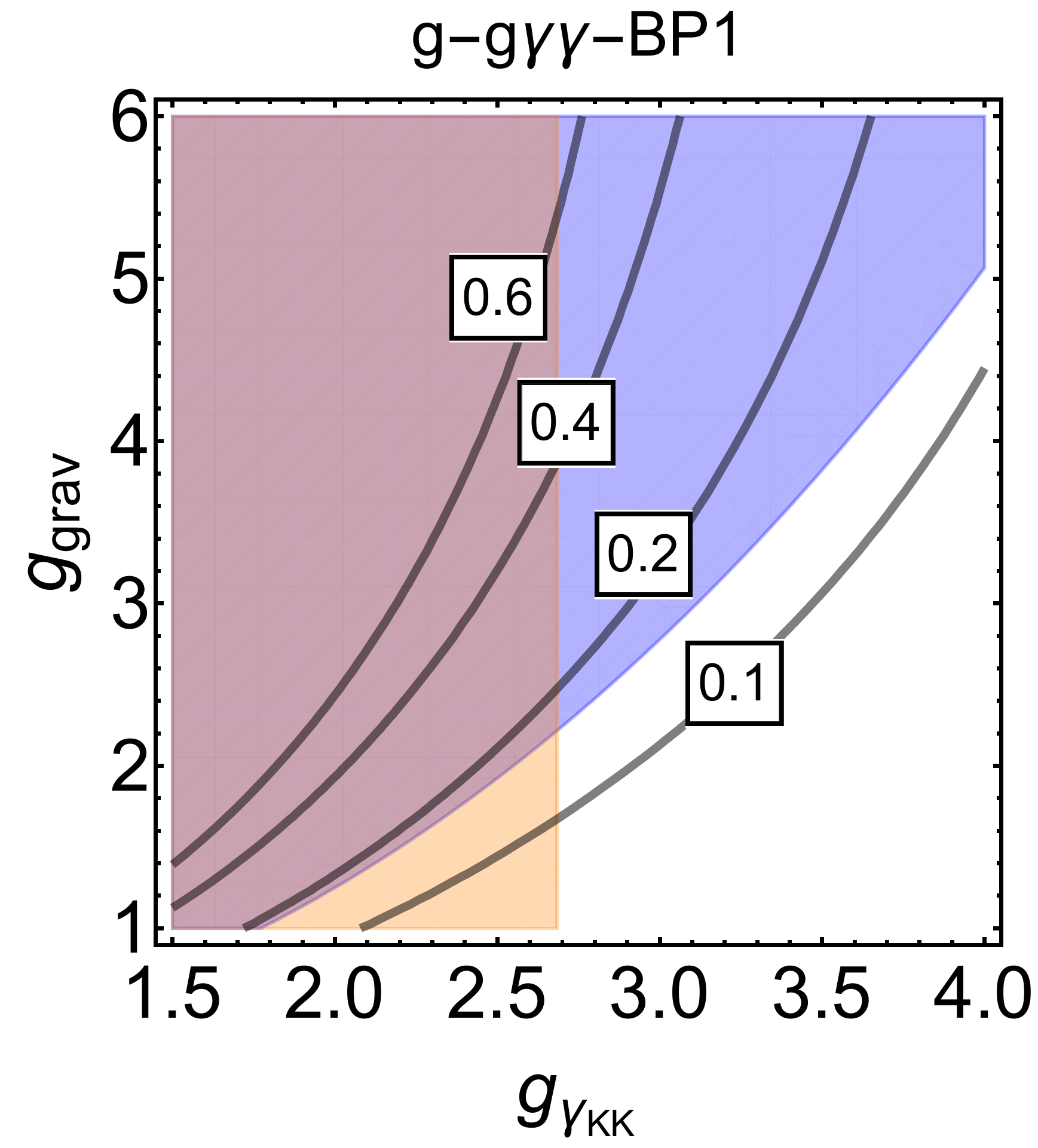}\\
\includegraphics[width=0.25\linewidth]{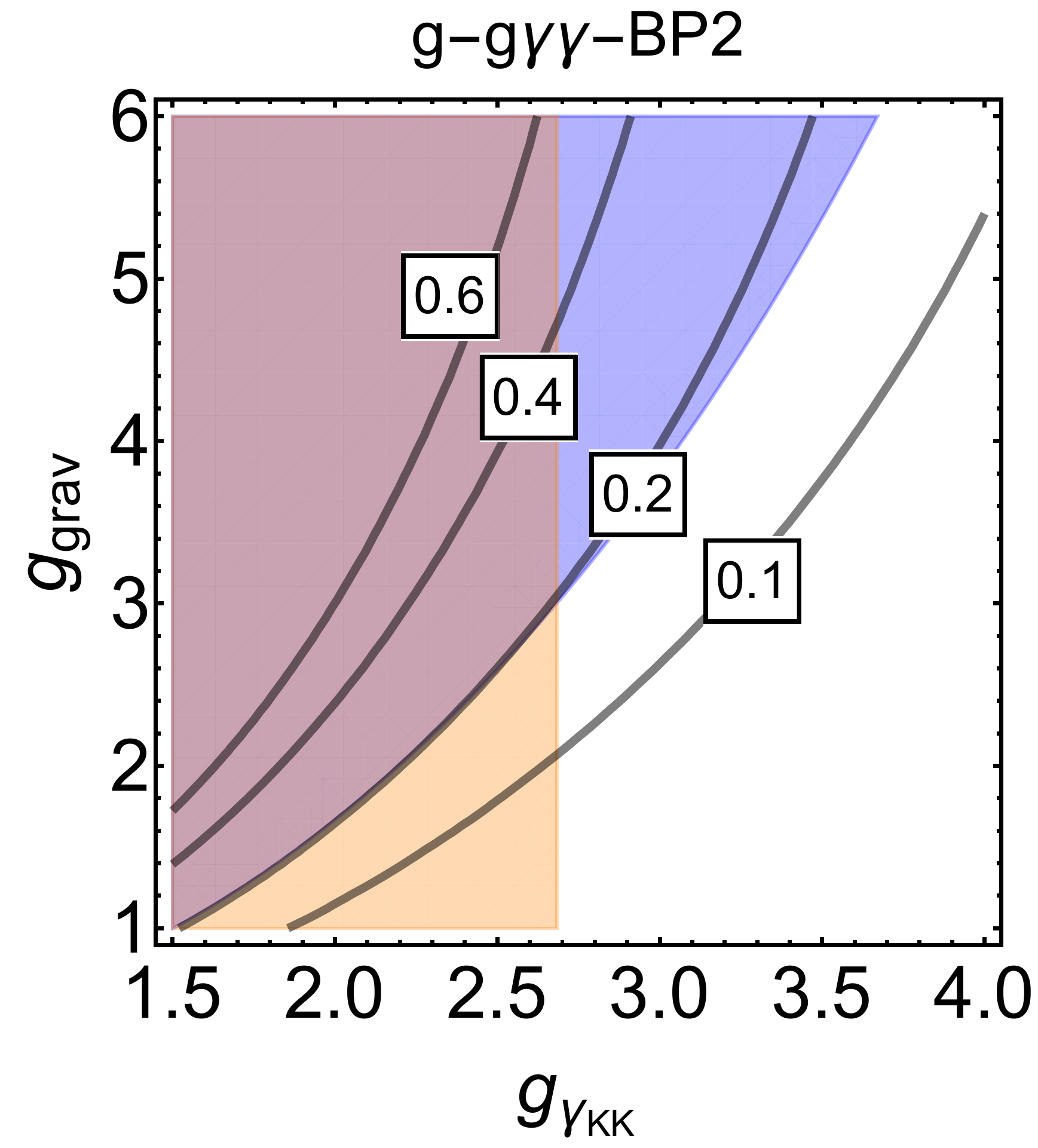}
\includegraphics[width=0.25\linewidth]{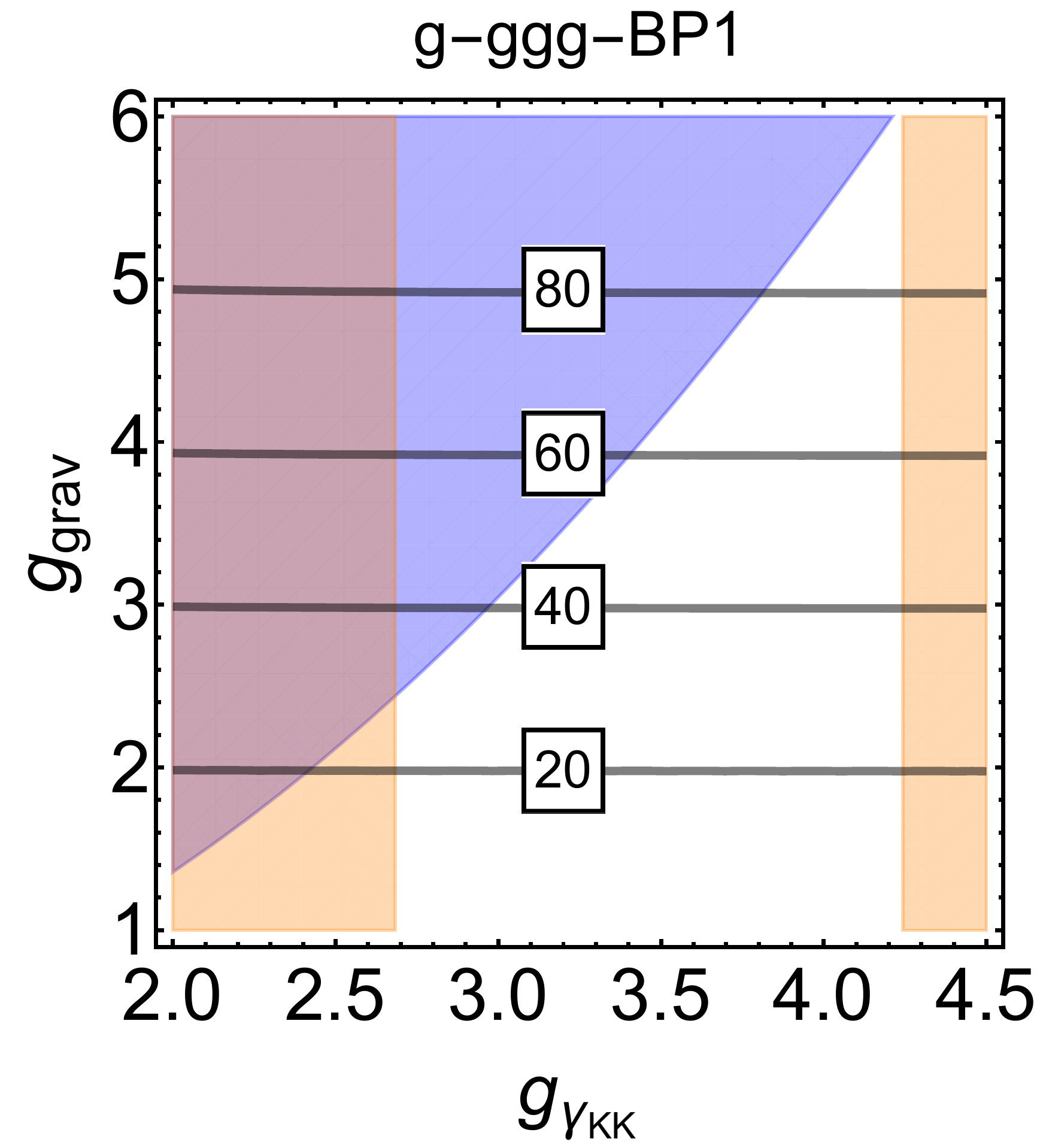}
\includegraphics[width=0.25\linewidth]{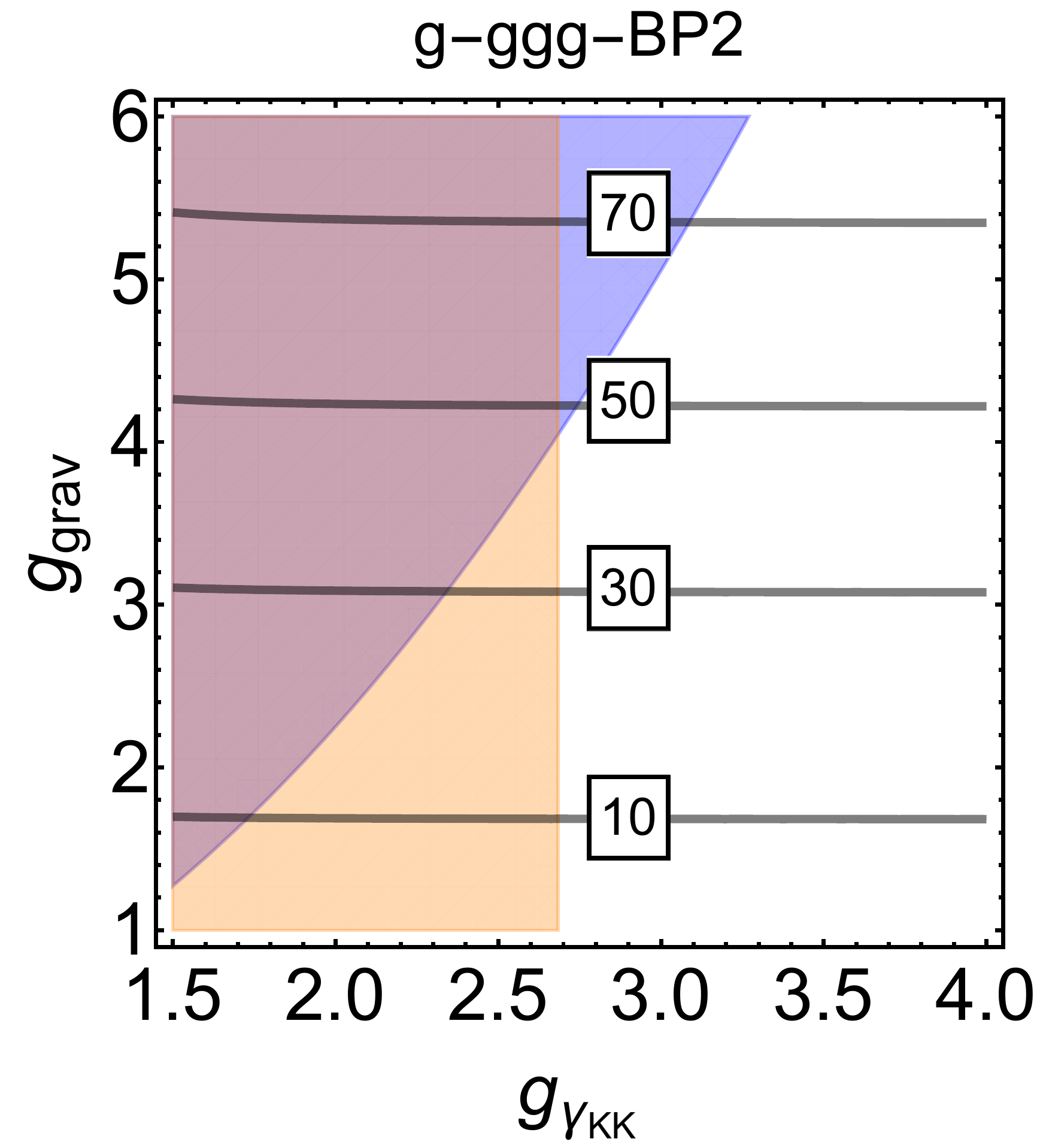}\\
\includegraphics[width=0.25\linewidth]{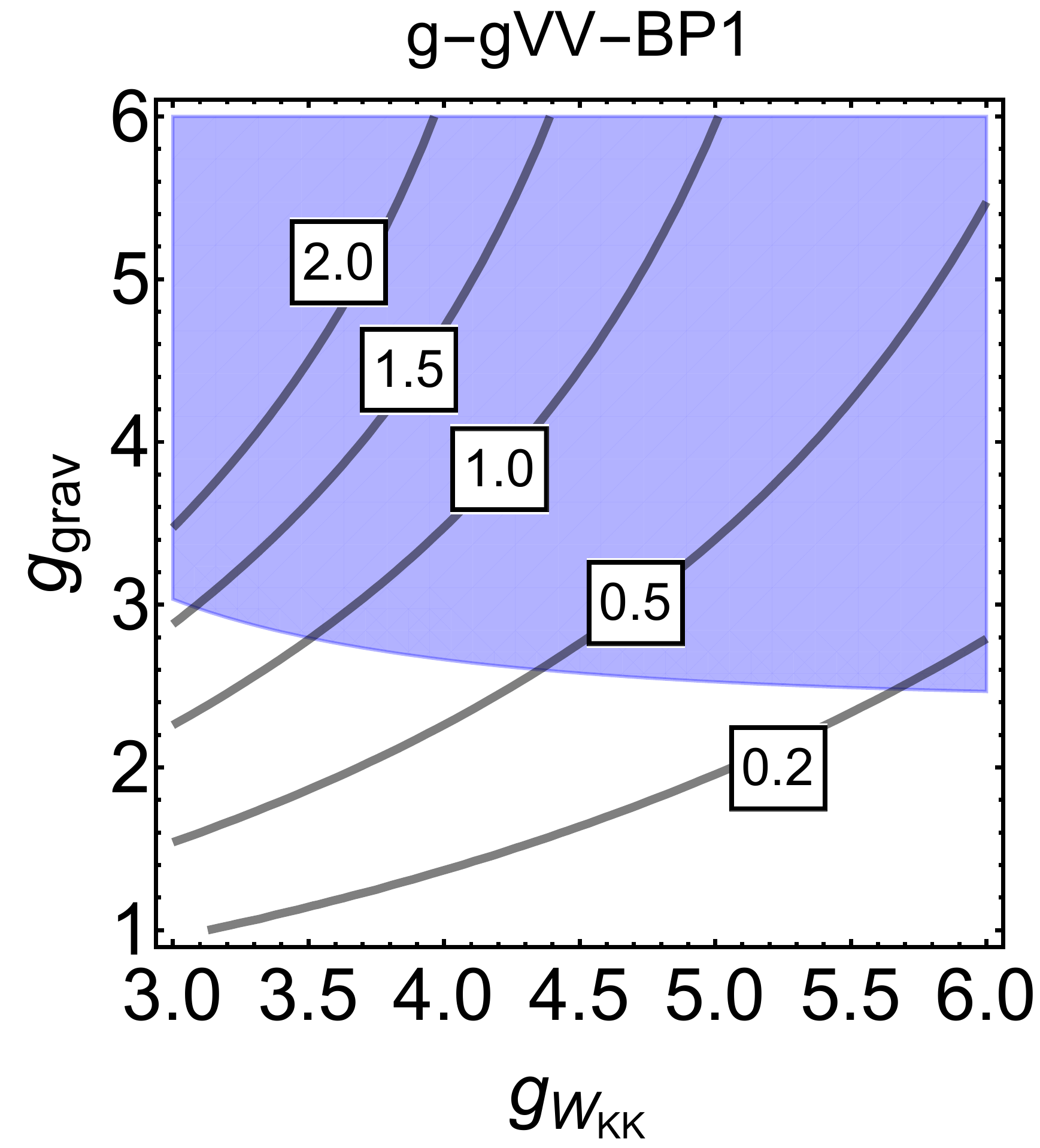}
\includegraphics[width=0.25\linewidth]{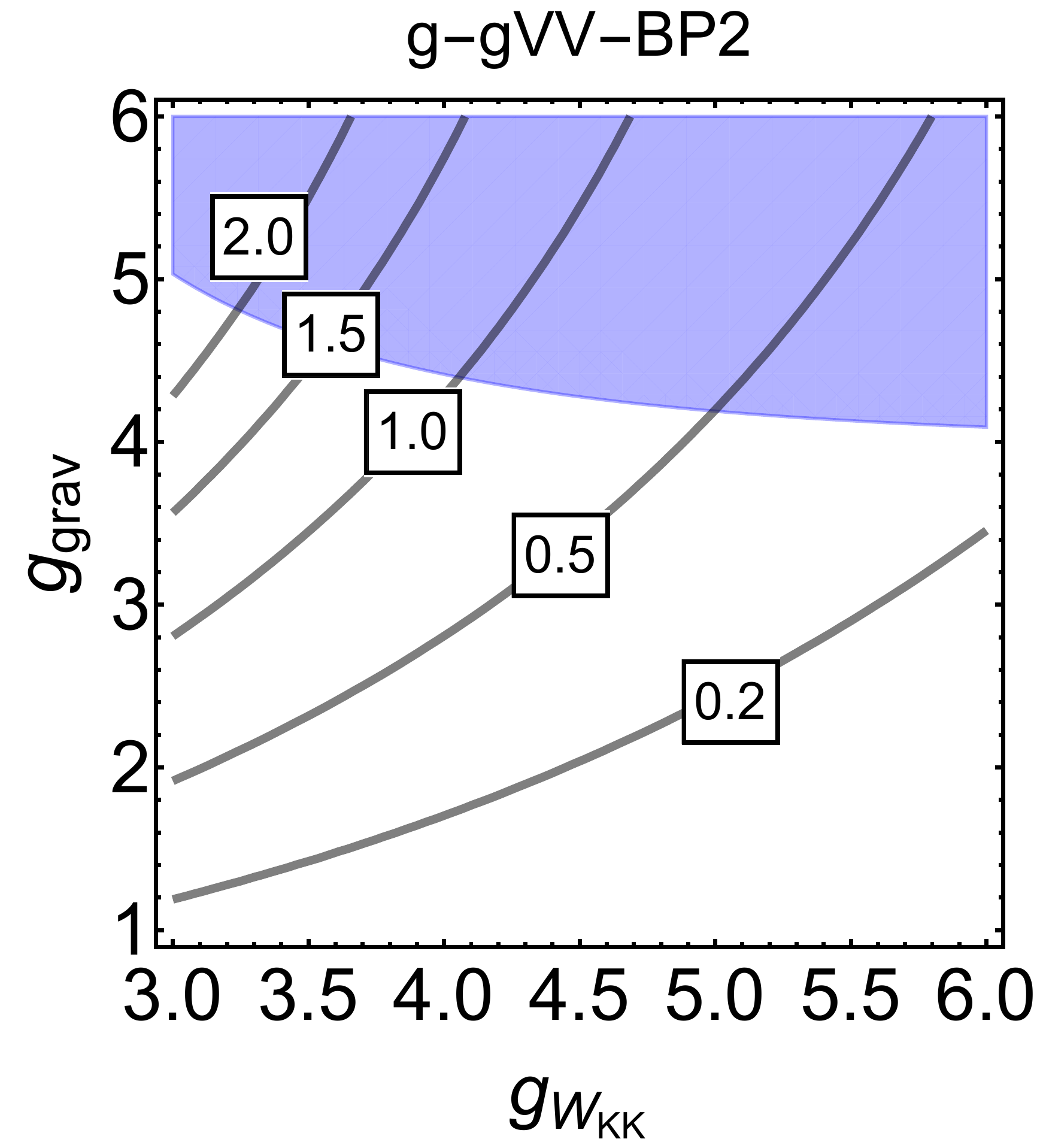}\\
\includegraphics[width=0.25\linewidth]{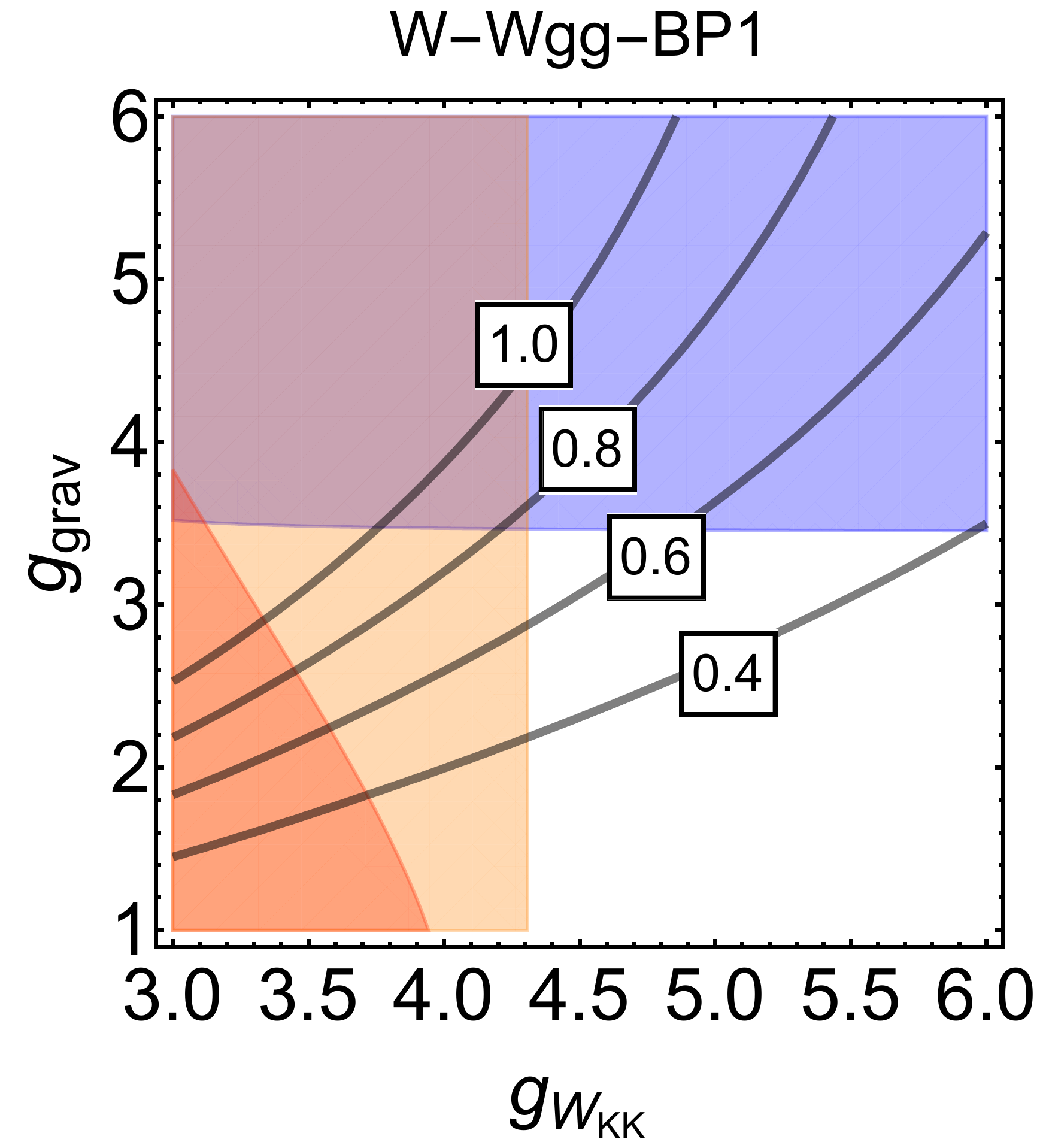}
\includegraphics[width=0.25\linewidth]{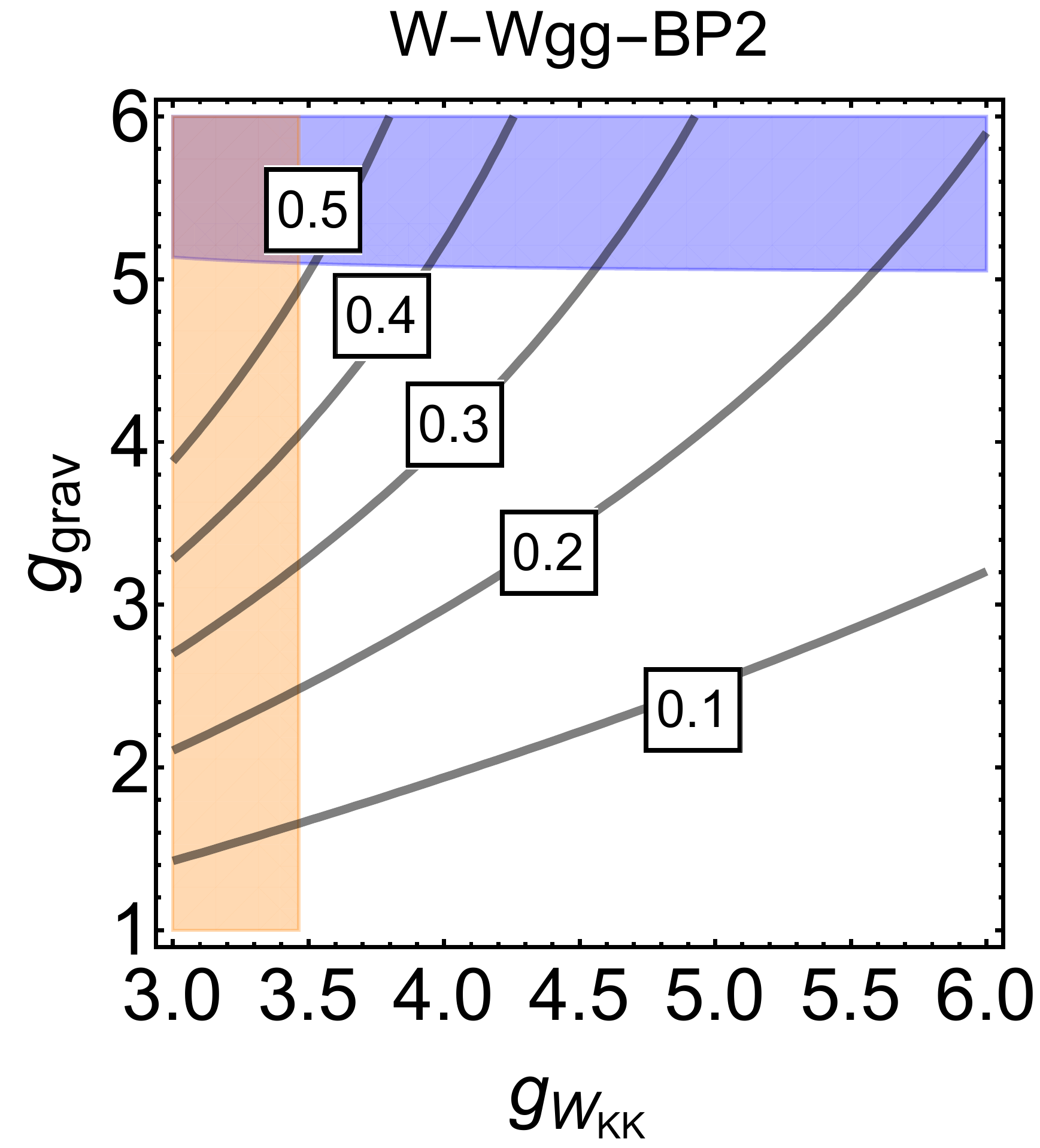}

\caption{ 
Contour plots of the cross sections for ten benchmark points in the plane of $g_{\gamma_{\rm KK}}$ (first six panels) or  $g_{W_{\rm KK}}$ (last four panels) vs. $g_{\rm grav}$.
All cross sections are in fb, and the input radion masses are either 1 TeV (BP1) or 1.5 TeV (BP2).
The blue (red) regions are excluded by diphoton ($W_\kk$ leptonic decay) bounds. 
The orange regions are forbidden due to $g_{B_{\rm KK}}\notin[3,6]$.  Each plot is labelled by the associated benchmark point.
The other parameters which are not specified in each contour plot are chosen to be the same as those in the associated benchmark point of Table~\ref{tab:BPtable}. }

\label{fig:allBP}
\end{figure}

We show
%
%
the contour plots of estimated signal cross sections for all ten benchmark points in the plane of $g_{\gamma_{\rm KK}}$ (first six panels) or  $g_{W_{\rm KK}}$ (last four panels) vs. $g_{\rm grav}$ in Fig.~\ref{fig:allBP}.
All cross sections are reported in fb, and the input radion masses are either 1 TeV (BP1) or 1.5 TeV (BP2). 
The other parameters unspecified in each panel are chosen to be the same as those in the associated benchmark point of Table~\ref{tab:BPtable}.

We remark that diphoton bounds constrain {\it any} radion decaying to a pair of photons.
Two sources for radion production are affected by the diphoton constraint:
one is direct production via gluon fusion, and the other  is from KK gauge boson decays. 
All diphoton bounds displayed in Fig.~\ref{fig:allBP} by blue regions result from taking these two sources into consideration. 
For channels involving $W_\kk$, we also consider the bound from the leptonic decay of $W_\kk$ (red regions).
There is another strong theoretical constraint applied to all channels, which demands $g_{B_\kk}\in [3,6]$ as discussed near relation~\eqref{eq:paramranges} (orange regions).
We clearly see from the contour plots that all our benchmark points are not ruled out.


\section{\label{EventSimulation} Collider Study  }

Armed with the benchmark points defined in the previous section, we now 
%
%
discuss 
our strategy for their collider studies.
We begin by explaining how we 
%
%
conduct Monte Carlo simulation and reconstruct/identify objects out of the simulated data. As some of the signal channels 
%
%
include
$W/Z$ gauge boson-induced jets in addition to 
%
%
the quark/gluon 
jets, we briefly review the jet substructure technique that we employ here. 
Moving onto data analyses, we discuss key mass variables allowing us to suppress background events significantly, thus increase signal sensitivity. 

\subsection{Event simulation}

Simulated event samples are used to model signal predictions in various channels discussed in the previous section and estimate SM background processes associated with each of the signal processes. 
For more realistic Monte Carlo simulation, we take into consideration various effects such as parton shower, hadronization/fragmentation, and detector responses.
To this end, we employ a sequence of simulation tools. 
We begin with creating our model files with \textsc{FeynRules}~\cite{Alloul:2013bka} and plug the outputs into a Monte Carlo event generator \textsc{MG5@aMC}~\cite{Alwall:2014hca} with parton distribution functions parameterized by \textsc{NN23LO1}~\cite{Ball:2012cx}. 
All the simulations are performed with a $\sqrt{s}=14$ TeV $pp$ collider at the leading order. 
The generated events are then streamlined to \textsc{Pythia 6.4}~\cite{Sjostrand:2006za} for taking care of showering and hadronization/fragmentation. 

\begin{table}[t]
\centering
\begin{tabular}{c|c|c}
\hline
 & Regular jets & Merged jets \\
\hline \hline
$R$ & 0.4 (anti-$k_t$ jet) & 0.8 (Cambridge-Achen jet) \\
\hline
$\mu_*$ & -- & 1.0 \\ 
$y_*$ & -- & 0.04 \\
$R_*$ & -- & 0.3 \\
\hline
\end{tabular}
\caption{\label{tab:param} Jet parameters for regular jets (second column) and merged jets (third column). }
\end{table}

As some of our signal processes accompany boosted gauge bosons in the final state, our scheme to find jets depends whether or not we require merged jets. 
For the channels involving only regular jets, we feed the output from \textsc{Pythia 6.4} into \textsc{Delphes 3}~\cite{deFavereau:2013fsa} interfaced with \textsc{FastJet}~\cite{Cacciari:2005hq,Cacciari:2011ma} for describing the detector effects and forming jets. 
The jets are constructed with the anti-$k_t$ algorithm~\cite{Cacciari:2011ma} with a radius parameter $R=0.4$ (see also Table~\ref{tab:param}).

For merged jets, we begin with the Cambridge-Achen jet algorithm~\cite{Dokshitzer:1997in,Wobisch:1998wt} to cluster particles from hadronically decaying $W/Z$ bosons.   
Tagging $W/Z$-induced jets is done by a jet substructure technique. 
In our analysis, we employ the Mass Drop Tagger (MDT)~\cite{Butterworth:2008iy}. 
The MDT essentially traces back the clustering sequences of a C/A jet and attempts to find subjets satisfying appropriate conditions.
We briefly summarize an MDT procedure below.
\begin{itemize}
\item[(1)] Clustering: We cluster energy deposits in the calorimeters using the C/A jet algorithm together with a jet radius parameter $R=0.8$ in order to capture all decay products from a boosted gauge boson.
\item[(2)] Splitting: We rewind the last clustering sequence of a jet $j$, denoting two subjets as $j_1$ and $j_2$ by the order of decreasing mass.
\item[(3)] Checking symmetry conditions: We set an upper bound $\mu_*$ and a lower bound $y_*$ on MDT parameters $\mu$ and $y$ as follows:
\bea
\mu \equiv \frac{m_{j_1}}{m_j} <\mu_*\,, \hspace{0.5cm} y \equiv \frac{\min\left[p_{T,j_1}^2,\, p_{T,j_2}^2\right]}{m_j^2}\Delta R_{j_1 j_2}^2 > y_*\,.
\eea  
If subjets fail in satisfying the above conditions, the MDT procedure redefines $j_1$ as $j$ and repeats the step described in (2). 
Our choice of $\mu_*$ and $y_*$ are tabulated in Table~\ref{tab:param}.\footnote{Detailed values for the C/A-jet radius $R$ and $\mu_*$ do not affect the $W/Z$-jet tagging efficiency substantially, as it is mostly dictated by $y_*$~\cite{Lim:2016ivn}.
}
\end{itemize}
Once the MDT finds a signal merged jet and identify two prongs in it, the MDT attempts to get rid of QCD contamination in subjets by reclustering energy deposits in the merged jet again employing the C/A jet algorithm of a smaller jet radius $R_{\rm filt}$.
\begin{itemize}
\item[(4)] Filtering: We recluster the merged jet constituents with the C/A jet algorithm of radius, 
\bea
R_{\rm filt}=\min \left(R_*,\,\frac{\Delta R_{j_1j_2}}{2}\right)
\eea
to obtain $n$ new subjets $\lbrace s_1,\cdots,s_n\rbrace$ sorted in decreasing $p_T$.  $R_*$ denotes the maximum allowed size for subjets to minimize the QCD contamination. The MDT considers an $\mathcal{O}(\alpha_s)$ correction from hard emission, by accepting at most three subjets in redefining a merged jet as
\bea
p_{\rm merged\, jet}^{\mu}=\sum_{i=1}^{\min(n,3)}p_{s_i}^{\mu}\,.
\eea
\item[(5)] The sum of these $n \leq 3$ subjets is taken as a groomed merged jet for further analysis. 
\end{itemize}
%
%
For candidate merged jets obtained by the MDT procedure, we retain the ones which satisfy a jet mass window requirement around the vector boson mass, and sort them by their hardness in $p_T$.

Vector boson jet candidates are required to satisfy two additional substructure requirements. First, we require a selection based on the $D_2^{(\beta=1)}$ energy correlation function calculated from the groomed jet \cite{Larkoski:2014gra,Larkoski:2015kga} which is useful for discriminating two-pronged structures from QCD jets, with $W$-jets tending to have smaller values and $Z$-jets larger values. The $D_2^{(\beta=1)}$ distribution for $W$-jets is $p_T$ dependent, requiring a cut which also varies with $p_T$. The $D_2$ cut required for 50\% efficiency of selecting a true $W$-jet is very close to linear for $250 \; \text{GeV} < p_T < 1500 \; \text{GeV}$ \cite{ATL-PHYS-PUB-2015-033}, motivating a cut:\footnote{This cut is derived from a rough linear fit to Fig.~(8c) of Ref.~\cite{ATL-PHYS-PUB-2015-033}. While their analysis used different jet clustering and grooming techniques, the analysis presented here should not be strongly affected by modest changes to this cut.}
\begin{equation}D_2< 1 + \left(p_T - 250 \; \text{GeV}\right)\times 7.7\times 10^{-4} \; \text{GeV}^{-1}.\end{equation}
The second jet tagging requirement is a cut on the number of tracks in the $W$-jet, which is typically smaller than in QCD quark or gluon jets with $p_T \sim 1 \; \text{TeV}$. We therefore require
\begin{equation}N_\text{trk} \leq 30,\end{equation}
where this is counted from the constituents of the ungroomed merged jet.

Finally, if the signal channel of interest accompanies $N$ boosted $W/Z$ gauge bosons in the final state, we take $N$ hardest merged jets as our $W/Z$-induced jets. 


\subsection{Mass variables \label{sec:massvariables}}

In this section, we discuss several key mass variables which enable us to separate signal events from relevant background ones. 
We remark that some of our signal channels contain $W$ or/and $Z$ gauge bosons in the final state and they are either boosted or semi-invisible. 
In the semi-invisible cases, we are interested in the signal processes where only one $W$ decays leptonically, and thus, we can reconstruct the neutrino momentum, hence the $W$ momentum.
If we regard each the massive SM gauge bosons as a single object along this line, every signal process in our study can be understood as a two-step cascade decay of a KK gauge boson into three visible particles via an on-shell intermediary state, radion:
\bea
A_\kk \rightarrow v_a \varphi \rightarrow v_a v_b v_c\,,
\eea
where $v_{a/b/c}$ denote the visible particles which will be either $\gamma$, $g$, or $W/Z$ in our collider analyses. 
Since the mass spectra for our benchmark points listed in the preceding section suggest that massive electroweak gauge bosons are highly boosted, we assume that $v_{a/b/c}$ are (at least, effectively) massless for convenience of the subsequent argument. 

We here and henceforth denote any reconstructed mass quantity by the upper-case $M$. Two invariant mass variables are readily available, which are reconstructed masses $M_{bc}$ $(=\sqrt{(p_b+p_c)^2})$ and $M_{abc}$ $(=\sqrt{(p_a+p_b+p_c)^2})$ which are supposed to be the same as $m_\varphi$ and $m_{\kk}$, respectively. 
Assuming that the decay widths for $A_\kk$ and $\varphi$ are negligible, we see that they are very powerful in suppressing relevant SM backgrounds.  
Another set of mass variables are $M_{ab}$ $(=\sqrt{(p_a+p_b)^2})$ and $M_{ac}$ $(=\sqrt{(p_a+p_c)^2})$.
Without considerable spin correlation, their differential distributions develop the famous triangular shape spanning from 0 to the kinematic endpoint
\bea
M_{ab}^{\max}=M_{ac}^{\max}=\sqrt{m_\kk^2-m_\varphi^2}\,.
\eea
However, both $M_{ab}$ and $M_{ac}$ provide useful handles orthogonal to $M_{bc}$ and $M_{abc}$ because energy-momentum conservation $p_{A_\kk} = p_a +p_b+p_c$ implies the following sum rule:
\bea
M_{abc}^2 = M_{ab}^2+M_{bc}^2+M_{ac}^2\,, \label{eq:sumrule}
\eea
where we again assume massless visible particles in the equality.
Indeed, $M_{abc}$ and $M_{bc}$ enforce us to select ``signal-like'' background events in terms of both the mass spectrum and the underlying event topology. 
So, one may argue that they are sufficient to reduce background events and we do not benefit from additional mass variables. 
It turns out that still the extra invariant mass variables are beneficial in the sense that they enable us to access the remaining difference between the signal and the background processes, which is encoded in the shapes of their distributions.
This point will be explicitly demonstrated in the context of concrete signal channels in the next section. 

Finally, it is noteworthy that we have implicitly assumed that the three visible particles $v_a$, $v_b$, and $v_c$ are perfectly distinguishable although combinatorial ambiguity often arises in more realistic situations. 
Unfortunately, all signal channels of ours summarized in Table~\ref{tab:BPtable} face this issue, motivating us to devise appropriate prescriptions. 
Two types of combinatorial ambiguity are possible. 
\begin{itemize}
\item Type I: $v_b$ and $v_c$ are indistinguishable while $v_a$ is distinguishable from the others,
\item Type II: $v_a$, $v_b$, and $v_c$ all are indistinguishable.
\end{itemize}
The channel of three-gluon final state falls into Type II, while the others are categorized to Type I. 

For Type I, there is no ambiguity in $M_{bc}$ and $M_{abc}$, whereas some recipe is needed for $M_{ab}$ and $M_{ac}$. 
Denoting indistinguishable $v_b$ and $v_c$ by $j$, we consider two sets of experimental observables.
\bea
\hbox{Set I.1:}&& M_{aj_h},\, M_{aj_s} \label{eq:setI1}\\
\hbox{Set I.2:}&& M_{aj(\textrm{high})}\equiv \max[M_{ab}, M_{ac}],\, M_{aj(\textrm{low})}\equiv \min[M_{ab}, M_{ac}] \label{eq:setI2}
\eea
In Set I.1, we first rank $v_b$ and $v_c$ by their $p_T$-hardness (i.e., $j_{h(s)}=$ the harder (softer) of the two) and form the respective invariant mass variables, while in Set I.2, we rank the two possible invariant masses by their magnitude~\cite{Lester:2005je,Lester:2006cf,Burns:2009zi,Dev:2015kca,Kim:2015bnd}.
Which one is superior to the other is beyond the scope of this paper, and their usefulness can be discussed in the context of specific signal channels. 

On the other hand, for Type II, all two-body invariant mass variables, $M_{ab}$, $M_{bc}$, and $M_{ac}$, are {\it not} experimental observables.
Again denoting indistinguishable $v_a$, $v_b$, and $v_c$ by $j$, we propose two possible prescriptions.
\bea
\hbox{Set II.1:}&& M_{\tilde{a}j_h},\, M_{\tilde{a}j_s} \\
\hbox{Set II.2:}&& M_{jj(\textrm{high})}\equiv \max[M_{ab}, M_{bc}, M_{ac}],\, M_{jj(\textrm{mid})}\equiv \textrm{med}[M_{ab}, M_{bc}, M_{ac}],\nonumber \\
&&M_{jj(\textrm{low})}\equiv \min[M_{ab}, M_{bc},  M_{ac}]
\eea
For Set II.1, we guess $v_a$ among the three particles by, for example, their $p_T$-hardness and repeat the same procedure as in Set I.1 with $\tilde{a}$ symbolizing the conjectured $v_a$. 
In Set II.2, we rank all three invariant masses by their magnitude followed by constructing invariant mass distributions in the maximum, the median, and the minimum~\cite{Kim:2015bnd}.
Again the discussion on their actual performance will be available in the context of concrete signal processes.


\section{Results for LHC Signals}
\label{results}
In this section, we study the LHC signals for the model discussed in Sec.~\ref{Sec:overview}. We focus on the production and dominant decay channels of the lightest KK particles corresponding to the SM gauge bosons, 
employing the representative benchmark points presented in Table~\ref{tab:BPtable}.
For each channel, we take two benchmark points, which correspond to two values of the radion mass: 1.0 TeV and 1.5 TeV. 
We present our results in the order of KK photon, KK gluon, and KK $W/Z$ channels.

\subsection{KK photon: photon $+$ dijet}
\label{sec:akk}
We begin by considering the production and decay of KK photons in our model. 
As discussed before, the final state particles in the {\it dominant} decay channel are a SM photon and two jets. 
We will find that indeed a small rate in this signal process limits the associated discovery potential. 
As the other decay modes do not have a large enough rate, we simply focus on the photon $+$ dijet channel via two benchmark points $\gamma$-$\gamma gg$-BP1 and $\gamma$-$\gamma gg$-BP2, defined in Table~\ref{tab:BPtable}.
Given the final state particles, the dominant SM background is a single photon plus two QCD jets.




Before proceeding into the detailed analysis, we remark that it is useful to impose parton-level ``pre''-selection cuts to generate signal and background events in the appropriate region of phase space.
These parton-level cuts are chosen such that there is always a final analysis cut, much stronger than the corresponding parton-level pre-selection cut. 
This allows us to remain conservative about the detector smearing effects. 
The robust features of our signal in this channel, which are useful to discriminate against the background, are a high transverse momentum for each of the two jets and the photon,\footnote{An alternative approach to reject background events would be to apply the cut on the photon energy. Since the photon comes from the decay of KK photon which is singly-produced at the leading order, in the photon energy distribution, events are likely to populate near the fixed energy value which would have been measured in the rest frame of the KK photon~\cite{Agashe:2012bn,Chen:2014oha}.}
and a large invariant mass formed by the two jets. 
With these motivations, at the parton level, we apply $p_{T,j} > 150 \GeV$ for the two jets, $p_{T,\gamma} > 150 \GeV$ for the photon, and $M_{jj} > 500 \GeV$ for the jet pair. These cuts are presented in the cut flow Table~\ref{tab:AKKcutflow}. 
The effectiveness of these pre-selection cuts is reflected in their efficiency: the signal cross section reduces only marginally (65\% and 58\% for $\gamma$-$\gamma gg$-BP1 and $\gamma$-$\gamma gg$-BP2 data sets, respectively), while the $\gamma jj$ background gets reduced significantly (by $3.4\times 10^{-5}\%$).

After imposing these cuts,
we streamline the parton-level signal and background events to \textsc{Pythia} and \textsc{Delphes} as per our general simulation scheme.
As our simulation study is done at the detector level, we also consider three-jet events for which one of the jets is misidentified as an isolated photon.
The ATLAS Collaboration has reported the photon fake rate to be around $10^{-4}$~\cite{Aad:2009wy}. 
A typical source is high-$p_T$ neutral pions, which come from jets, decaying into two photons.
We use \textsc{Delphes} with the default setup, which yields a similar fake rate.
We find that most of the three-jet background can be removed by our choice of cuts, without affecting the final results significantly. 

\begin{figure}[t]
    \centering

    \includegraphics[width = 7 cm]{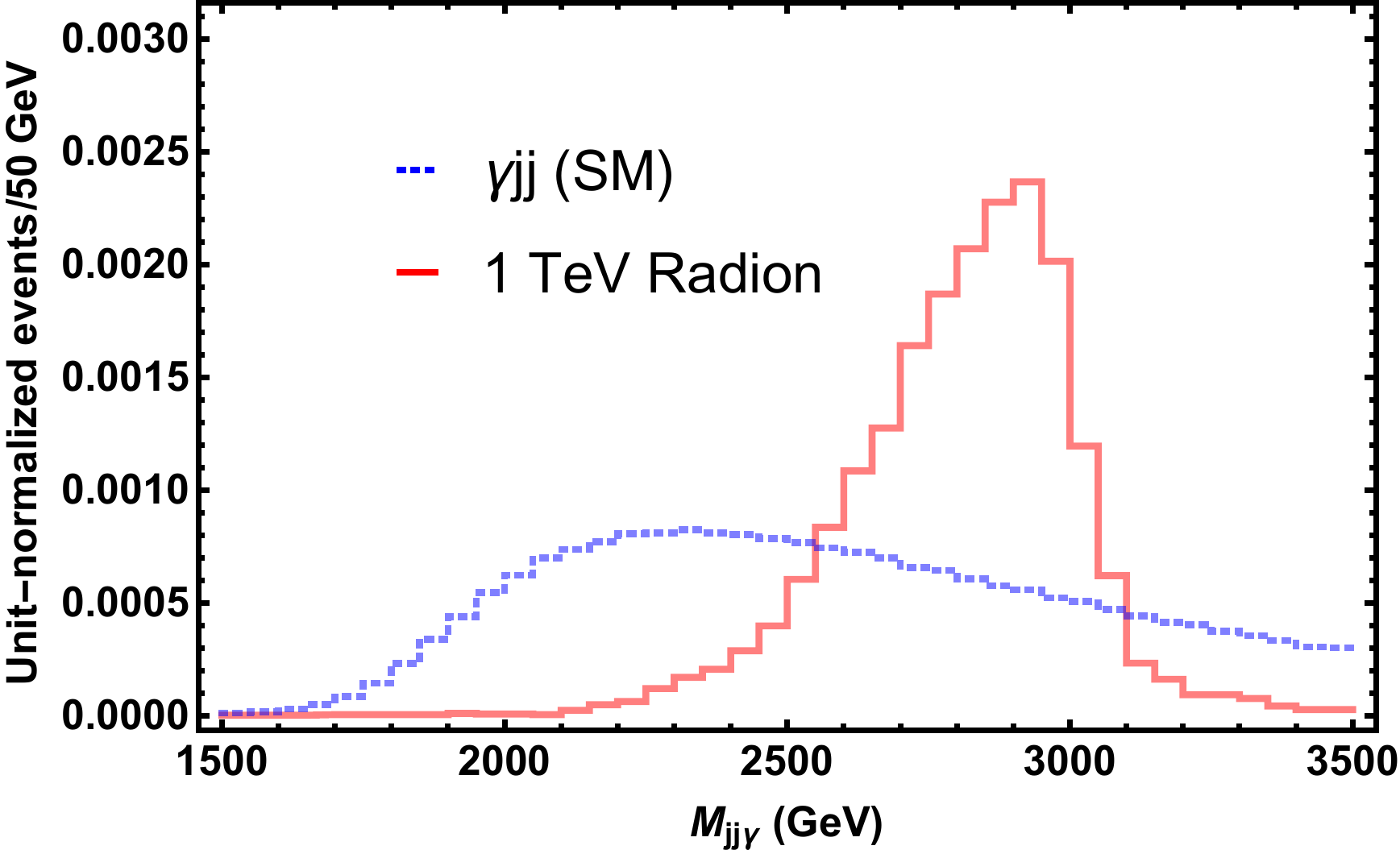}
    \includegraphics[width = 7 cm]{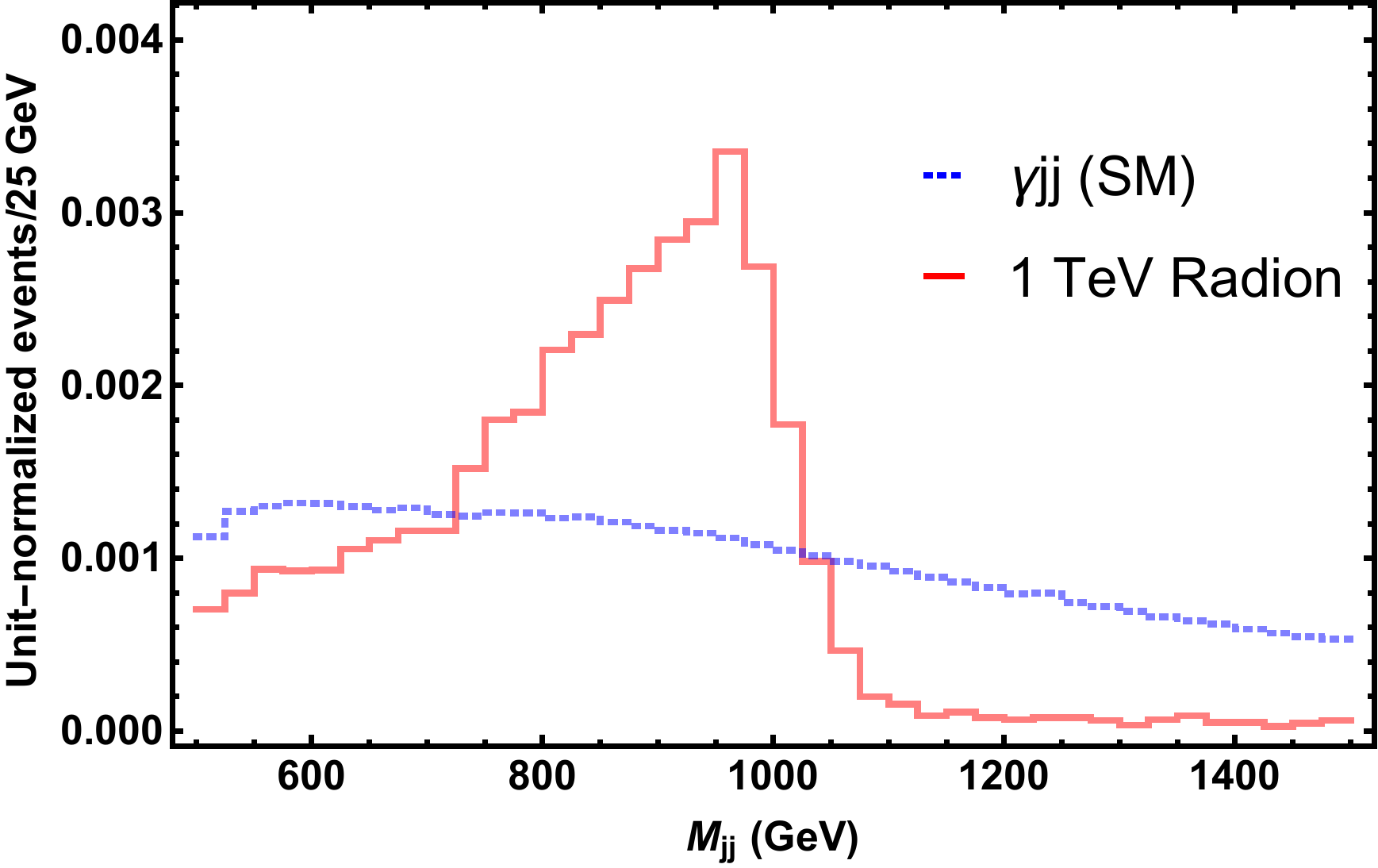}
    \includegraphics[width = 7 cm]{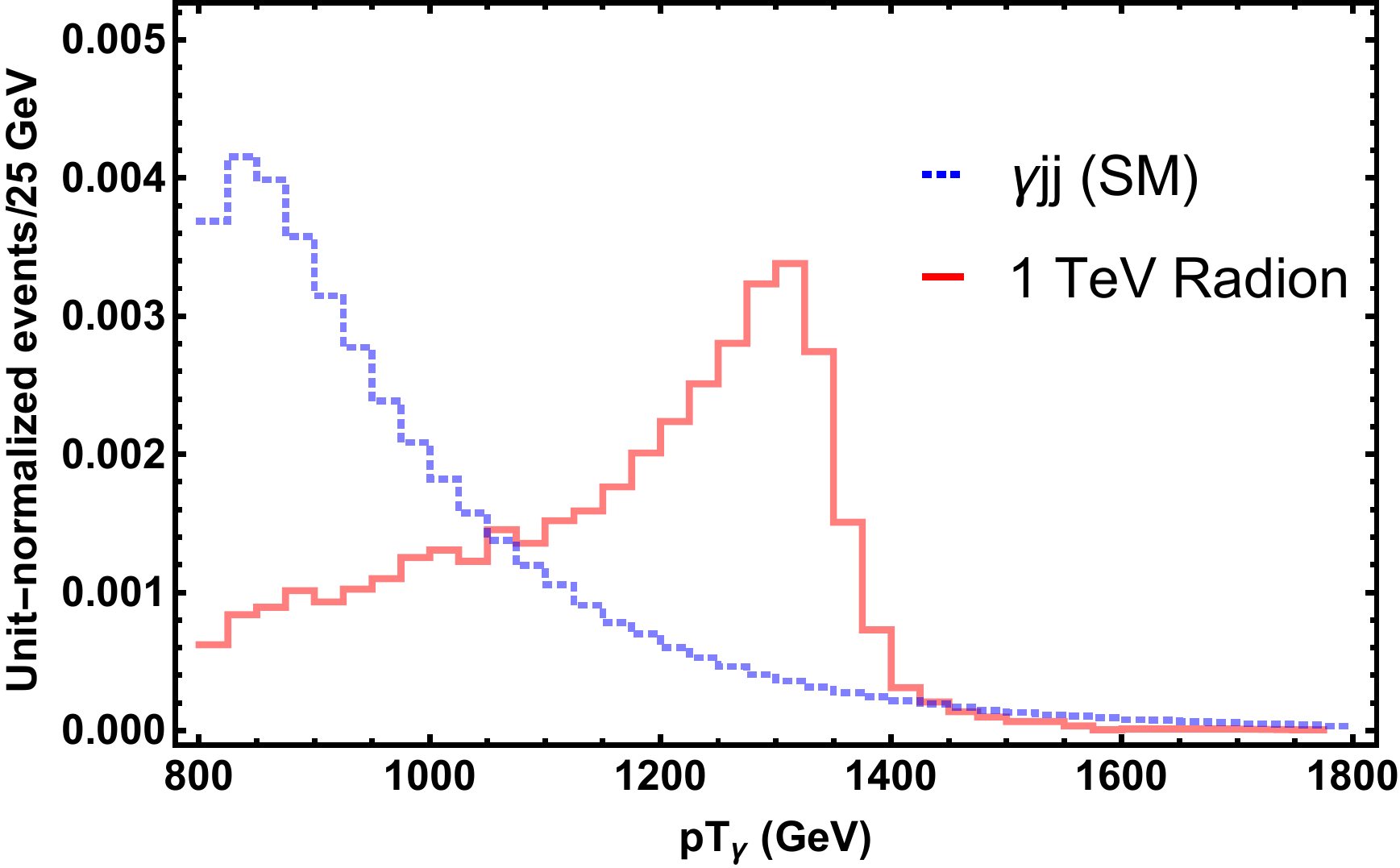}
    \includegraphics[width = 7 cm]{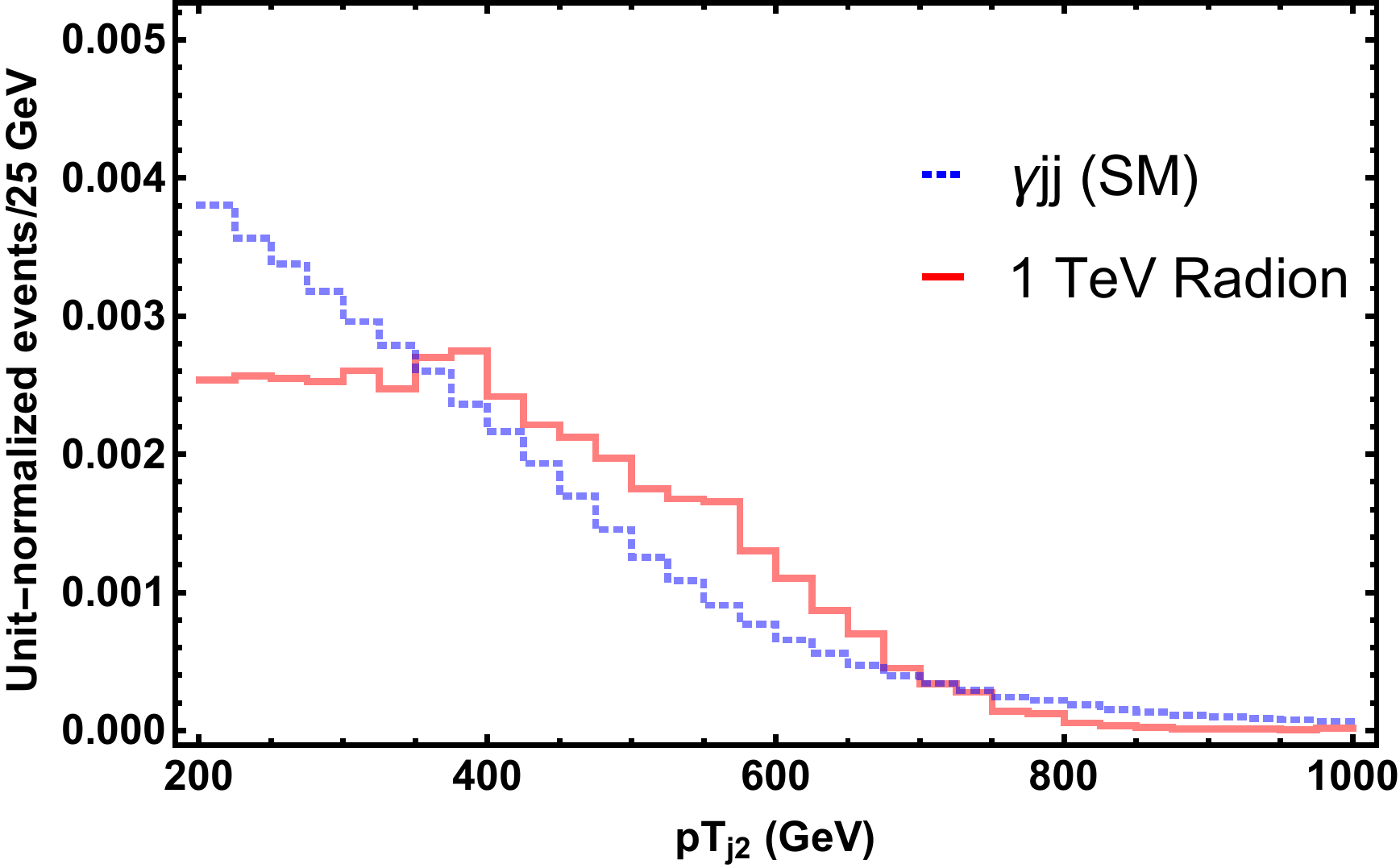}
    \includegraphics[width = 7 cm]{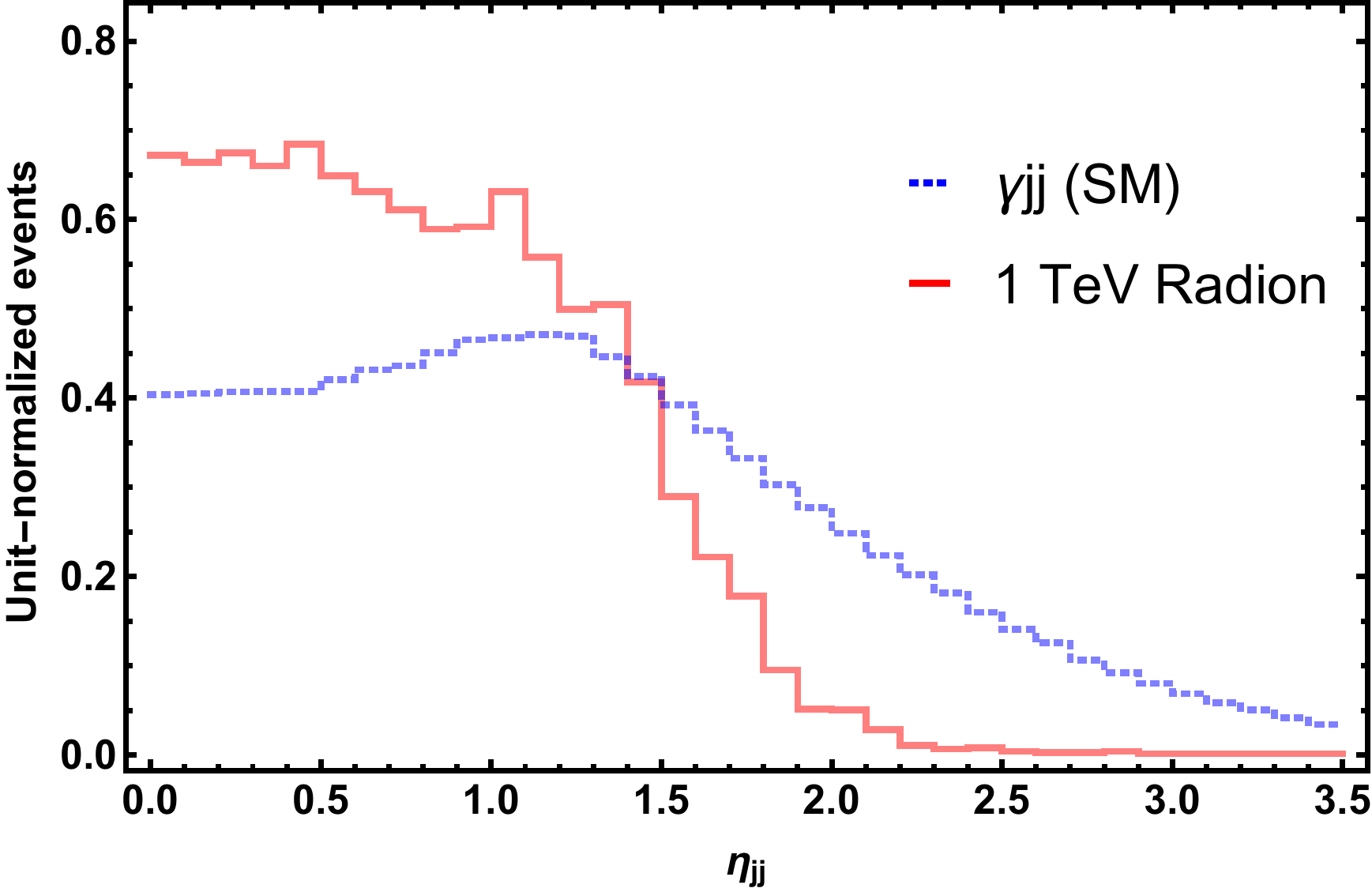}
    \caption{
    $\gamma$-$\gamma g g$-BP1 benchmark point: Distributions of variables: $M_{jj\gamma}$ (top-left), $M_{jj}$ (top-right), $p_{T,\gamma}$ (mid-left), $p_{T,j_2}$ (mid-right) and $\eta_{jj}$ (bottom) for signal (red solid) and background (blue dashed). 
    }
    \label{fig:akk_RadMass1tev}
\end{figure}

\begin{figure}[t]
    \centering

    \includegraphics[width = 7 cm]{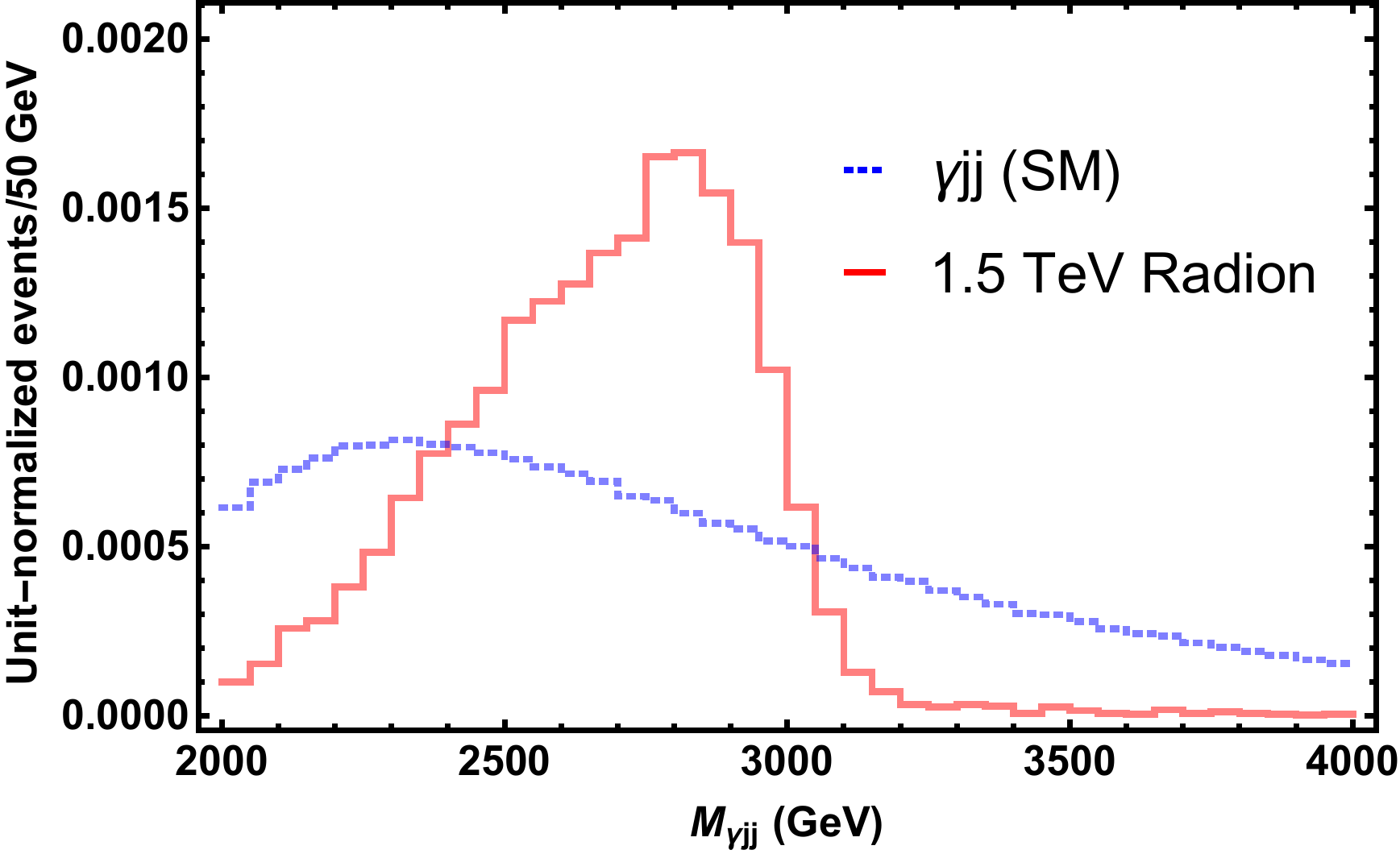}
    \includegraphics[width = 7 cm]{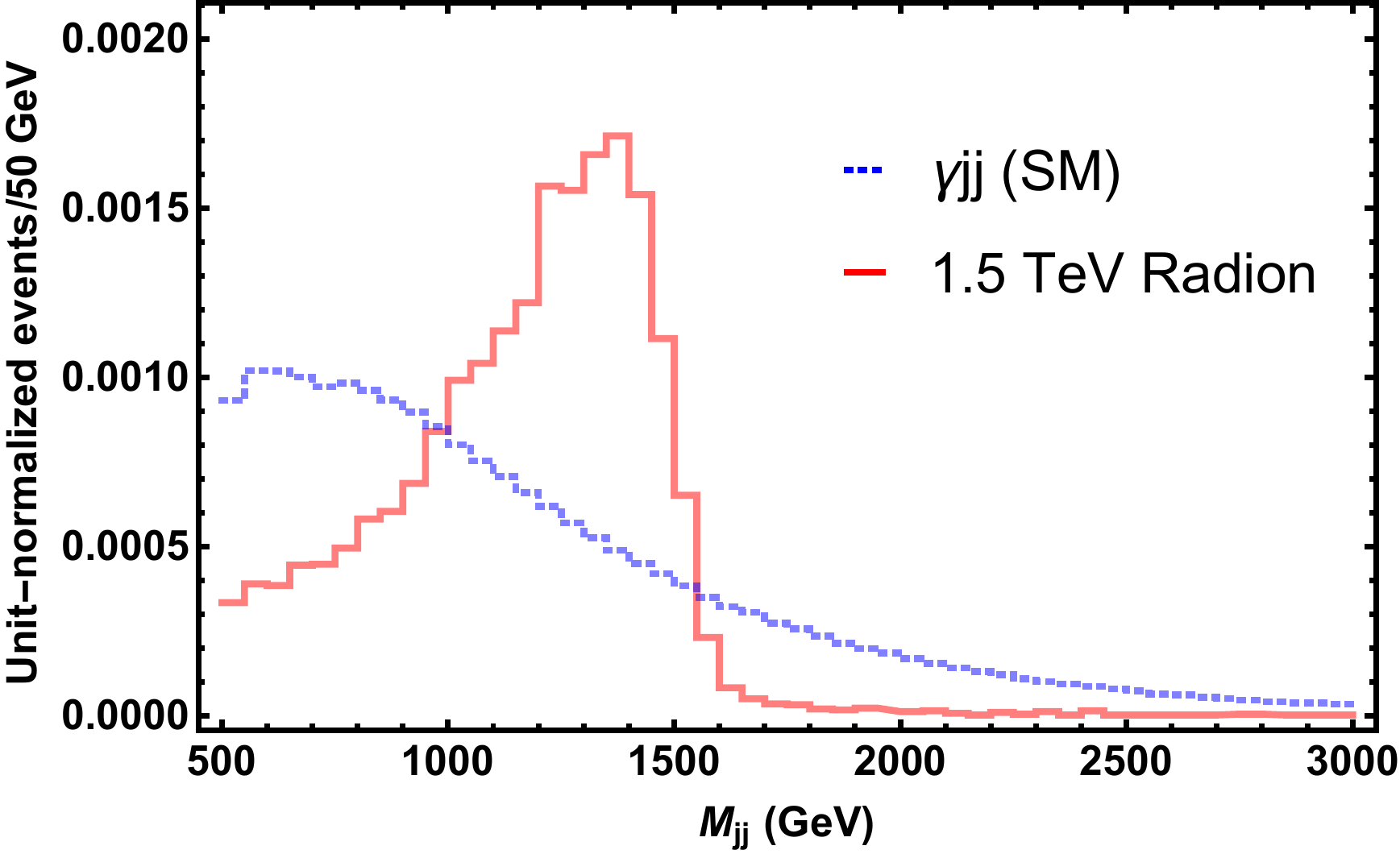}
    \includegraphics[width = 7 cm]{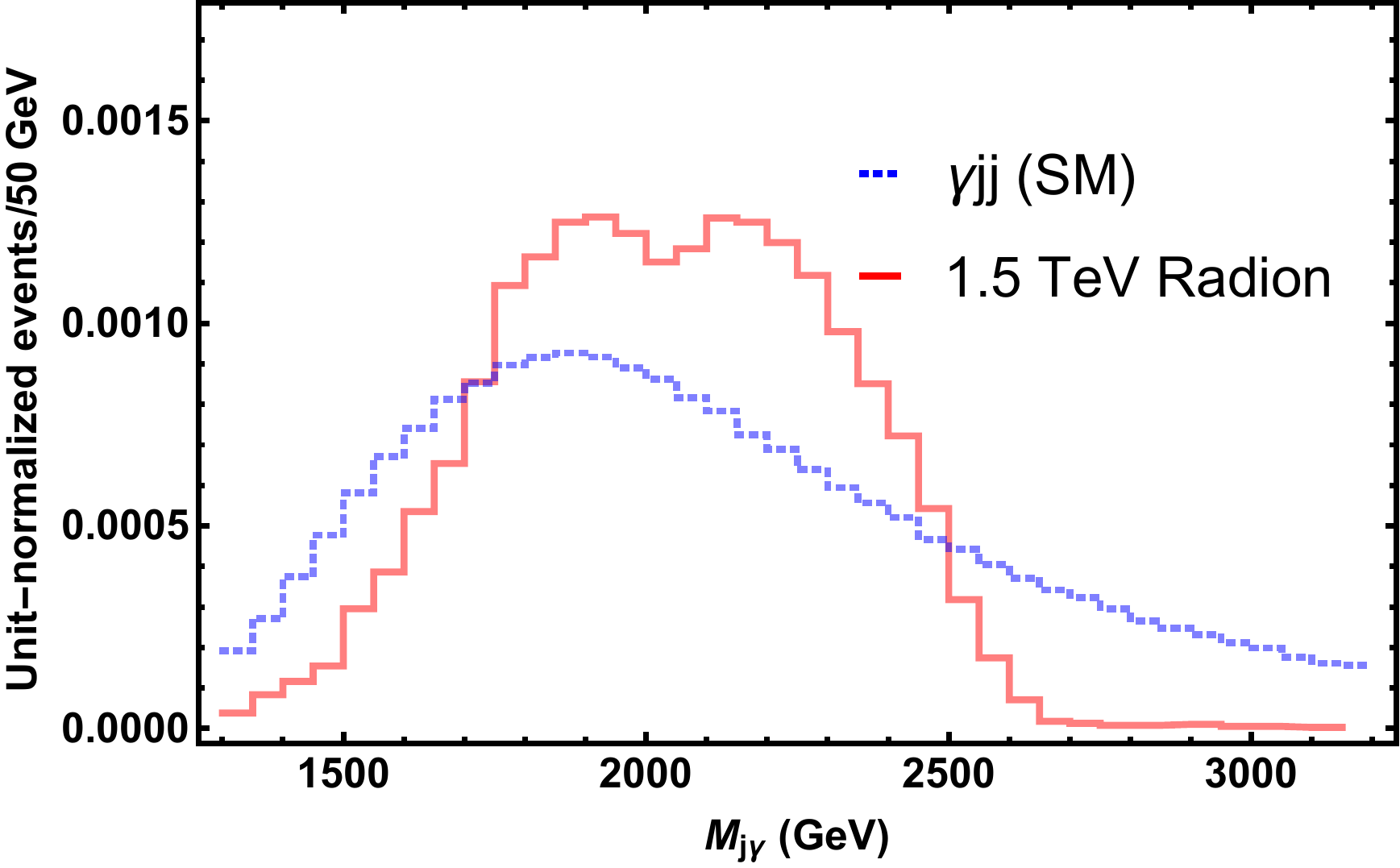}
    \includegraphics[width = 7 cm]{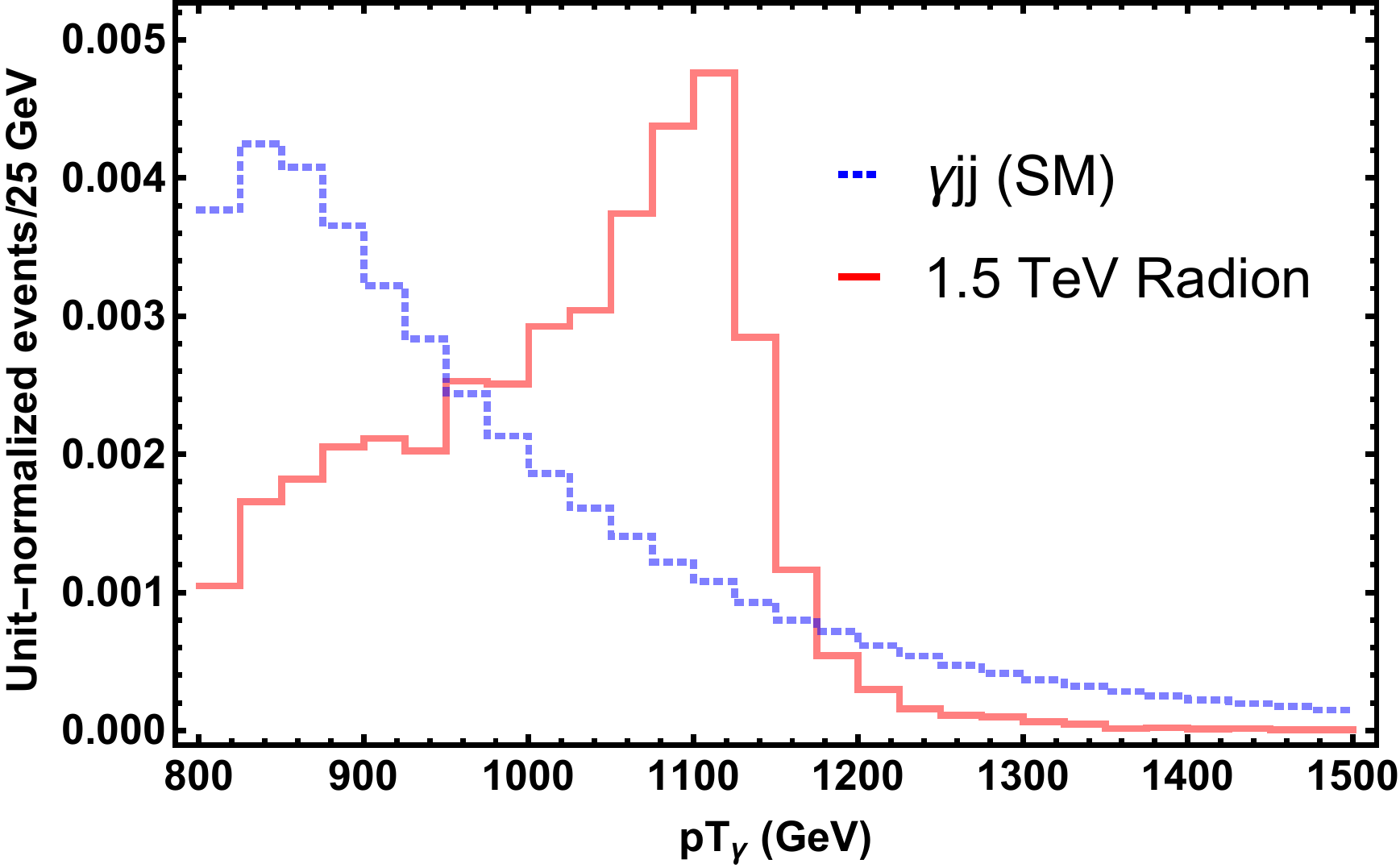}
    \includegraphics[width = 7 cm]{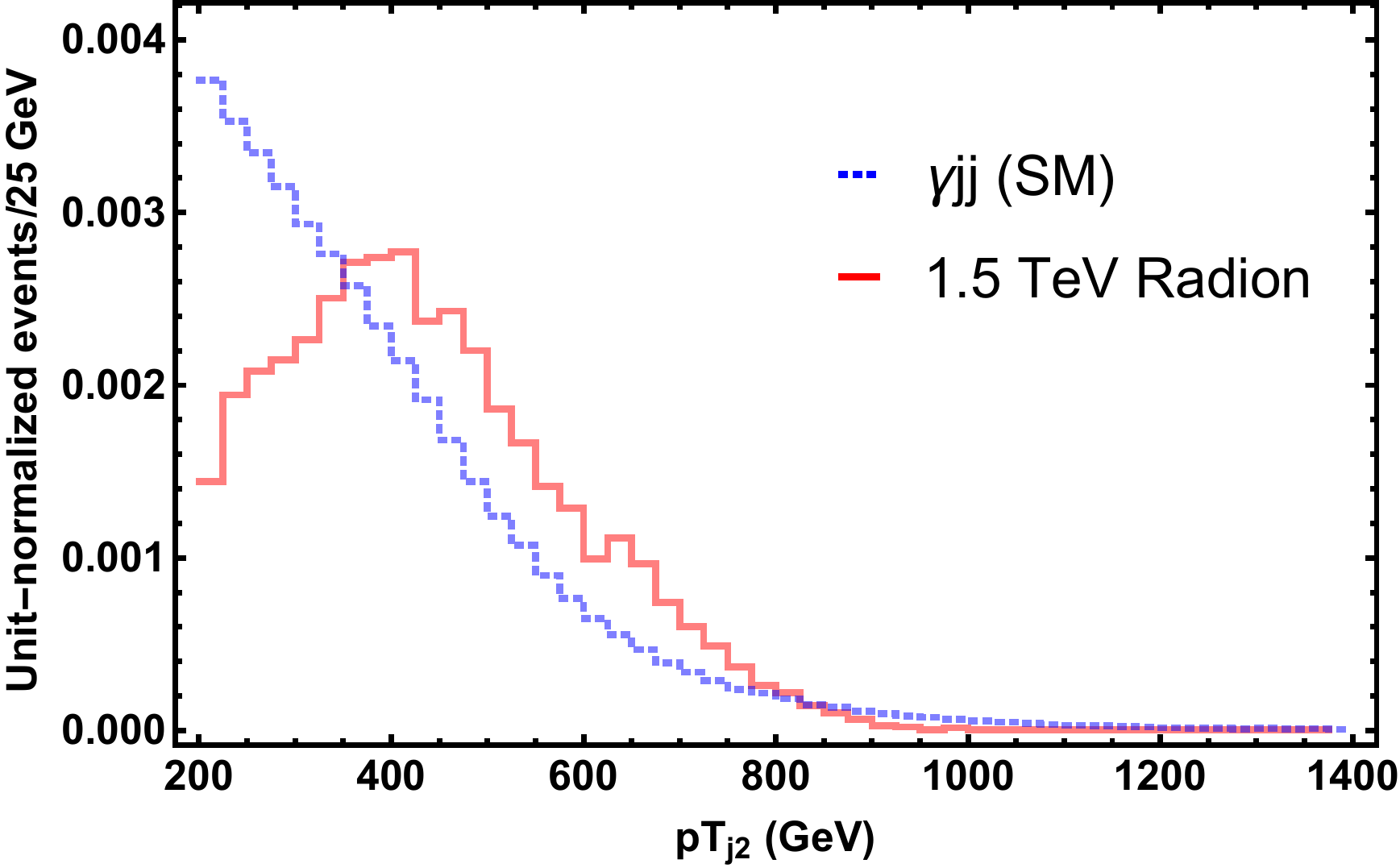}
    \caption{Distributions of variables for $\gamma$-$\gamma g g$-BP2: $M_{jj\gamma}$ (top left), $M_{jj}$ (top right), $M_{j\gamma}$ (mid left), $p_{T,\gamma}$ (mid right) and $p_{T,j_2}$ (bottom)  for signal (red) and background (blue).
    }
    \label{fig:akk_RadMass1.5tev}
\end{figure}

Defining $N_\gamma$ and  $N_j$ as the number of photons and jets in the event, respectively, we only consider detector events with 
\bea
N_\gamma \geq 1,\,\,\, N_j \geq 2\,,
\eea
in order to focus on the relevant background events. 
As motivated earlier, the signal events have \textit{two} invariant mass variables that can be used to get rid of the background events. 
They essentially give rise to the masses of KK photon and radion, and we denote them by
$M_{jj\gamma}$ and $M_{jj}$, respectively. 
In addition, the transverse momentum of the photon $p_{T\gamma}$ in the signal is very hard, allowing another independent way to suppress the background events. 
We also find that the transverse momentum of the second hardest jet, $p_{T,j_{2}}$, the absolute rapidity distance between the two hardest jets, $\eta_{jj}$, and the maximum of the invariant mass between a photon and the two energetic jets, 
$M_{j\gamma({\rm high})} (= \max[M_{j_{1}\gamma}, M_{j_{2}\gamma}])$ are further useful in reducing the background.
In Figs.~\ref{fig:akk_RadMass1tev} ($\gamma$-$\gamma g g$-BP1) and~\ref{fig:akk_RadMass1.5tev} ($\gamma$-$\gamma g g$-BP2), we exhibit the unit-normalized distributions of signal and background events in the variables discussed so far.
These events are after imposing the pre-selection cuts. 


We then provide the final flow of signal and background events according to the cuts discussed so far in Table~\ref{tab:AKKcutflow} in terms of their respective cross sections. 
From the cut flow, we observe that all the cuts are almost equally important to reduce the background. 
Defining our (statistical) significance as $S/\sqrt{B}$ with $S$ and $B$ being the number of signal and background events, respectively, we find that a moderate significance of 0.98$\sigma$ (0.97$\sigma$) can be achieved for $\gamma$-$\gamma gg$-BP1 ($\gamma$-$\gamma gg$-BP2) at an integrated luminosity of 300 fb$^{-1}$. 
This small significance results primarily from a small rate for the signal as well as the presence of jets, which are smeared substantially, restricting the efficiencies of the invariant mass window cuts. 
Nevertheless, once we increase the statistics by a factor of 10, i.e., an integrated luminosity of 3000 fb$^{-1}$, we may achieve 3.10$\sigma$ and 3.09$\sigma$ for the two benchmark points. 

\begin{table}[t]
\centering
\begin{tabular}{|c|c|c|c|}
\hline 
Cuts & $\gamma$-$\gamma gg$-BP1 & $\gamma$-$\gamma gg$-BP2 & $\gamma j j$ \\
\hline \hline
No cuts & 0.20 & 0.29 & ($8.65\times 10^7$) \\
$N_\gamma \geq 1, N_j \geq 2$, pre-selection cuts & 0.10 & 0.13 & 18.56  \\
\hline
$ M_{jj\gamma} \in [2600, 3100] \GeV$ & 0.08 & -- & 4.55 \\
$ M_{jj} \in [700, 1050] \GeV$ & 0.06 & -- & 1.22 \\
$p_{T,\gamma} \leq 1200 \GeV$ & 0.03 & -- & 0.28 \\
\hline
$ M_{jj\gamma} \in [2200, 3200] \GeV$ & -- & 0.12 & 10.03 \\
$ M_{jj} \in [1000, 1600] \GeV$ & -- & 0.10 & 4.55 \\
$ p_{T,\gamma} \in [950, 1200] \GeV$ & -- & 0.07 & 1.66 \\
$M_{j\gamma({\rm high})} \leq 2600 \GeV$ & -- & 0.07 & 1.54 \\
\hline
$S/B$ & 0.10 & 0.05 & -- \\
$S/\sqrt{B}$ ($\mathcal{L}=300$ fb$^{-1}$) & 0.98 & 0.97 & -- \\
$S/\sqrt{B}$ ($\mathcal{L}=3000$ fb$^{-1}$) & 3.10 & 3.09 & -- \\
\hline
\end{tabular}
\caption{Cut flows for signal and major background events in terms of their cross sections (in fb). The number in the parentheses for $\gamma j j$ is obtained with basic cuts ($p_{T,j} > 20 \GeV$, $p_{T,\gamma} > 10 \GeV$, $|\eta_j| < 5$, $|\eta_\gamma| < 2.5$, $\Delta R_{jj} > 0.4$, $\Delta R_{j\gamma} > 0.4$, $\Delta R_{\gamma\gamma} > 0.4$) at the generation level to avoid divergence. The pre-selection cuts ($p_{T,j} > 150 \GeV$, $p_{T,\gamma} > 800 \GeV$, $M_{jj} > 500 \GeV$) are imposed at parton level to generate events in the relevant phase space, and are reimposed at the detector level.}
\label{tab:AKKcutflow}
\end{table} 

\subsection{KK gluon}
We next consider the production and decay of KK gluons in our model. Due to a higher production cross section, there are multiple decay channels here that become relevant phenomenologically: the trijet, jet $+$ diphoton, and jet $+$ diboson ($W/Z$) decay modes.
In all three, one regular jet comes from the decay of the KK gluon.
The different radion decay modes give rise to the other two objects in the final state. Since the hierarchy in radion decay modes are dictated largely by the hierarchy in SM gauge couplings and multiplicity factors, the decay modes of the radion, in decreasing order of magnitude, are to $jj$, $W$/$Z$, and $\gamma\gamma$ final states.


For the case of radion decay to $W$/$Z$, there are multiple final states to consider. 
Since $\text{BR}(\varphi \to ZZ) \propto g_{Z_{\rm KK}}^{-4}$ as in eq.~\eqref{eq:Radion_BR} and $g_{Z_{\rm KK}} = g_{W_{\rm KK}}/\sqrt{1-g_{W_{\rm KK}}^2/g_{\gamma_{\rm KK}}^2}$ is necessarily larger than $g_{W_{\rm KK}}$, the $ZZ$ mode is heavily suppressed compared to the $WW$ mode. We therefore focus on the $WWj$ channel. There are three final states of potential interest: fully hadronic $JJj$ (where $J$ denotes a hadronic merged jet coming from a boosted $W$), semileptonic $\ell \nu J j$, and fully leptonic $\ell \nu \ell \nu j$. The fully leptonic final state is the cleanest, but has the smallest branching ratio.
Moreover, it contains {\it two in}visible neutrinos manifesting themselves as a missing transverse energy, so it is impossible to reconstruct the radion resonance mass which is one of the crucial handles to suppress relevant background events.\footnote{One could instead try transverse mass variables (e.g., $M_T$, $M_{T2}$~\cite{Lester:1999tx} or their variants) or even (3+1)-dimensional variables (e.g., $M_2$~\cite{Mahbubani:2012kx,Cho:2014naa}), but we do not pursue our study in this direction.}
The fully hadronic and semi-leptonic channels have similar branching fractions, and in existing LHC searches for simple diboson resonances have comparable sensitivity in this mass range~\cite{Aaboud:2016okv}.
We focus on the fully hadronic channel which allows a rather sharp feature in the reconstructed radion mass, which we shall demonstrate shortly. 

\subsubsection{Decay to trijet }
\label{sec:trijet}
We consider two benchmark points $g$-$ggg$-BP1 and $g$-$ggg$-BP2, and their model parameters are summarized in Table~\ref{tab:BPtable}. Obviously, the dominant SM background comes from the three-jet QCD process.  
%
Again we need to consider pre-selection cuts to generate signal and background events in the relevant part of phase space. 
Our choice of mass spectra enforces the three jets to come with high transverse momenta and the three two-jet invariant masses to be large.
We use these signal features, which are distinctive from typical background events, to establish the pre-selection cuts.
With these motivations, at the parton level, we choose $p_{T,j} > 150 \GeV$ for the three jets and $M_{jj} > 500 \GeV$ for the three combinations of jet pairing.
%
%
Their effectiveness is reflected in the efficiency: the signal cross section is reduced only marginally (88\% and 92\% for $g$-$ggg$-BP1 and $g$-$ggg$-BP2 benchmark points, respectively), whereas the three-jet QCD background is significantly suppressed by $5.5\times 10^{-3}\%$.

\begin{figure}[t]
    \centering

    \includegraphics[width = 7 cm]{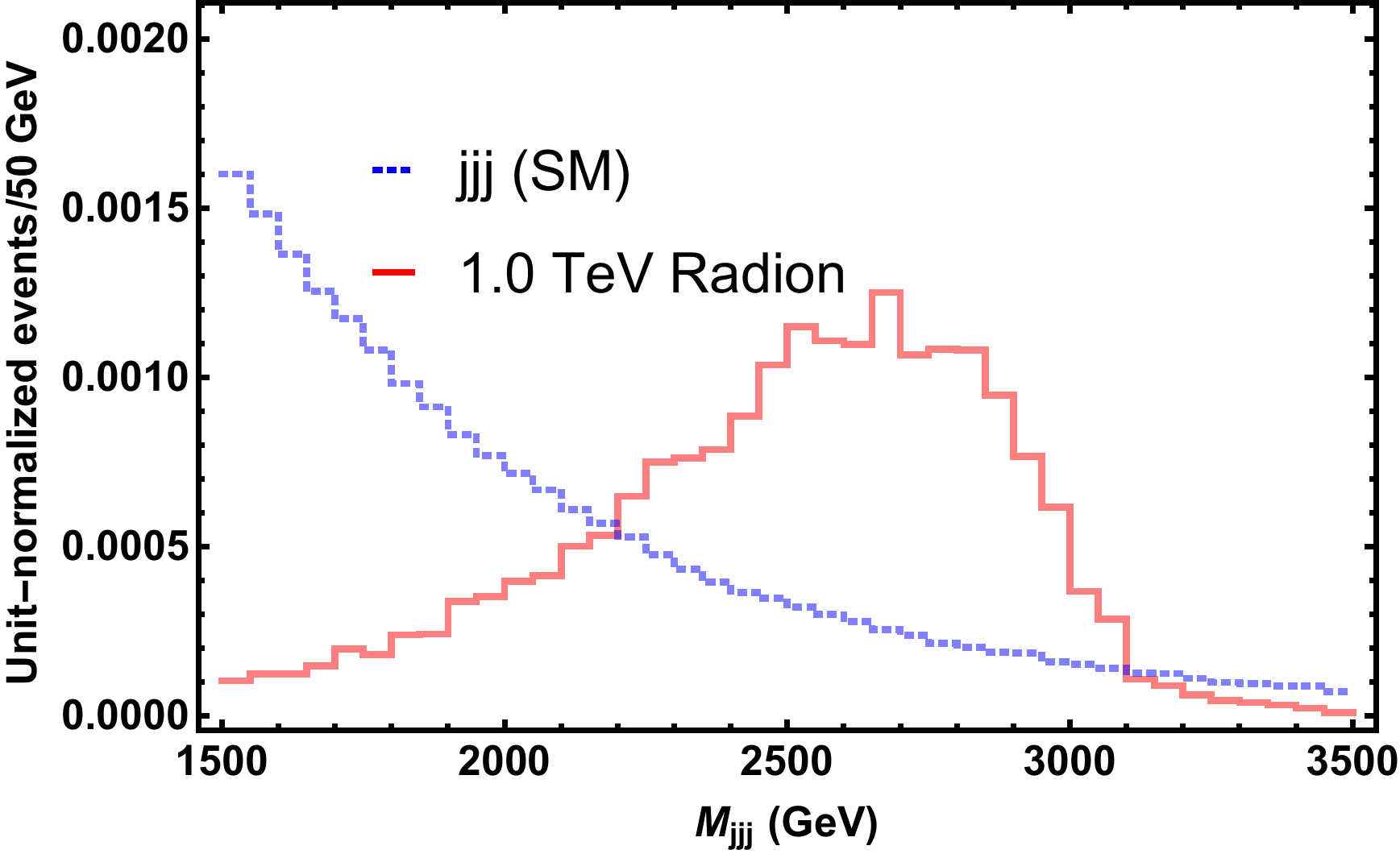}
    \includegraphics[width = 7 cm]{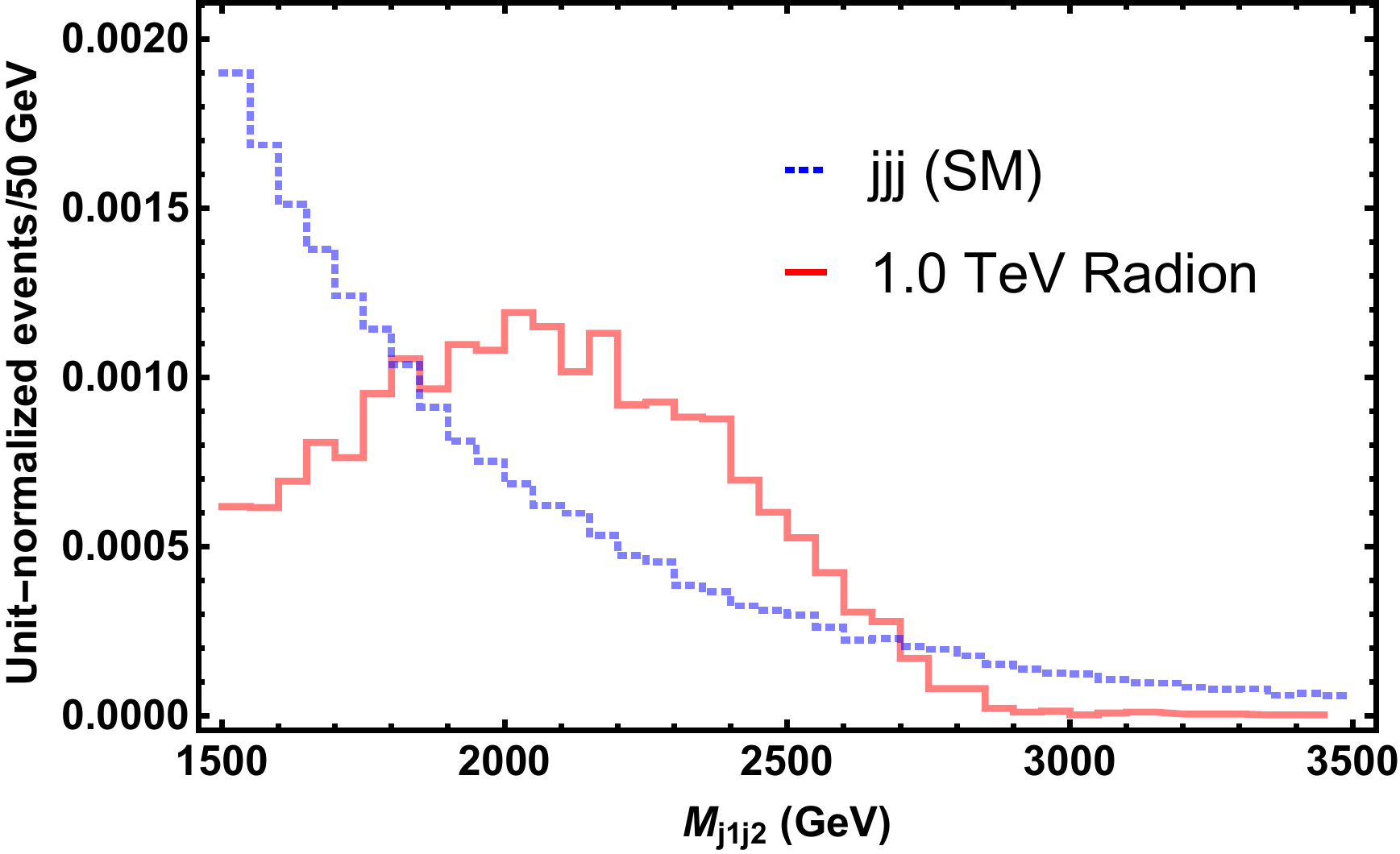}
    \includegraphics[width = 7 cm]{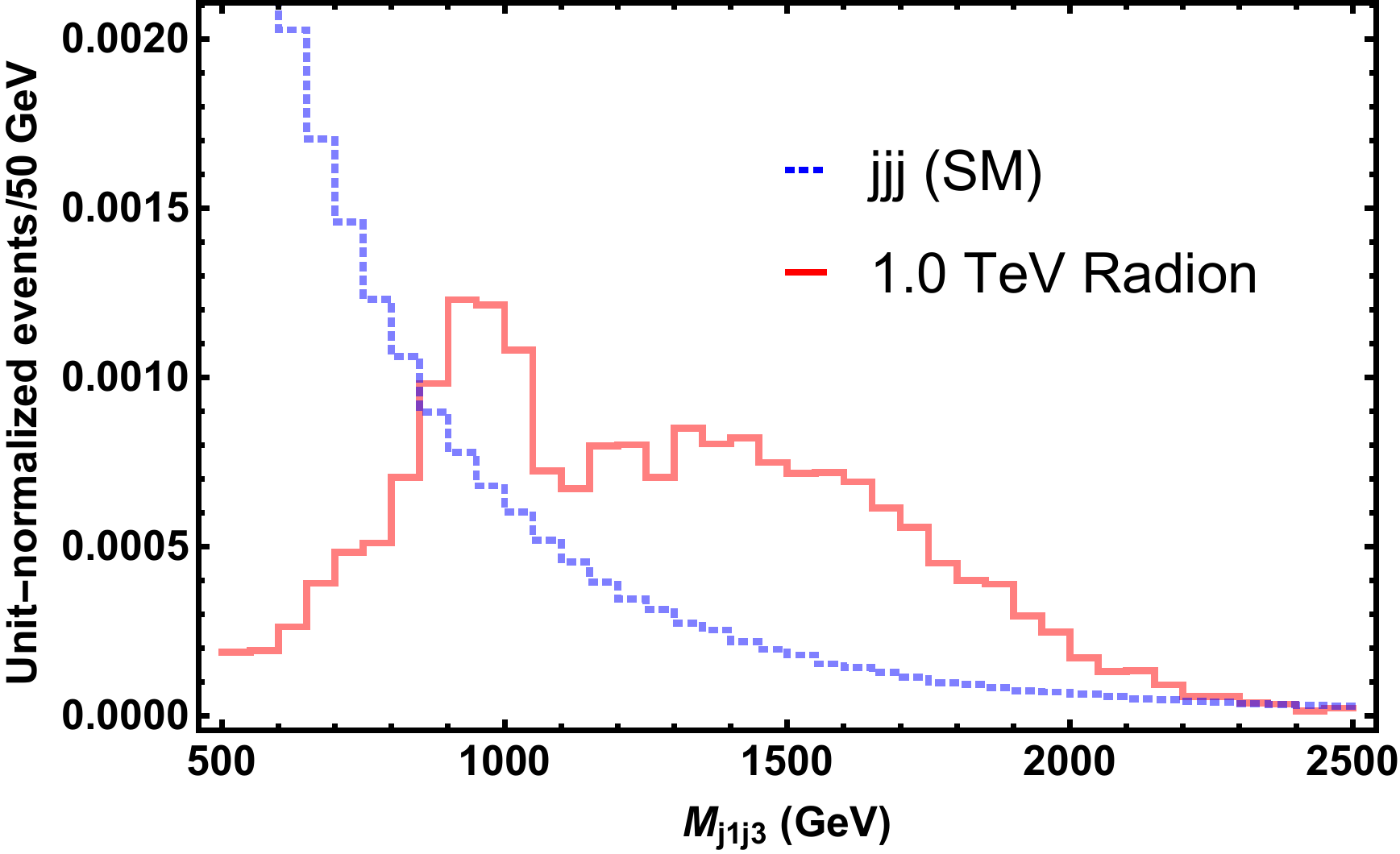}
    \includegraphics[width = 7 cm]{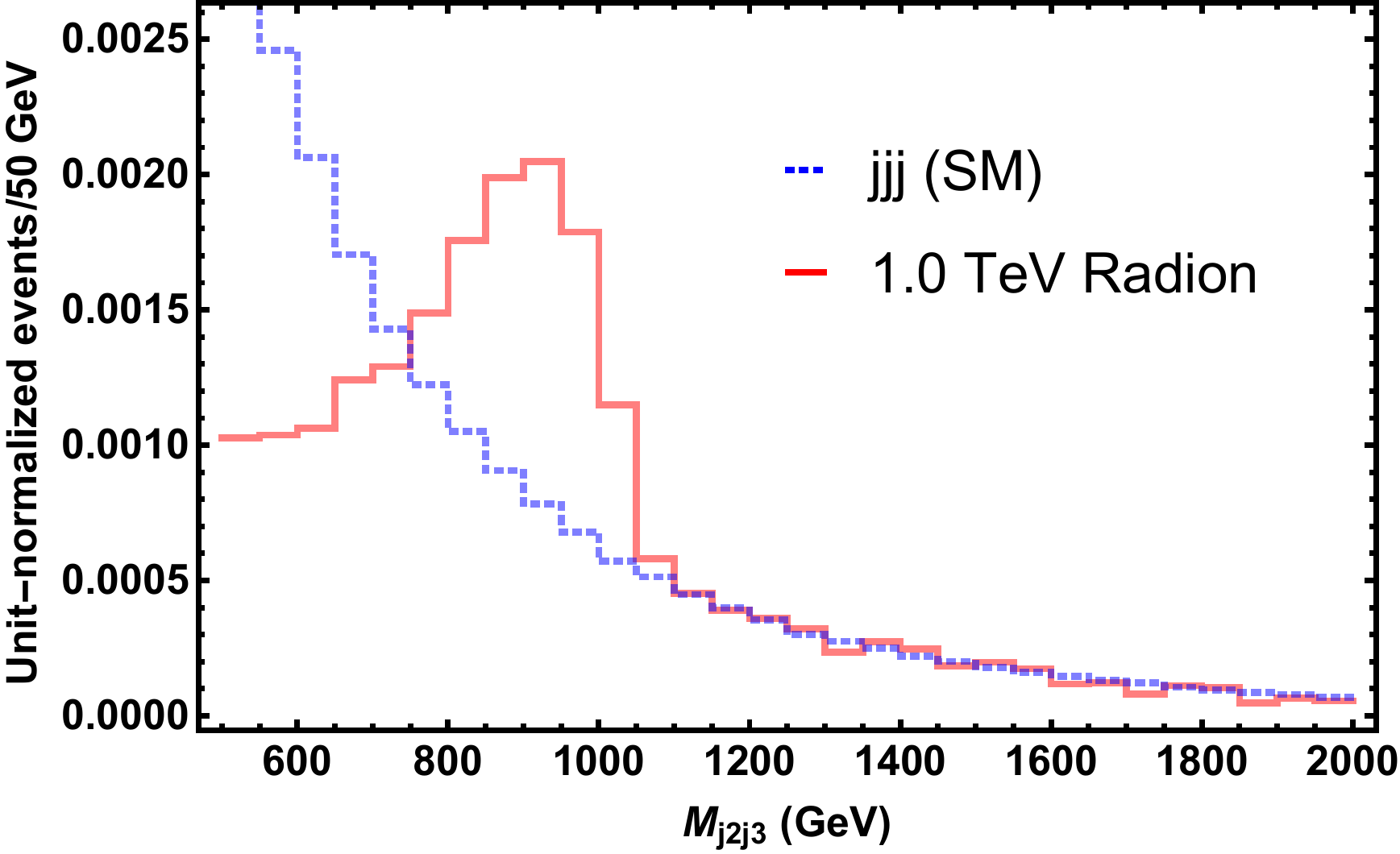}
    \includegraphics[width = 7 cm]{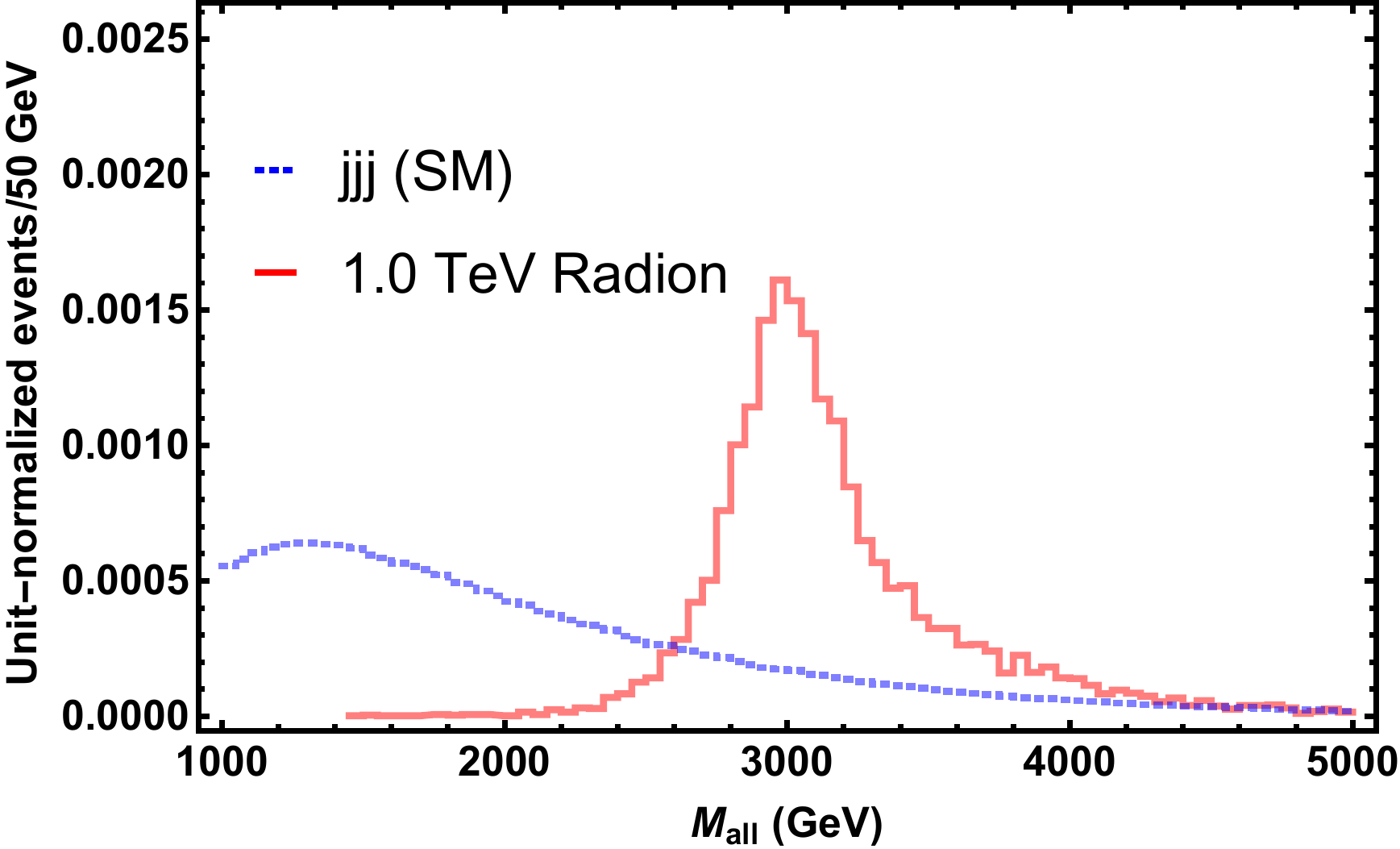}
    \includegraphics[width = 7 cm]{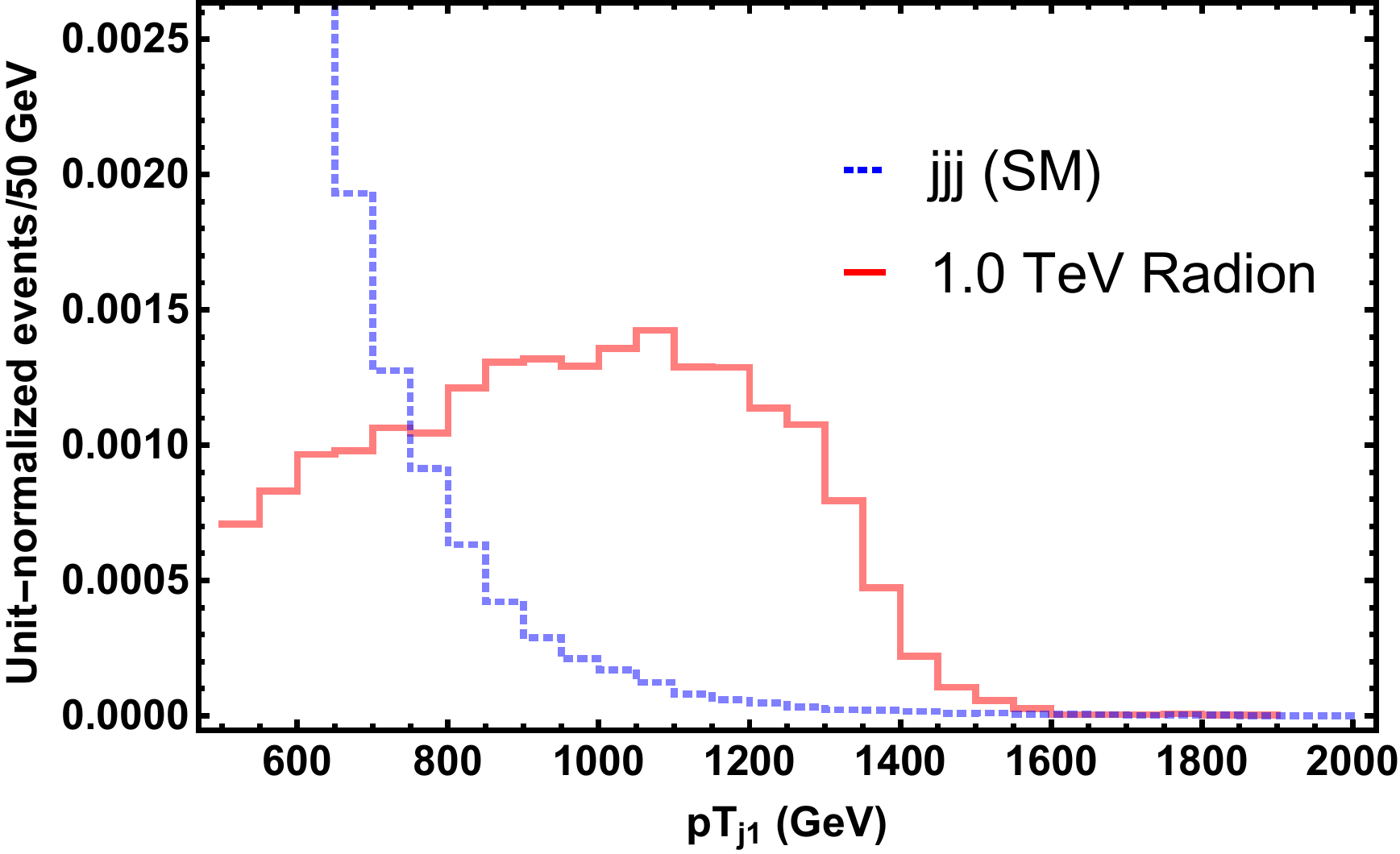}
    \caption{
    $g$-$ggg$-BP1 benchmark point: Distributions of variables: $M_{jjj}$ (upper-left), $M_{j_1j_2}$ (upper-right), $M_{j_1j_3}$ (middle-left), $M_{j_2j_3}$ (middle-right), $M_{all}$ (lower-left), $p_{T,j_1}$ (lower-right) for signal (red solid histograms) and background (blue dashed histograms).}
    \label{fig:gkk_jjj_RadMass1tev}
\end{figure}

\begin{figure}[t]
    \centering

    \includegraphics[width = 7 cm]{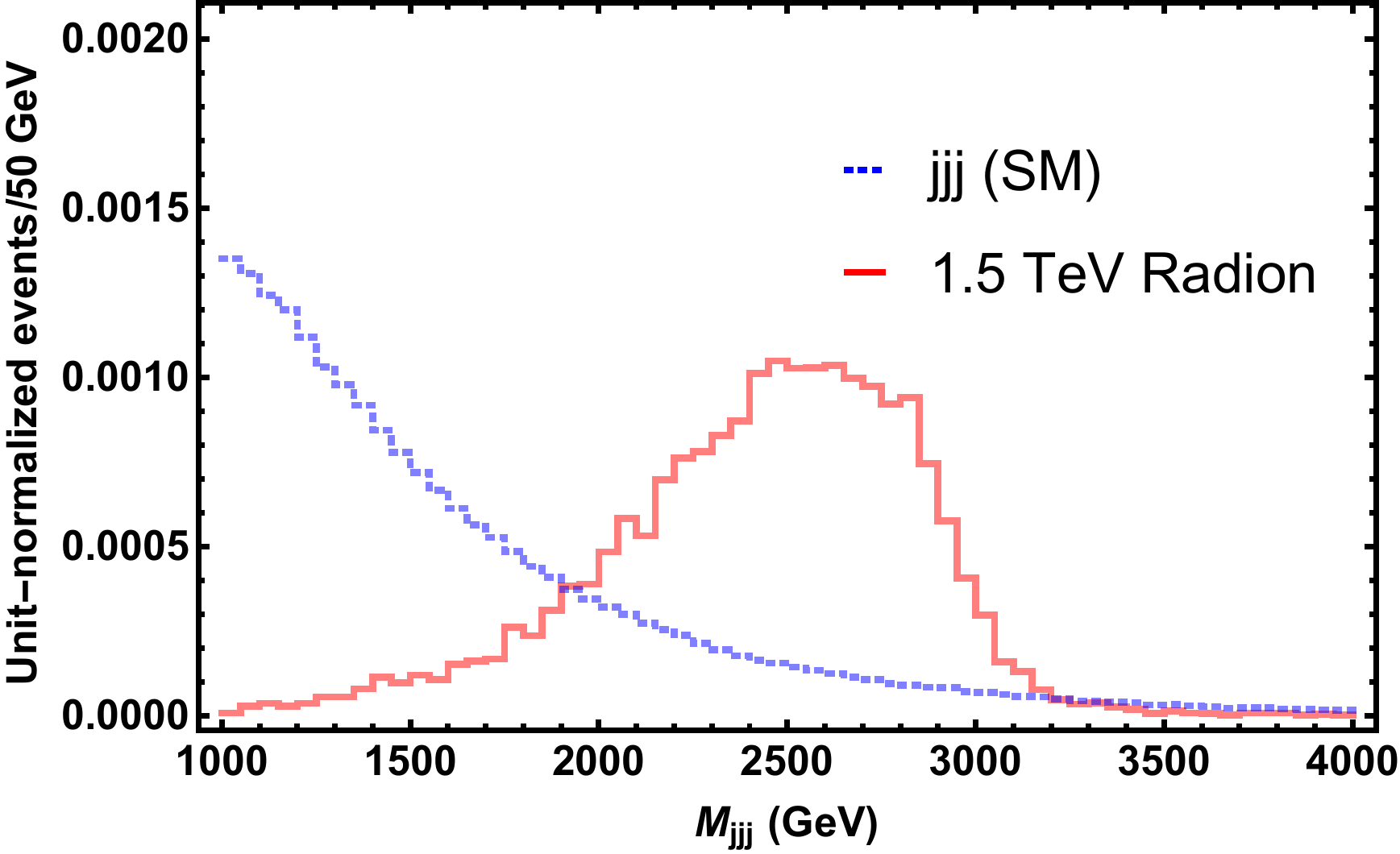}
    \includegraphics[width = 7 cm]{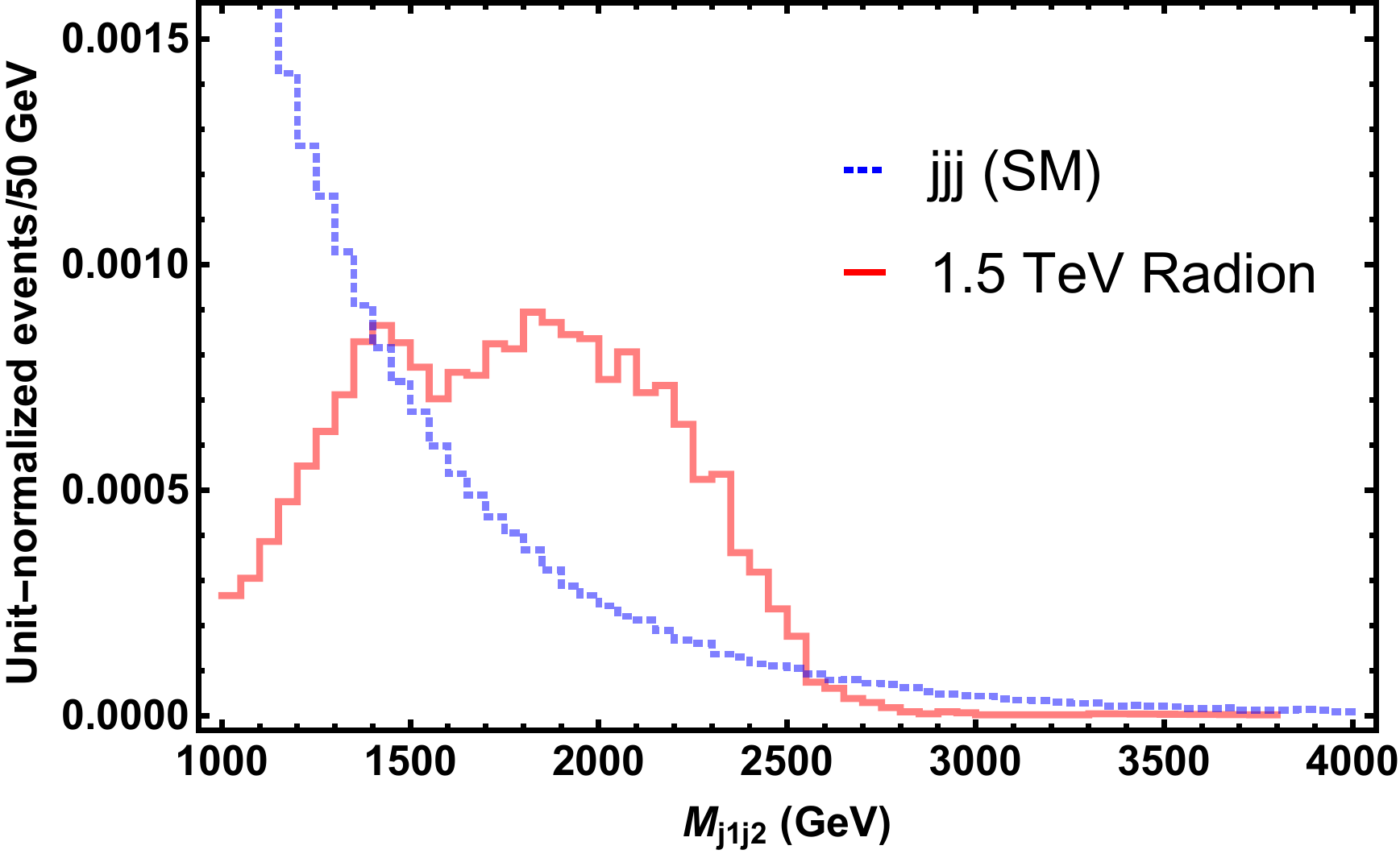}
    \includegraphics[width = 7 cm]{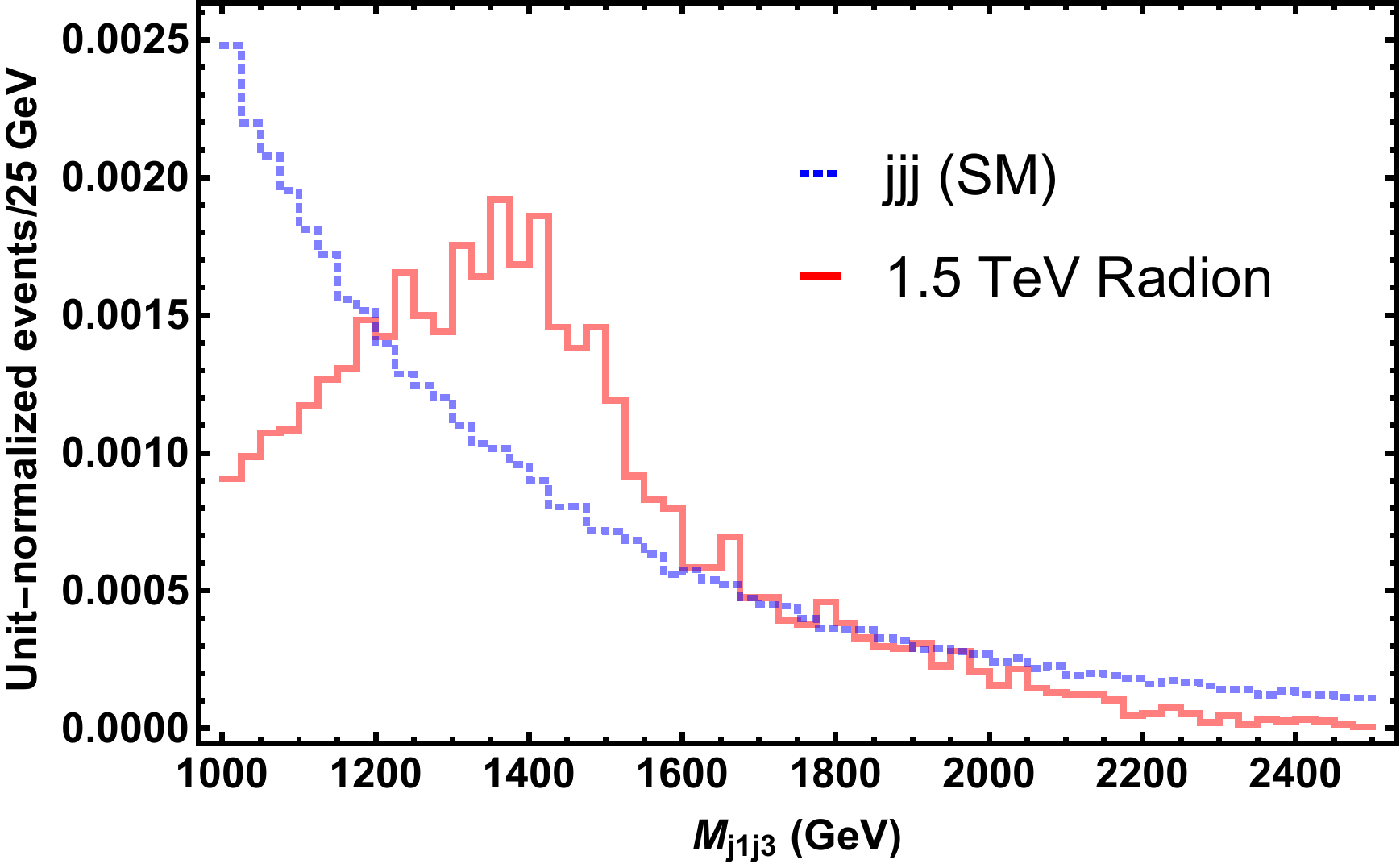}
    \includegraphics[width = 7 cm]{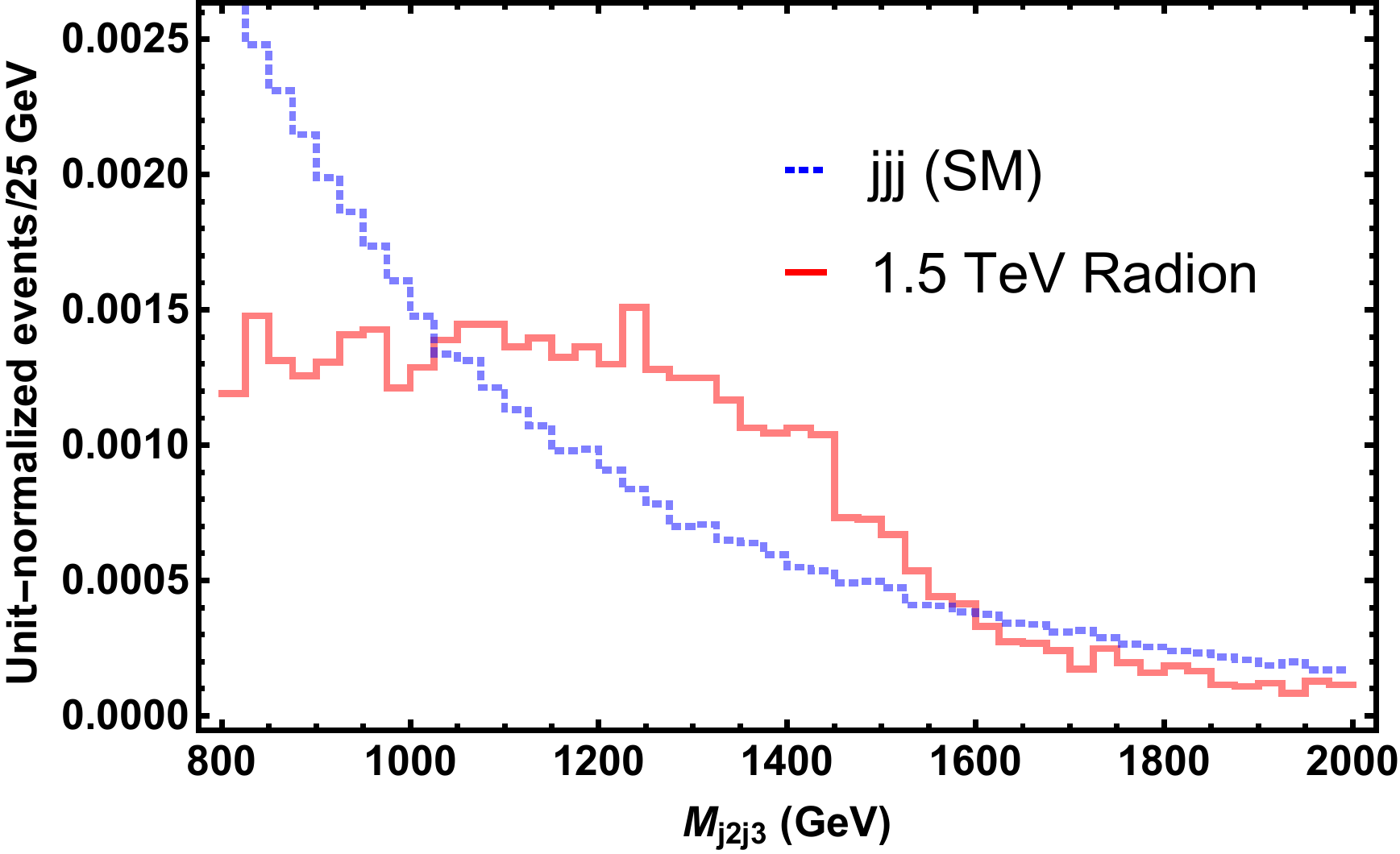}
    \includegraphics[width = 7 cm]{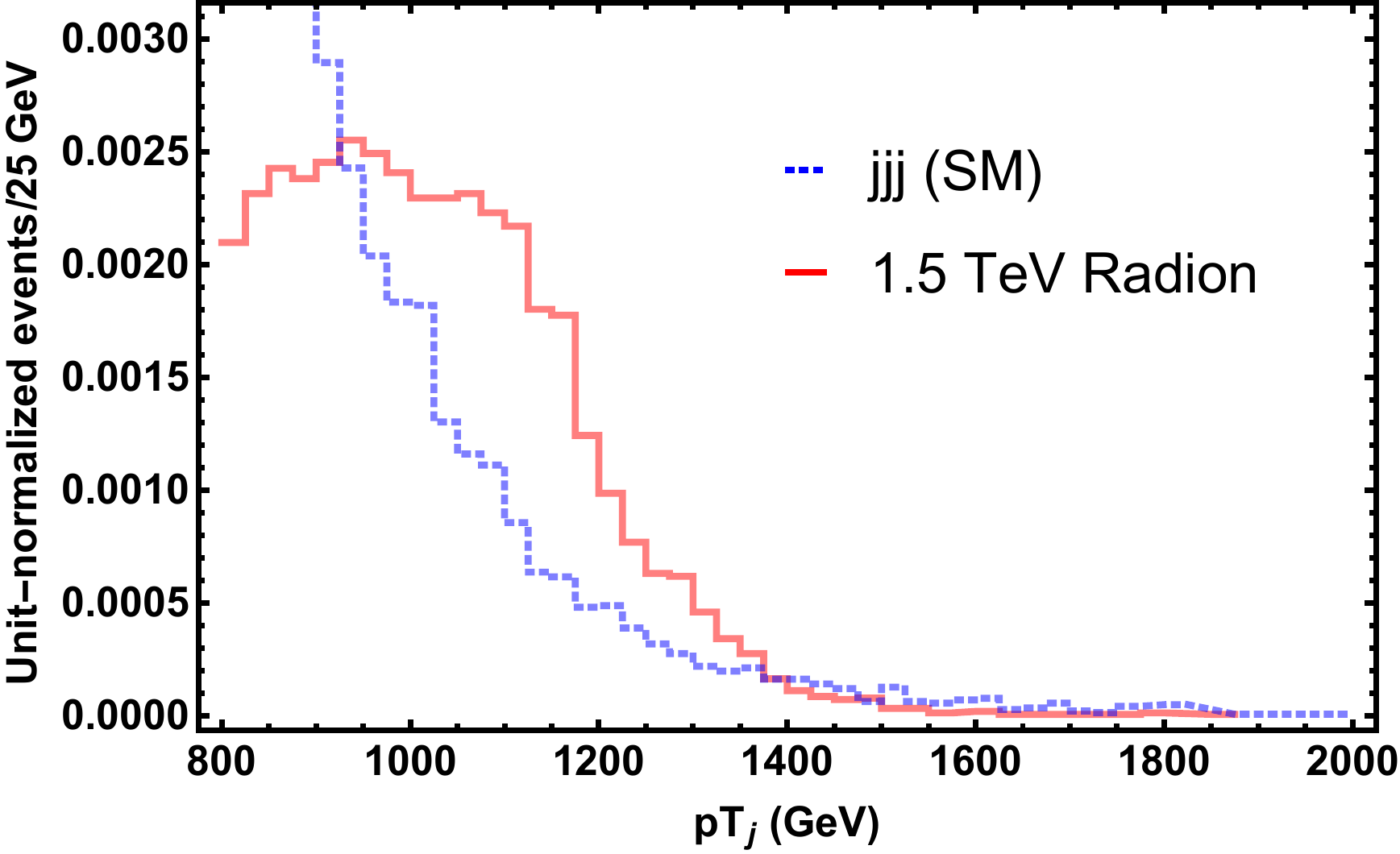}
    \includegraphics[width = 7 cm]{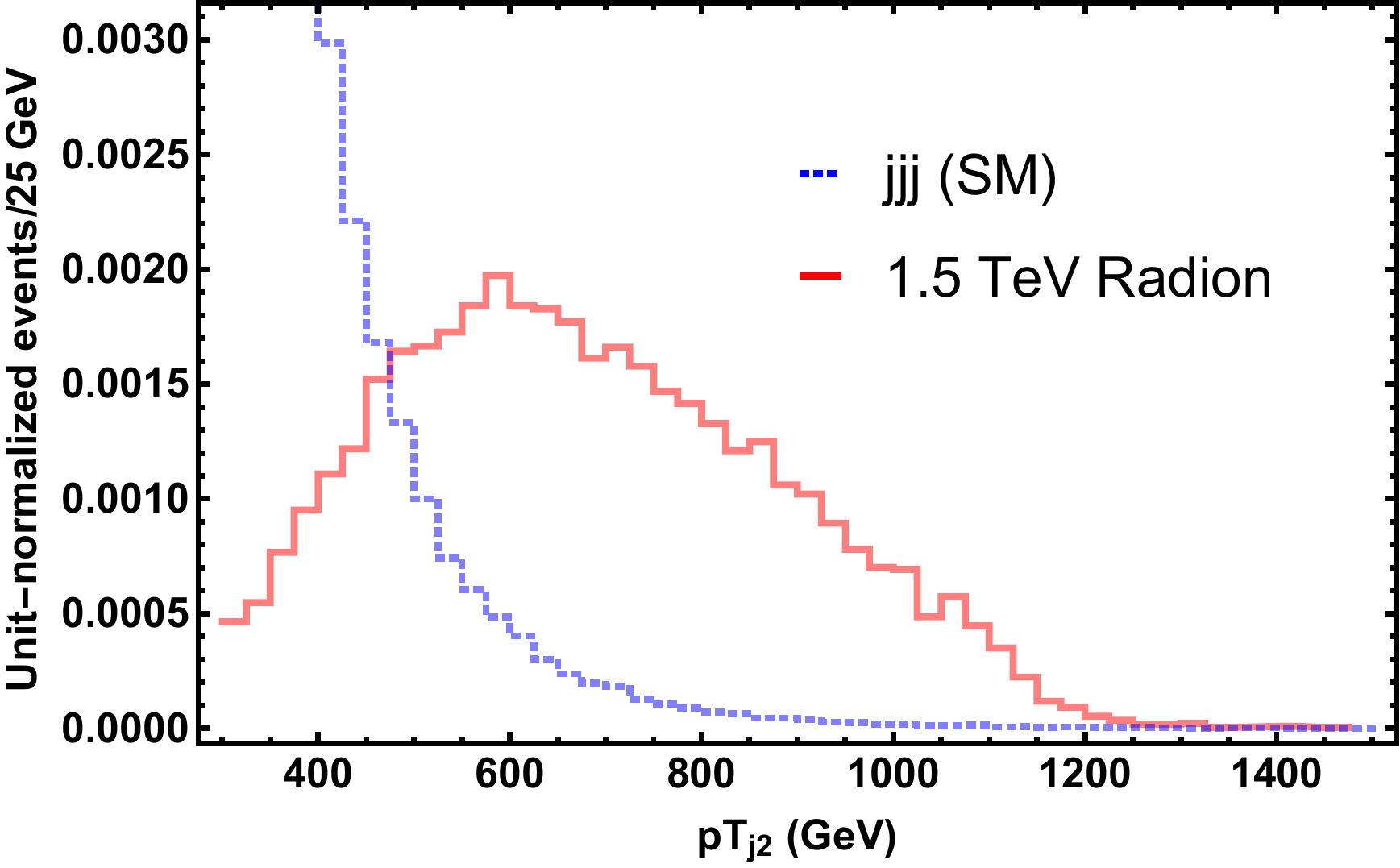}
    \includegraphics[width = 7 cm]{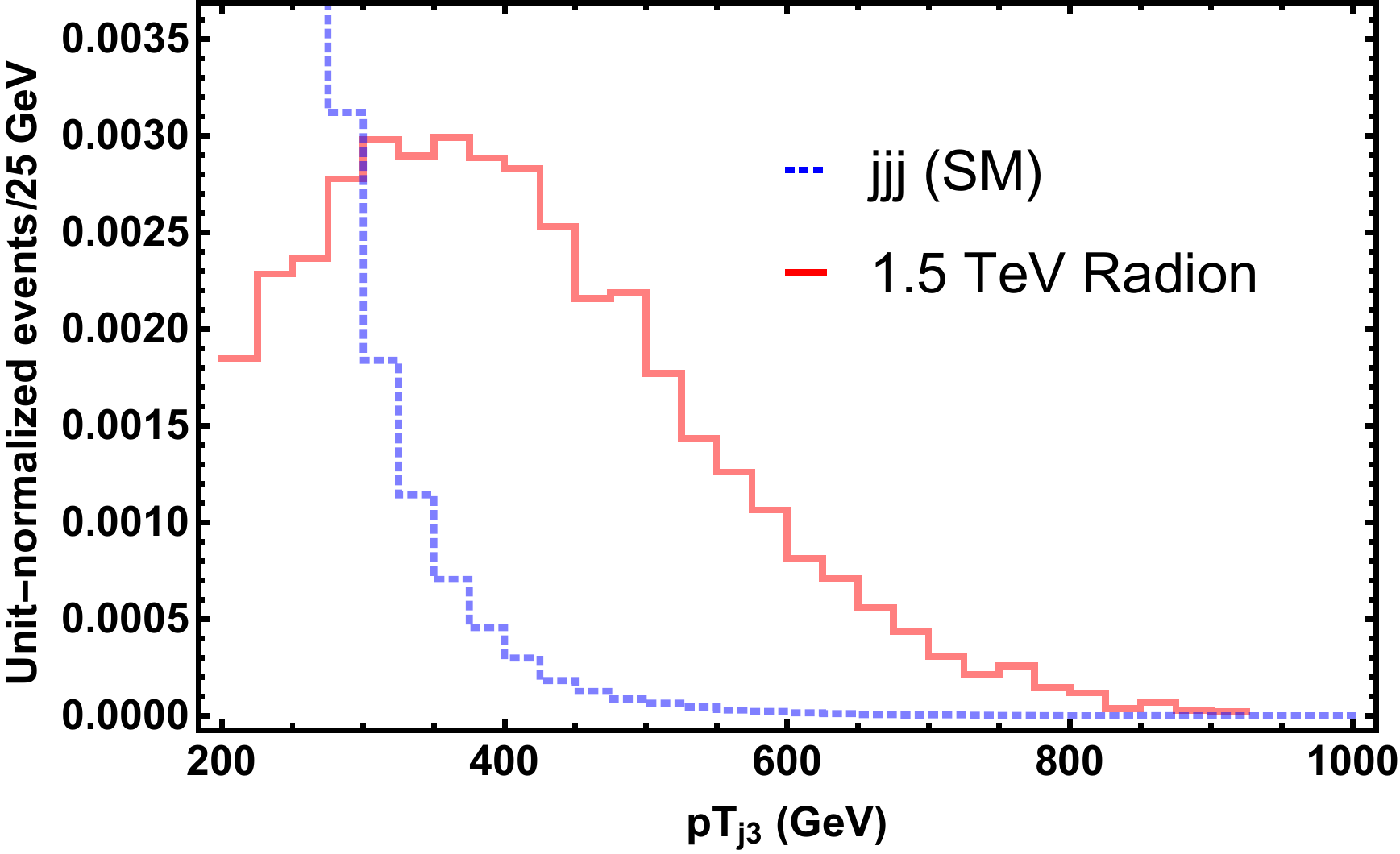}
    \includegraphics[width = 7 cm]{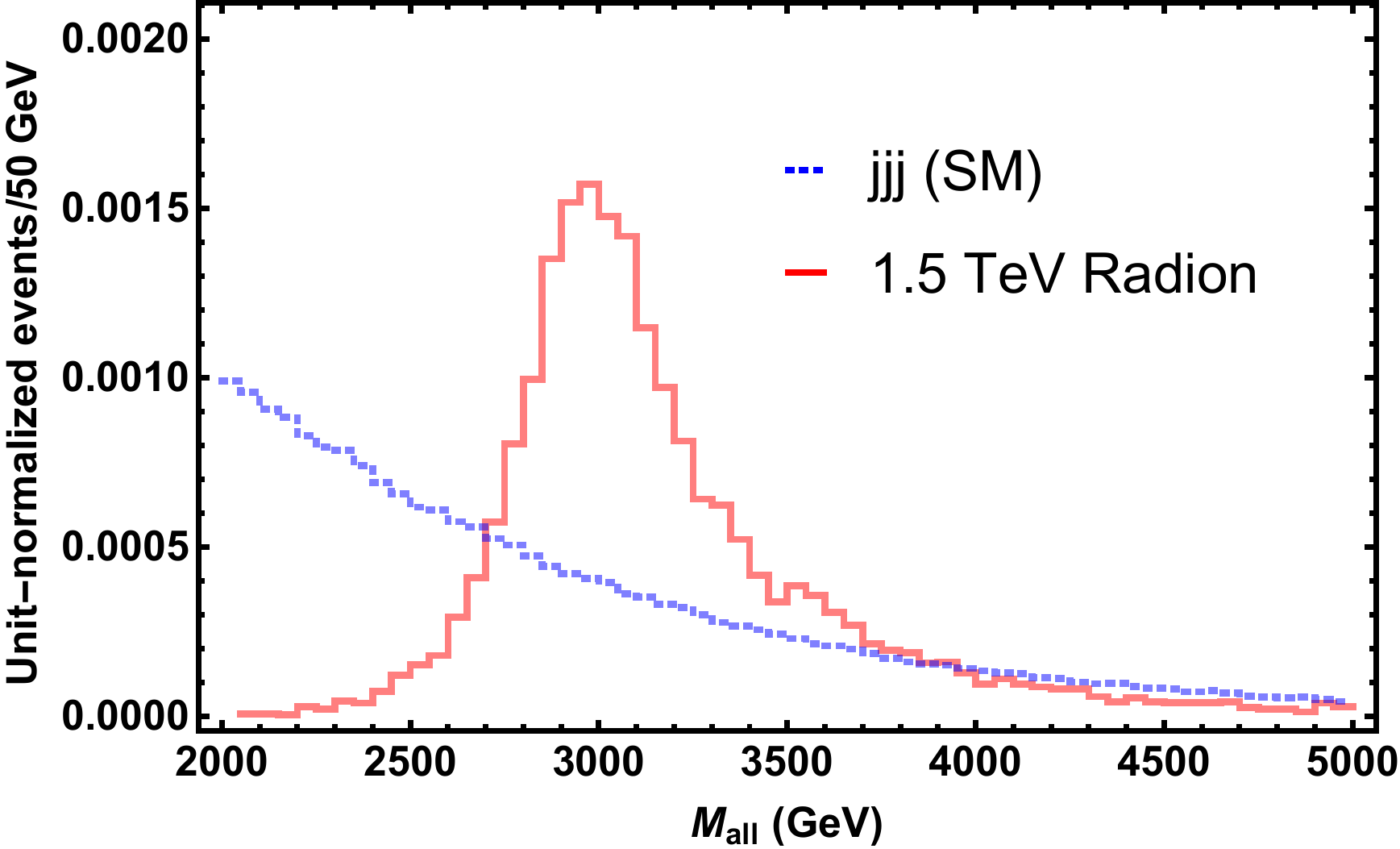}
    \caption{
    $g$-$ggg$-BP2 benchmark point: Distributions of variables: $M_{jjj}$ (top-left), $M_{j_1j_2}$ (top-right), $M_{j_1j_3}$ (second row left), $M_{j_2j_3}$ (second row right), $p_{T,j_1}$ (third row left), $p_{T,j_2}$ (third row right), $p_{T,j_3}$ (bottom-left) and $M_{all}$ (bottom-right) (red solid histograms) and background (blue dashed histograms).}
    \label{fig:gkk_jjj_RadMass1.5tev}
\end{figure}

After generating parton-level signal and background events along with the pre-selection cuts, we feed them to \textsc{Pythia} and \textsc{Delphes} as before.
As the signal process of interest accompanies three jets in the final state, we select the events having
\bea
N_j \geq 3\,.
\eea
As discussed earlier, mass variables are useful in discriminating the signal events from the background ones. 
Let us first denote the three hardest jets (in decreasing order of $p_T$) as $j_1$, $j_2$, and $j_3$.
We emphasize here that depending on the benchmark points, different invariant mass combinations carry different potential in distinguishing the signal from the background.
For example, in the case with radion mass of 1 TeV,
the hardest jet in $p_T$ mostly comes from the direct decay of the KK gluon
because the mass gap between the radion and the KK gluon is quite large.
As a result, $M_{j_2j_3}$ has an invariant mass peak feature, 
corresponding to $m_\varphi$. 
On the contrary, the situation becomes completely reversed in the case of radion mass of 1.5 TeV. 
We find that $M_{j_1j_3}$ and $M_{j_1j_2}$ 
(partially) develop a resonance-like feature in their distributions because $j_1$ is mostly from the decay of the radion while the jet from the KK gluon decay can be either $j_2$ or $j_3$ event-by-event. 
Therefore, $M_{j_2j_3}$ shows quite a broad distribution.
On top of these mass variables,
the transverse momenta of the three jets are also useful in signal identification. 
In addition, the total invariant mass formed by all the visible particles, which we denote as $M_{all}$, also turns out to be beneficial in distinguishing the signal from the background. 
We show the unit-normalized distributions of signal and background events in the variables discussed thus far in Figs.~\ref{fig:gkk_jjj_RadMass1tev} ($g$-$ggg$-BP1) and~\ref{fig:gkk_jjj_RadMass1.5tev} ($g$-$ggg$-BP1), from which we develop our intuition for choosing a set of cuts for each benchmark point. These events are after imposing the pre-selection cuts.  
 
As before, we provide our cut flow results for signal and background events in Table~~\ref{tab:GKK-jjj-cutflow}. 
Our data analysis suggests that we may achieve higher statistical significances of $3.49\sigma$ and $5.25\sigma$ for $g$-$ggg$-BP1 and $g$-$ggg$-BP2 benchmark points respectively, even at an integrated luminosity of 300 fb$^{-1}$.
Definitely, the numbers here are greater than those for the KK photon in the previous section. 
This is expected mainly due to an increased rate for the signal (i.e., QCD coupling vs. QED coupling in the KK photon case), even though the three-jet background renders signal isolation challenging (compared to the 2 jets $+$ photon background). 

\begin{table}[t]
\centering
\begin{tabular}{|c|c|c|c|}
\hline 
Cuts & $g$-$ggg$-BP1 & $g$-$ggg$-BP2 & $j j j$ \\
\hline \hline
No cuts & 29.33 & 46.60 & ($7.7 \times 10^7$) \\
$N_j \geq 3$ with pre-selection cuts & 23.23 & 40.05 & $1.9\times 10^6$  \\
\hline
$ M_{jjj} \in [2500, 3100] \GeV$ & 12.20 & -- & $7.9 \times 10^4$  \\
$ M_{{j_1}{j_2}} \in [1700, 2900] \GeV$ & 11.12 & -- & $3.9 \times 10^4$  \\
$ M_{{j_1}{j_3}} \in [850, 2100] \GeV$ & 9.96 & -- & $1.9 \times 10^4$  \\
$ M_{{j_2}{j_3}} \in [800, 1050] \GeV$ & 5.12 & -- & $2015.28$  \\
$ p_{T,{j_1}} \geq 1100 \GeV $ & 2.73 & -- & 266.41 \\
$ M_{all} \leq 3300 \GeV $ & 1.98 & -- & 94.53 \\
\hline
$ M_{jjj} \in [2400, 3100] \GeV$ & -- & 22.31 & $1.0 \times 10^5$  \\
$ M_{{j_1}{j_2}} \in [1300, 2400] \GeV$ & -- & 19.57 & $4.8 \times 10^4$  \\
$ M_{{j_1}{j_3}} \in [1100, 1700] \GeV$ & -- & 13.82 & $1.0 \times 10^4$  \\
$ M_{{j_2}{j_3}} \in [900, 1550] \GeV$ & -- & 8.81 & 1564  \\
$ p_{T,{j_1}} \geq 900 \GeV $ & -- & 6.79 & 807.83 \\
$ p_{T,{j_2}} \geq 600 \GeV $ & -- & 6.20 & 644.54 \\
$ p_{T,{j_3}} \geq 300 \GeV $ & -- & 5.44 & 464.07 \\
$ M_{all} \in [2800, 3300] \GeV$ & -- & 3.43 & 124.61 \\
\hline
$S/B$ & 0.02 & 0.03 & --\\
$S/\sqrt{B}$ ($\mathcal{L}=300$ fb$^{-1}$) & 3.49 & 5.25 & -- \\
$S/\sqrt{B}$ ($\mathcal{L}=3000$ fb$^{-1}$) & 11.03 & 16.60 &-- \\
\hline
\end{tabular}
\caption{Cut flows for signal and major background events in terms of their cross sections (in fb). The number in the parentheses for $jjj$ is obtained with basic cuts ($p_{T,j} > 20 \GeV$, $p_{T,\gamma} > 10 \GeV$, $|\eta_j| < 5$, $|\eta_\gamma| < 2.5$, $\Delta R_{jj} > 0.4$, $\Delta R_{j\gamma} > 0.4$, $\Delta R_{\gamma\gamma} > 0.4$) at the generation level to avoid divergence. The pre-selection cuts ($p_{T,j} > 150 \GeV$, $M_{jj} > 300 \GeV$) are imposed at the parton level as well to generate events in the relevant phase space, and are reimposed at the detector level.}
\label{tab:GKK-jjj-cutflow}
\end{table}

\subsubsection{Decay to jet and diphoton }
\label{sec:jetdiphoton}
We next move our focus onto the jet $+$ diphoton decay mode of the KK gluon, where the two photons come from the radion decay. 
As usual, we consider two representative benchmark points denoted by $g$-$g\gamma \gamma$-BP1 and $g$-$g\gamma \gamma$-BP2 (see Table~\ref{tab:BPtable} for model parameters).
The dominant SM background for this decay mode comes from the $j\gamma\gamma$ process. 
However, it becomes important to take the effect of jet-faking photons at the detector level. 
\dk{To this end,}
we simulate the $jj\gamma$ process as well, and impose the same set of cuts to estimate its contribution to the total 
background.\footnote{We expect that the contribution from three-jet QCD events are small enough to be neglected, considering that two jets are simultaneously misidentified as photons, in combination with the set of selection cuts that we apply.}
%
We once again need to employ pre-selection cuts to generate signal and background events in the relevant part of the phase-space. 
Motivated by our considerations earlier, we impose selections on the transverse momenta of the final objects. 
We require $p_T > 200 \GeV$ for the jet and $p_T > 200 \GeV$ for the photons at the parton level. 
We also impose a selection on the invariant mass of the two photons at parton level, requiring $M_{\gamma\gamma} > 750 \GeV $. 
The values of these variables tend to be much higher for the signal events than those for the background ones, allowing a clean way to generate relevant events. 
%
%
Again, their effectiveness is reflected in the efficiency: the signal cross section is reduced only marginally (78.2\% and 87.8\% for $\gamma$-$\gamma gg$-BP1 and $\gamma$-$\gamma gg$-BP2 benchmark points,  respectively), while the jet $+$ diphoton background is significantly suppressed by $1.8\times 10^{-3}\%$.

\begin{figure}[t]
    \centering

    \includegraphics[width = 7 cm]{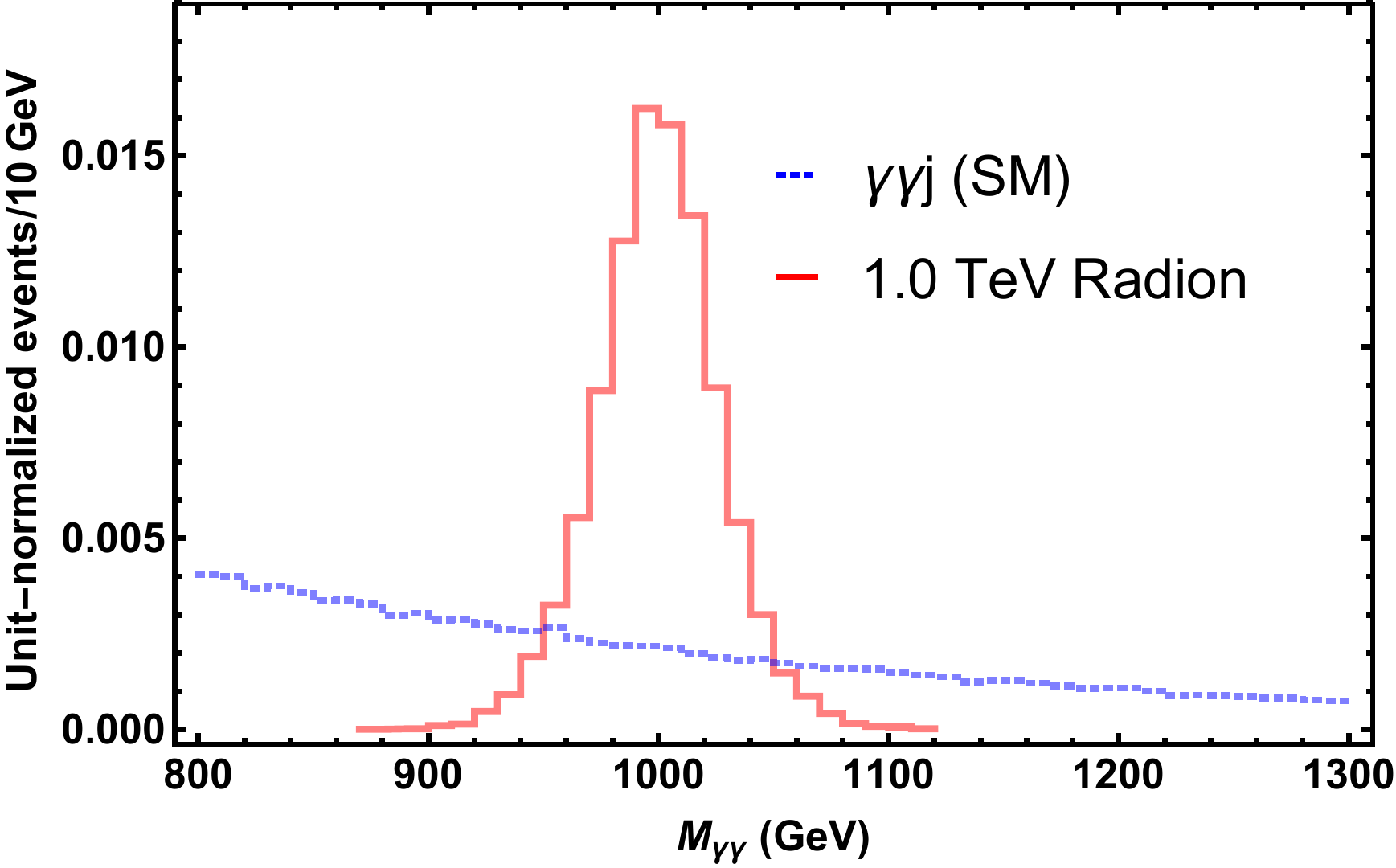}
    \includegraphics[width = 7 cm]{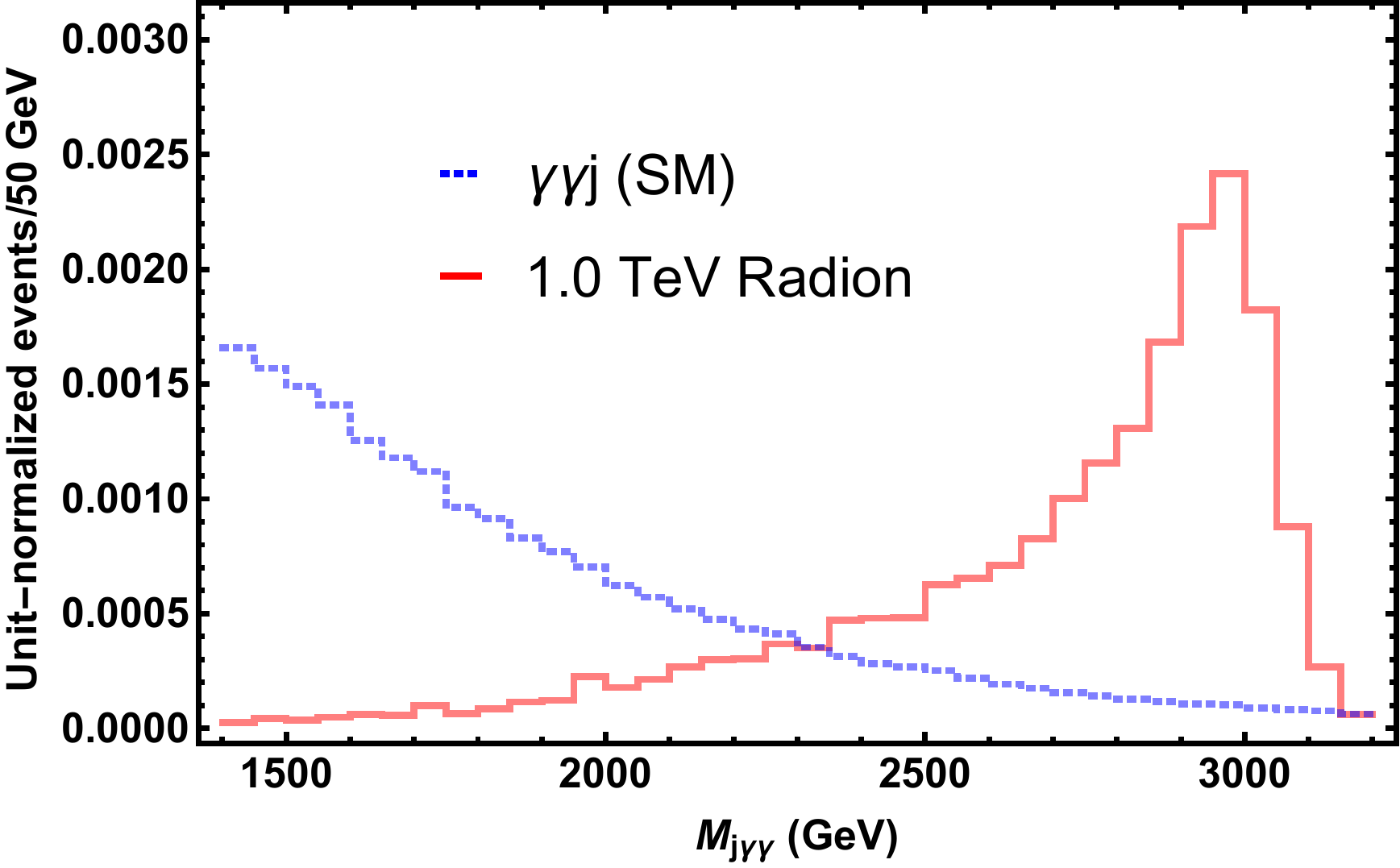}
    \caption{
    $g$-$g\gamma\gamma$-BP1 benchmark point: Distributions of variables: $M_{\gamma\gamma}$ (left) and $M_{j\gamma\gamma}$ (right) for signal (red solid histograms) and background (blue dashed histograms).
    \label{fig:gkk_jaa_RadMass1tev} } 
    \vspace{0.5cm}
%

    \includegraphics[width = 7 cm]{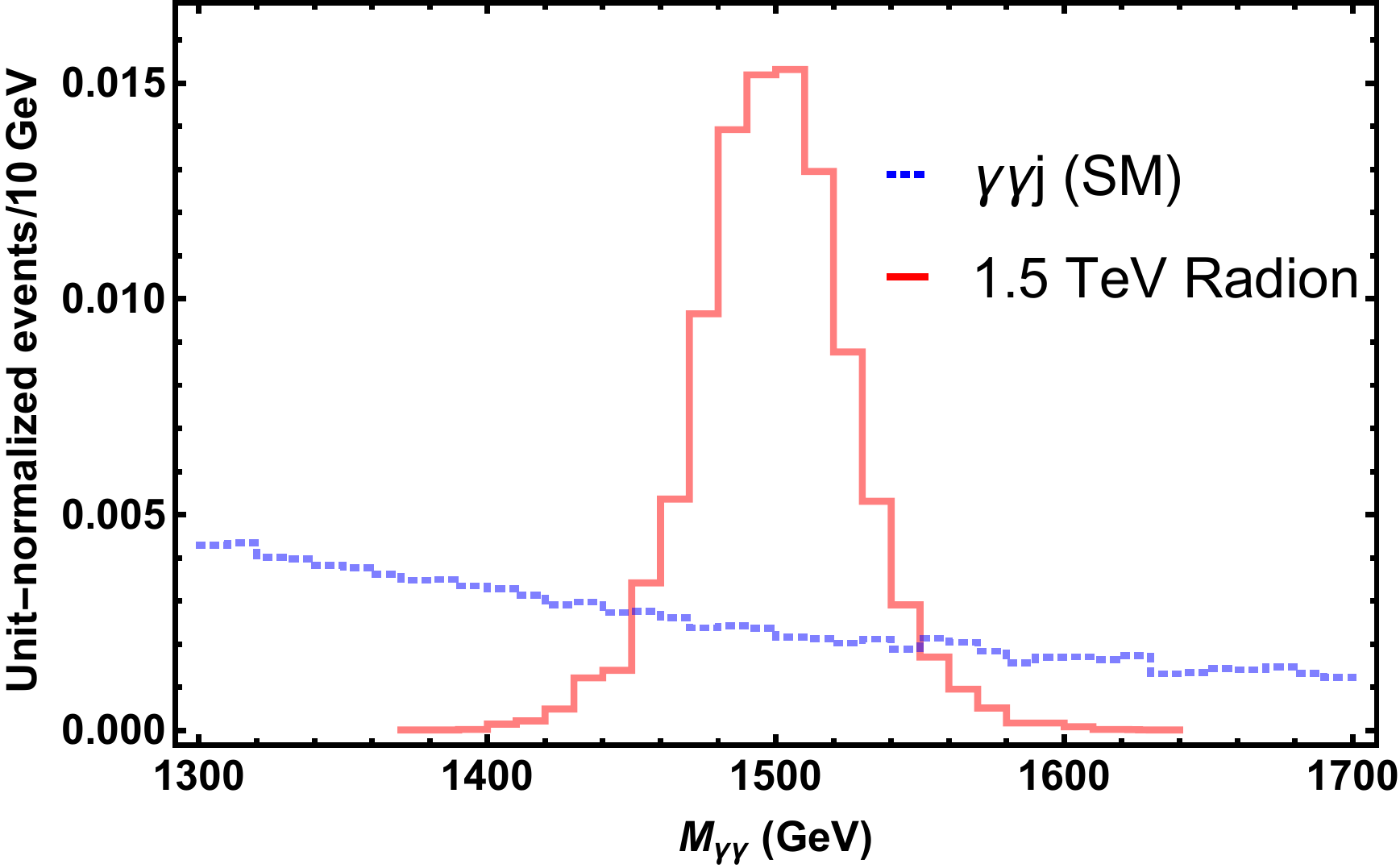}
    \includegraphics[width = 7 cm]{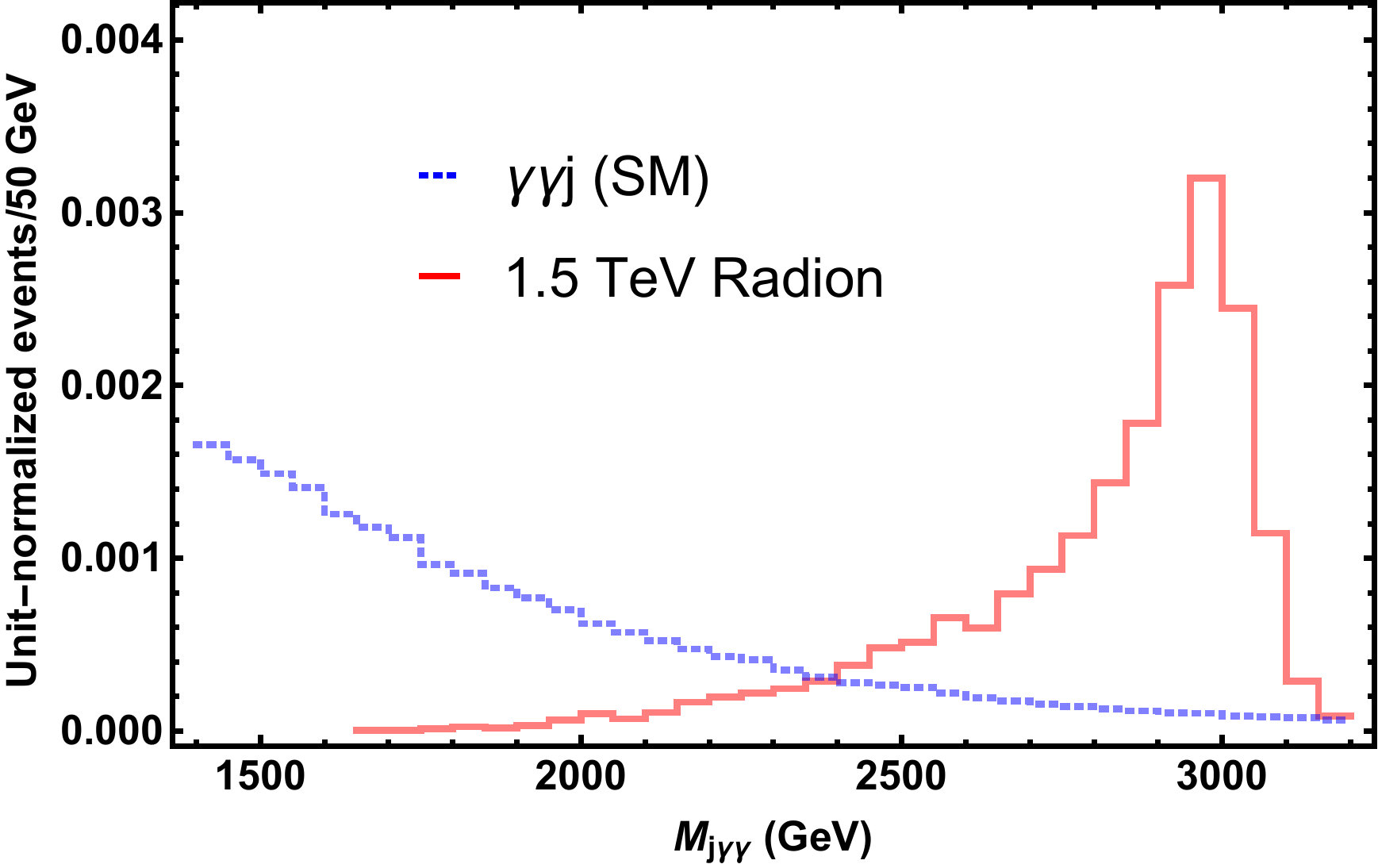}
    \caption{Distributions of variables for $g$-$g\gamma\gamma$-BP2 benchmark point: $M_{\gamma\gamma}$ (left) and $M_{j\gamma\gamma}$ (right) for signal (blue) and background (blue).
    \label{fig:gkk_jaa_RadMass1.5tev} }
\end{figure}

The parton-level signal and background events generated with the pre-selection cuts are fed into a sequence of \textsc{Pythia} and \textsc{Delphes}.
As mentioned before, we perform background simulation with $j\gamma\gamma$ and  $jj\gamma$ processes.
We find that the two background contribute to the total background at an equal level.
As a parton-level signal event contains two photons and a single jet, we restrict ourselves to the phase space involving
\bea
N_\gamma \geq 2,\,\,\, N_j \geq 1 \,.
\eea
The unsmeared nature of the two photons in the final state makes clean signal identification possible. 
This is clearly supported by a sharp peak in the diphoton invariant mass distribution (see the left panels in Figs.~\ref{fig:gkk_jaa_RadMass1tev} and~\ref{fig:gkk_jaa_RadMass1.5tev}).
The other (resonant) invariant mass variable $M_{j\gamma\gamma}$ is broadened primarily due to the jet involved, but still provides a strong handle to distinguish the signal events from the background ones (see the right panels in Figs.~\ref{fig:gkk_jaa_RadMass1tev} and~\ref{fig:gkk_jaa_RadMass1.5tev}).
 
Finally, the cut flow for this channel is presented in Table~\ref{tab:GKK-jaa-cutflow}. 
Since it turns out that the data is essentially signal dominated, we conservatively adopt $S/\sqrt{S+B}$ as our figure of merit to estimate the statistical significance.
We find a statistical significances of $4.3\sigma$ ($5.4\sigma$) for $g$-$g\gamma\gamma$-BP1 ($g$-$g\gamma\gamma$-BP2) benchmark point, even at an integrated luminosity of 300 fb$^{-1}$. This suggests that this could serve as the first discovery channel of gauge KK particles over the other ones in our study.
\begin{table}[t]
\centering
\hspace*{-0.4cm}
\begin{tabular}{|c|c|c|c|c|}
\hline 
Cuts & $g$-$g\gamma\gamma$-BP1 & $g$-$g\gamma\gamma$-BP2 & $j\gamma \gamma$ & $jj\gamma$ \\
\hline \hline
No cuts & 0.17 & 0.19 & ($1.07 \times 10^5$) & ($8.7 \times 10^7$)\\
$N_{j(\gamma)} \geq 1\, (2)$ with pre-selection cuts & 0.10 & 0.13 & 1.35 & 1.60\\
\hline
$ M_{\gamma\gamma} \in [950, 1350] \GeV$ & 0.10 & -- & 0.2 & 0.13 \\
$ M_{j\gamma\gamma} \in [2100, 3200] \GeV$ & 0.09 & -- & 0.02 & 0.02 \\
\hline
$ M_{\gamma\gamma} \in [1450, 1550] \GeV$ & -- & 0.12 & 0.04 & 0.04 \\
$ M_{j\gamma\gamma} \in [2500, 3150] \GeV$ & -- & 0.11 & 0.005 & 0.006 \\
\hline
$S/\sum{B}$ & 2.25 & 10.0 & -- & -- \\
$S/\sqrt{S+\sum{B}}$ ($\mathcal{L}=300$ fb$^{-1}$) & 4.3 & 5.4 & -- & -- \\
$S/\sqrt{S+\sum{B}}$ ($\mathcal{L}=3000$ fb$^{-1}$) & 13.6 & 17.1 & -- & -- \\
\hline
\end{tabular}
\caption{Cut flows for signal and major background events in terms of their cross sections (in fb). The numbers in the parentheses for $j\gamma\gamma$ and $jj\gamma$ are obtained with basic cuts ($p_{T,j} > 20 \GeV$, $p_{T,\gamma} > 10 \GeV$, $|\eta_j| < 5$, $|\eta_\gamma| < 2.5$, $\Delta R_{jj} > 0.4$, $\Delta R_{j\gamma} > 0.4$, $\Delta R_{\gamma\gamma} > 0.4$) at the generation level to avoid divergence. The pre-selection cuts ($p_{T,j} > 200 \GeV$, $p_{T,\gamma} > 200 \GeV$, $M_{\gamma\gamma} > 750 \GeV$) are imposed at the parton level to generate events in the relevant phase space, and are reimposed at the detector level.}
\label{tab:GKK-jaa-cutflow}
\end{table}


\subsubsection{$\textrm{jet}+\textrm{diboson}$ ($W/Z$-jets) \label{sec:jetdiboson}}

The fully hadronic analysis for $WWj$ proceeds by requiring two merged jets consistent with coming from boosted $W$'s and reconstructing the radion mass, and an additional gluon-induced jet which, combined with the reconstructed radion, reproduces a KK gluon mass peak. The dominant background is SM $jjj$ production, with two jets being mistagged as vector-boson jets.

The $W$-jets are selected according to the criteria described in Sec.~\ref{EventSimulation}, with a mass window requirement
\bea
65 \; \text{GeV} < M_W < 100 \; \text{GeV}\,.
\eea
Here we capitalize the mass symbol to distinguish it from the corresponding input mass.
A second jet collection is made using the anti-$k_t$ algorithm with radius parameter $R = 0.4$. 
Jets are kept if they have $|\eta| < 3$, are separated from $W$-candidates by $\Delta R > 0.8$, and both pairings of this jet with a $W$-candidate has invariant mass $M_{Jj} > 400 \; \text{GeV}$. The hardest remaining jet is the $g$-candidate.

\begin{figure}[t]
\includegraphics[width=7.2cm]{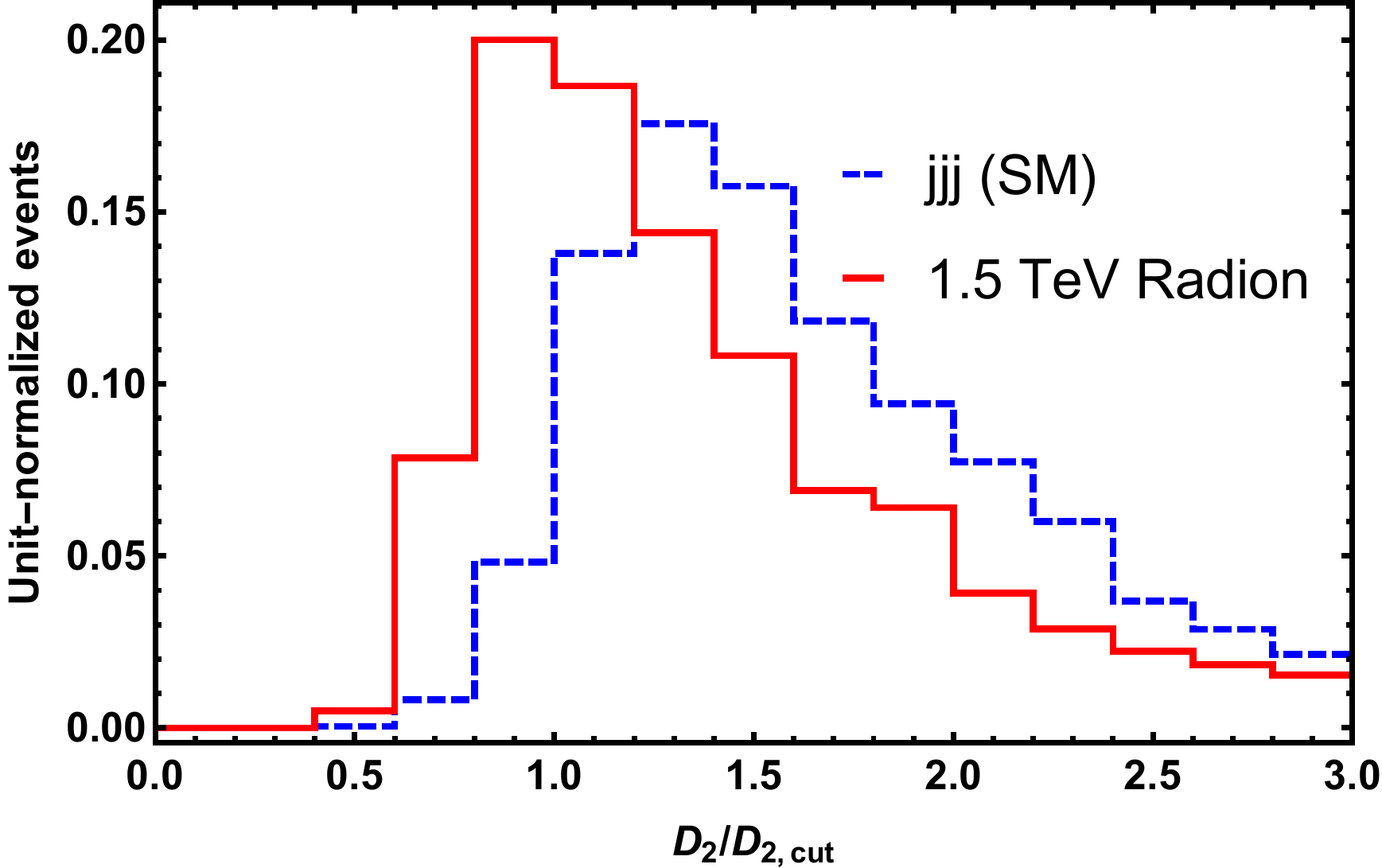} \hspace{0.2cm}
\includegraphics[width=7.2cm]{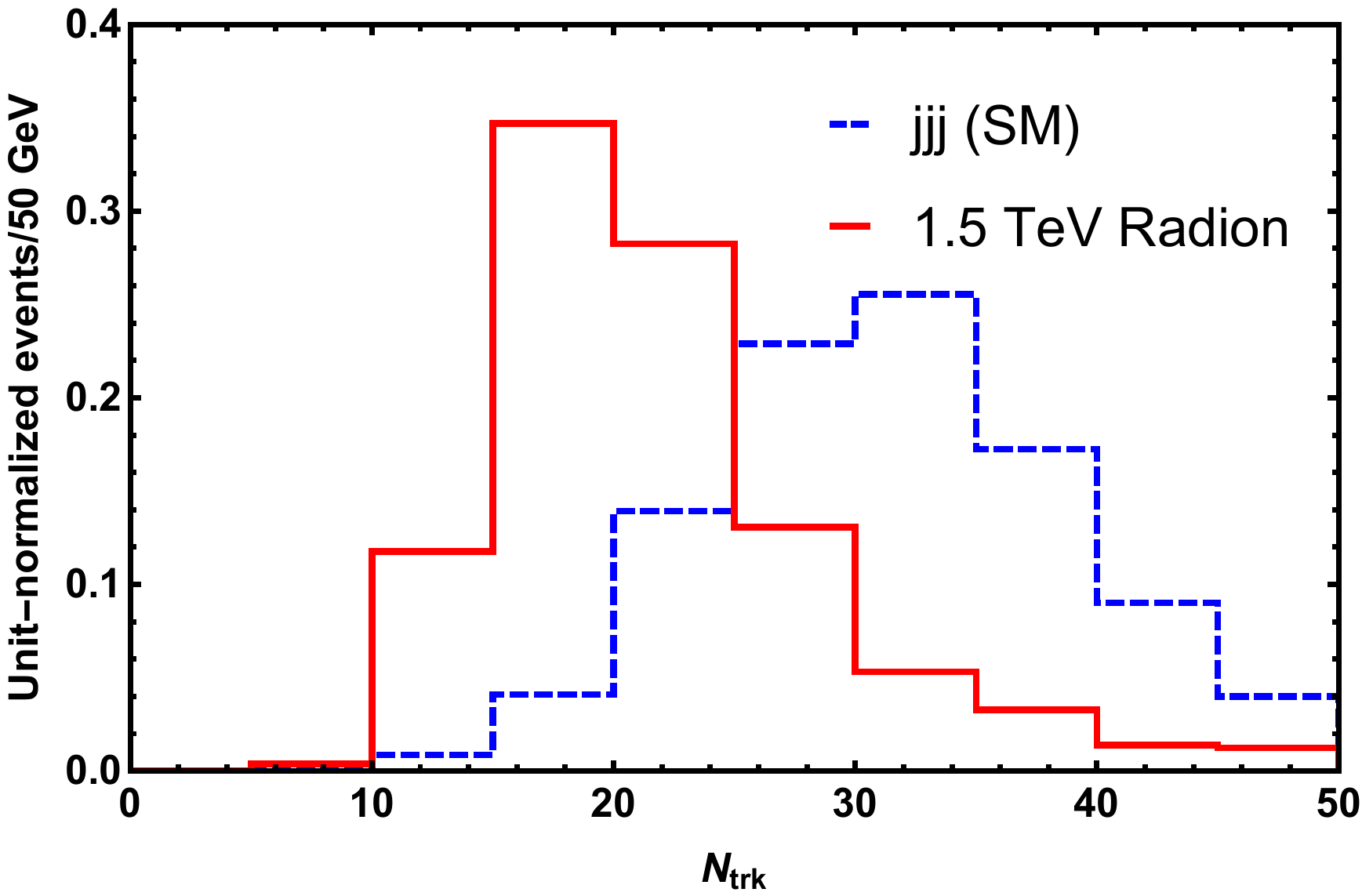}
\caption{Distributions in $N_\text{trk}$ (left) and $D_2$ (right) for signal (solid histograms) and background (dashed histograms) for radion mass of 1.5 TeV.\label{fig:JCshapedists} }
\end{figure}

Three-jet background events for the detector level analysis are simulated with the following parton level cuts: $p_{T, j} > 450 \; \text{GeV}$ on the leading jet, $p_{T, j} > 250 \; \text{GeV}$ on the remaining jets, $\Delta R_{jj} > 0.5$ between all jets, $M_{jj} > 250 \; \text{GeV}$ between all jets. We additionally require a strong cut $M_{jjj} > 2000 \; \text{GeV}$ in order to generate a sufficiently large sample of events in the signal region. 
In contrast, signal $WWj$ events are simulated with no parton-level cuts as before. 
In order to be consistent with the parton-level cuts applied to the background, events are retained for further analysis if they satisfy the following requirements:
\begin{align}p_{T,J_1} &> 600 \; \text{GeV},\\
p_{T,J_2} &> 300 \; \text{GeV},\\
p_{T,j} &> 600 \; \text{GeV},\\
M_{JJj} &> 2250 \; \text{GeV},\end{align}
where $J_1$ and $J_2$ are the leading and the subleading merged jets.
As discussed earlier, $D_2$ and $N_{\rm trk}$ are also useful in isolating signal from background. 

In order to illustrate the discriminating power of the $W$-jet tagging observables used in this analysis, we present in Fig.~\ref{fig:JCshapedists} the distributions of $D_2$ and $N_\text{trk}$ in selected events in event samples where the cuts on these distributions have not been applied.

\begin{figure}[t]
\centering
\includegraphics[width=7.2cm]{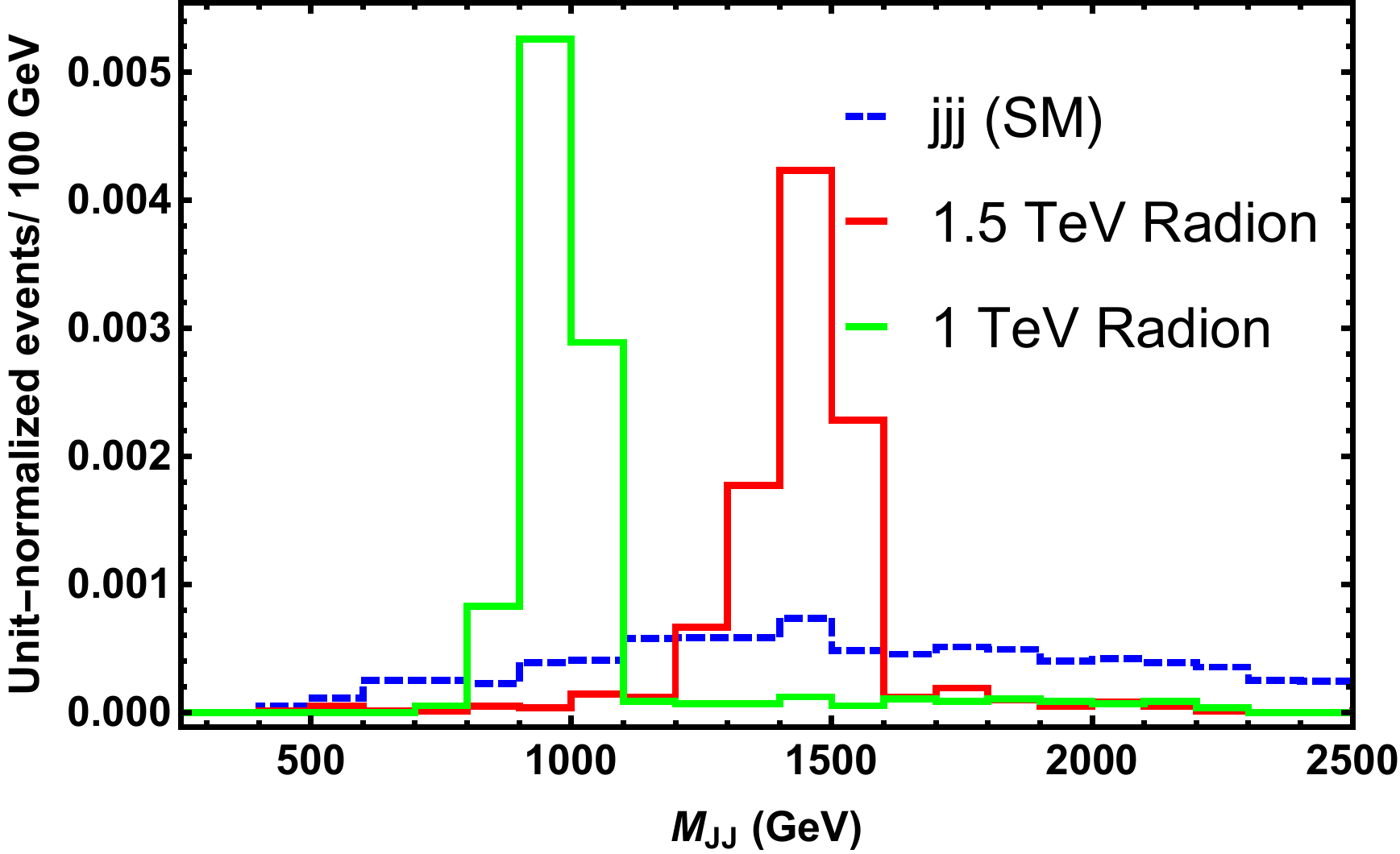} \hspace{0.2cm}
\includegraphics[width=7.2cm]{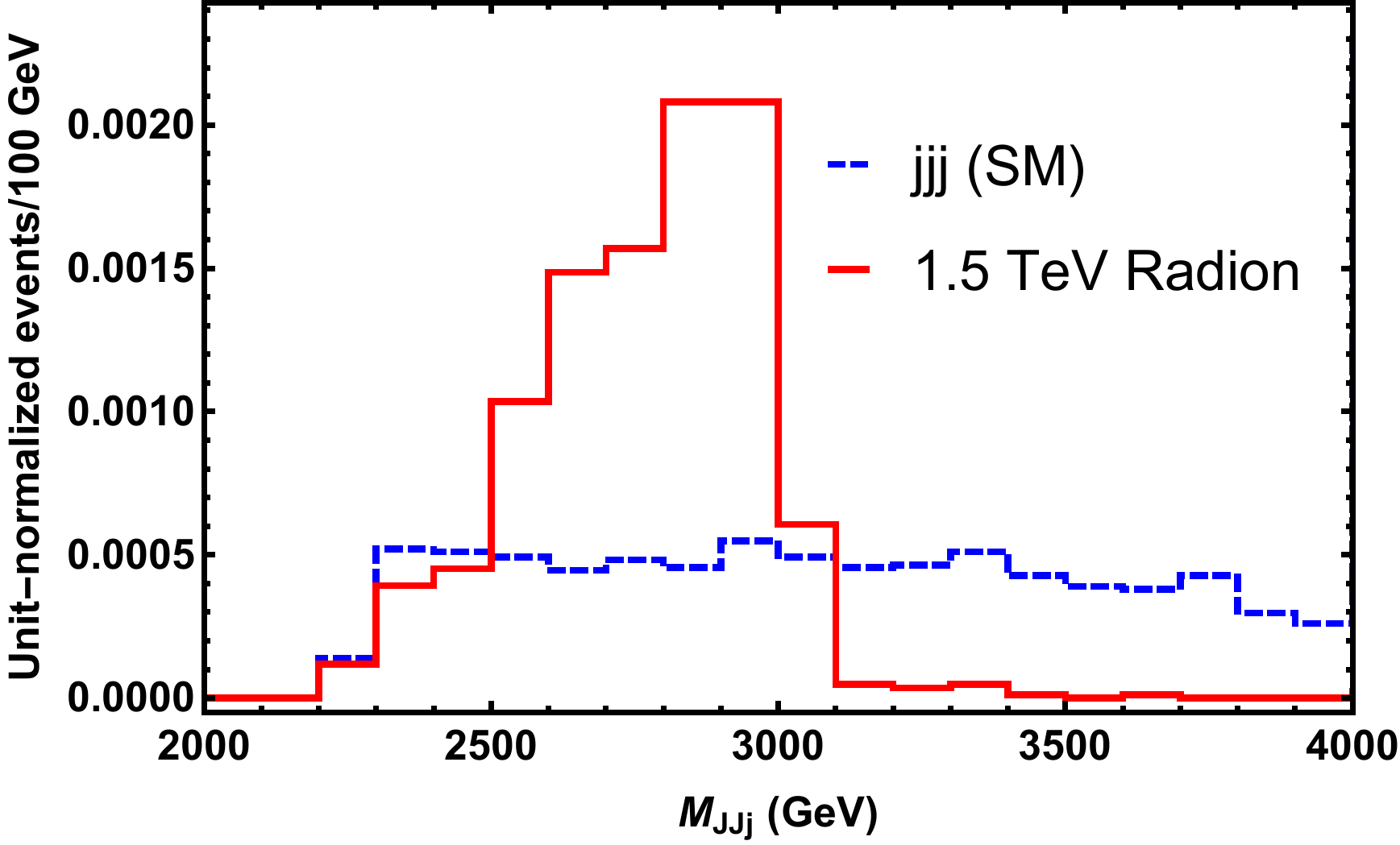}
\caption{Distributions in $M_{JJ}$ (left) and $M_{JJj}$ (right) for signal (solid histograms) and background (dashed histograms). \label{fig:JCmassdists} }
\end{figure}

In the left panel of Fig.~\ref{fig:JCmassdists}, we see that the diboson pairs reproduce a sharp invariant mass peak at around the input radion mass. 
This radion peak is sharper than that for the $ g_{\rm KK} \to j j j$ final state because in the latter case final state radiation from the gluons depletes them of energy, broadening the invariant mass peak. 
This effect is largely absent for the color-neutral $W$-jets. 
On the other hand, the $JJj$ invariant mass is partially smeared due to final state radiation from the gluon. After isolating the two mass peaks, we find that $5\sigma$ discovery is possible with an integrated luminosity of $3000 \; \text{fb}^{-1}$ for radion mass of $1.5 \; \text{TeV}$ (see also the cut flow in Table~\ref{tab:JCcutflow}).

\begin{table}[t]
\centering
\begin{tabular}{|c|c|c|c|}
\hline 
Cuts & $g$-$gVV$-BP1 & $g$-$gVV$--BP2 & $jjj$  \\
\hline \hline
No cuts & 1.6  & 2.6 & ($7.7 \times 10^7$) \\
Basic Cuts (with no $V$-tagging) & 0.59 & 1.4 & $1.3\times 10^4$  \\
$65 \; \text{GeV} < M_J < 100 \; \text{GeV}$ & 0.29 & 0.68 & 990 \\
$D_2$ cut & $7.5\times10^{-2}$ & 0.19 & 54 \\
$N_\text{trk} < 30$ & $6.0\times10^{-2}$  & 0.16 & 28 \\
$p_{T,j} > 600 \; \text{GeV}, p_{T,J_1} > 600 \; \text{GeV}$ & $4.6\times10^{-2}$ & 0.11 & 9.4 \\
$M_{JJj} \in [2500,\,3100]$ & $3.8\times 10^{-2}$ & $9.7\times 10^{-2}$ & 4.8 \\	
\hline
$M_{JJ} \in [900,\,1050]$ GeV & $3.1\times10^{-2}$  & -- & 0.17  \\
\hline
$M_{JJ} \in [1350,\,1600]$ GeV & --  & $7.8\times 10^{-2}$  &  0.56  \\
\hline
$S/B$ & 0.18 & 0.14 & --\\
$S/\sqrt{B}$ ($\mathcal{L}=300$ fb$^{-1}$) & 1.3 & 1.8 & -- \\
$S/\sqrt{B}$ ($\mathcal{L}=3000$ fb$^{-1}$) & 4.0 & 5.7 &-- \\
\hline
\end{tabular}
\caption{Cut flows for signal and major background events in terms their cross sections. The cross sections are in fb. The number in the parentheses for $jjj$ is obtained with basic cuts  ($p_{T,j} > 20 \GeV$, $p_{T,\gamma} > 10 \GeV$, $|\eta_j| < 5$, $|\eta_\gamma| < 2.5$, $\Delta R_{jj} > 0.4$, $\Delta R_{j\gamma} > 0.4$, $\Delta R_{\gamma\gamma} > 0.4$) at the generation level to avoid divergence. In the second row, the same basic cuts are imposed to both signal and background events. \label{tab:JCcutflow} }
\end{table}

\subsection{KK $W/Z$: leptonic $W+\textrm{dijet}$ \label{sec:kkwlepton}}

We finally consider the production of KK $W/Z$ gauge bosons and collider signatures from their decays. 
They are also featured by a decent production cross section, so that a few channels deserve to be investigated. 
To secure enough statistics at the 14 TeV LHC, we lay our focus on the processes in which the radion decays into a gluon pair. 
The remaining SM particle in the final state is either $W$ or $Z$ gauge boson from the decay of the corresponding KK particle. 
Obviously, the hadronic channels come with higher cross sections than the leptonic ones. 
As mentioned earlier, hadronic gauge bosons appear as ``single'' boosted merged jets, the dominant SM background is the three-jet QCD process. 
Even if the tagging of merged jets (by the procedure described in Sec.~\ref{EventSimulation}) suppress the background, its huge production cross section overwhelms the signal cross section even with posterior cuts, which are also confirmed by our simulation study. 
We therefore focus on the final state involving a leptonic $W$ and two jets in the rest of this section.\footnote{We do not consider the channel with a leptonic $Z$ since its associated cross section is too small to obtain enough statistics.}

The signal process of interest, mass spectra, and model parameter values are summarized in Table~\ref{tab:BPtable}. 
As obvious from the signal process, the resulting final state contains two hard jets and an isolated lepton $\ell (=e,\mu)$ at the leading order.
Defining $N_{\ell}$ and $N_j$ as the number of isolated leptons and {\it non}-$b$-tagged jets, respectively, we restrict ourselves to the events satisfying
\bea
N_{\ell} &=& 1 \hbox{ with }|\eta_{\ell}|<2.5, \label{eq:Nl}\\ 
 N_j&\geq& 2 \hbox{ with }|\eta_j|<4. \label{eq:Nj} 
\eea
It is clear that in our benchmark choices signal lepton and jets are rather hard, motivating us to further impose the following selection cuts for the lepton and the two hardest jets:
\bea
p_{T,\ell}&>&150 \hbox{ GeV,} \\
p_{T,j_{h(s)}}&>&400\,(200) \hbox{ GeV, }|\eta_{j_{h/s}}|<2.5\,,
\eea
where $h(s)$ stands for the harder (softer) jet out of the two hardest jets as before. 
The existence of an invisible neutrino yields a large missing transverse momentum $\met$ which is the opposite of the vectorial $p_T$ sum of reconstructed objects in the event, comprised of the jets with $p_T$ defined in~\eqref{eq:Nj}.
Since there is no other invisible particle, the unknown neutrino $z$-momentum can be reconstructed by requiring $W$ mass shell condition up to two-fold ambiguity.  
For the detector-level events, we scan the $W$ mass from 60 GeV to 100 GeV by an interval of 2 GeV and choose the solution whose input $W$ mass is closest to the nominal $W$ mass 80 GeV. 
Interestingly enough, we observe that both solutions yield the same values in relevant invariant mass variables, so we do not encounter two-fold ambiguity as long as invariant mass quantities are concerned. 
We later denote $w$ as the resulting reconstructed $W$ for notational brevity. 

Given the collider signature, the dominant SM background is irreducible $Wjj$, and $jjj$ and $t\bar{t}$ could be potentially comparable. 
For the pure QCD background, a three-jet event would appear as a background one if one of three jets is misidentified as a lepton and mismeasurement of jets gives rise to a sizable missing transverse momentum.
Although the associated process is featured by a huge production cross section, we anticipate that the tiny lepton-fake rate can significantly suppress this background.
To ensure further reduction, we impose a rather hard missing transverse momentum cut which is expected to reject events whose $\met$ is purely instrumental. In our analysis, we choose
\bea
\met > 200 \textrm{GeV}.
\eea 
In addition, we require each jet to sufficiently distant in the azimuthal angle $\phi$ from the direction defined by the missing transverse momentum $\mpt$. 
This enables us to select the events in which the measured $\met$ does not arise from mismeasured jets. 
For our study, we evaluate $\Delta \phi$ for the first two hardest jets and demand the same cut, following Ref.~\cite{Aad:2014kra} which studied a similar signature:
\bea
\Delta \phi(\mpt,j_{h(s)})>2.0\,(0.8)\,. \label{eq:cutdelta}
\eea
We therefore expect that the three-jet background is well under control and negligible compared to the irreducible background. 
It turns out that our simulation study also supports this expectation. 
To the best of our knowledge, none of experimental papers have explicitly reported the rate for the lepton-faking jets. 
We assess it by comparing the relevant cross section reported in Ref.~\cite{Khachatryan:2014fba} and our simulated event sample, and find that the rate is of order $10^{-4}$. 
Implementing the fake-object module into \textsc{Delphes}, we generate three-jet events which are significantly reduced to be negligible by a set of our selection cuts. 

\begin{figure}[t]
\centering
\includegraphics[width=7.2cm]{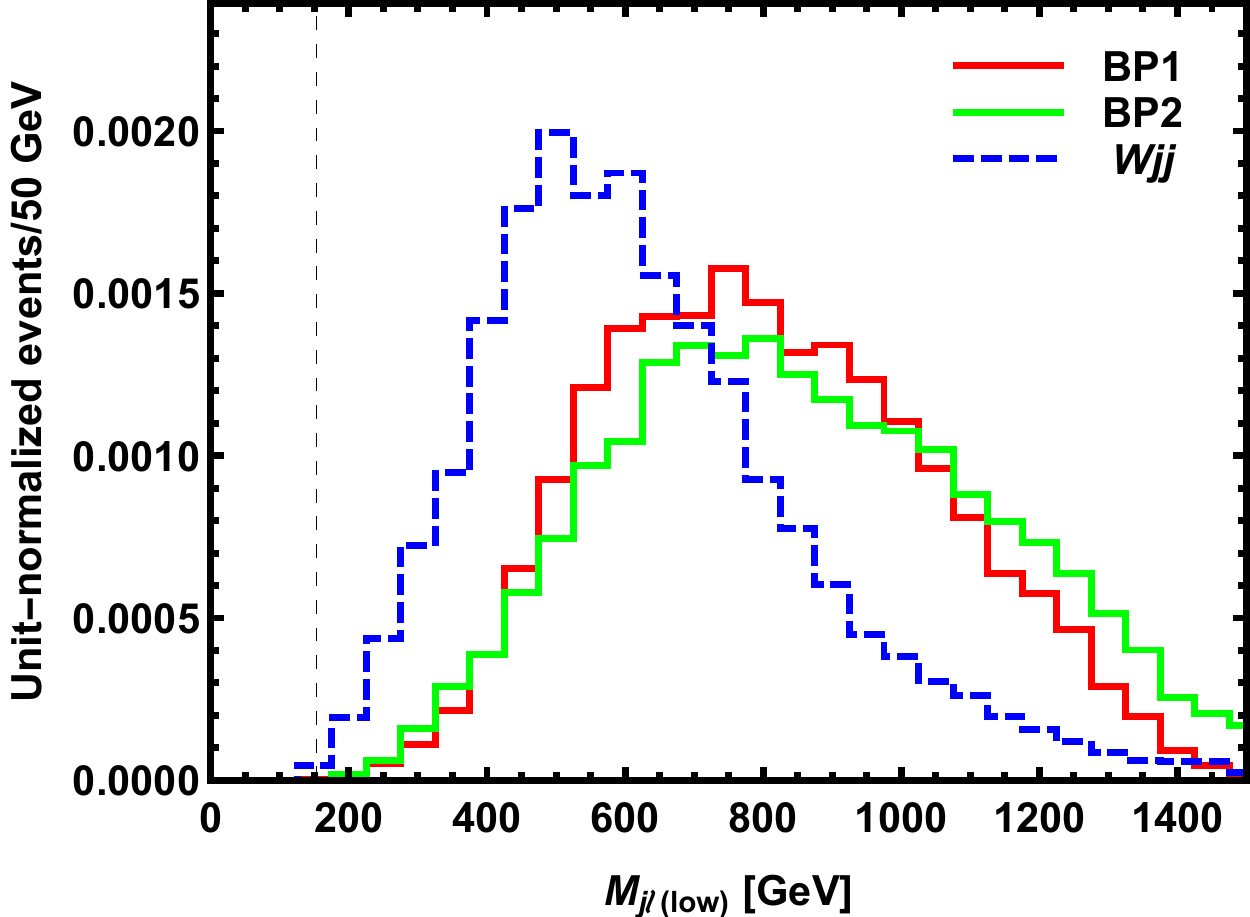} \hspace{0.2cm}
\includegraphics[width=7.2cm]{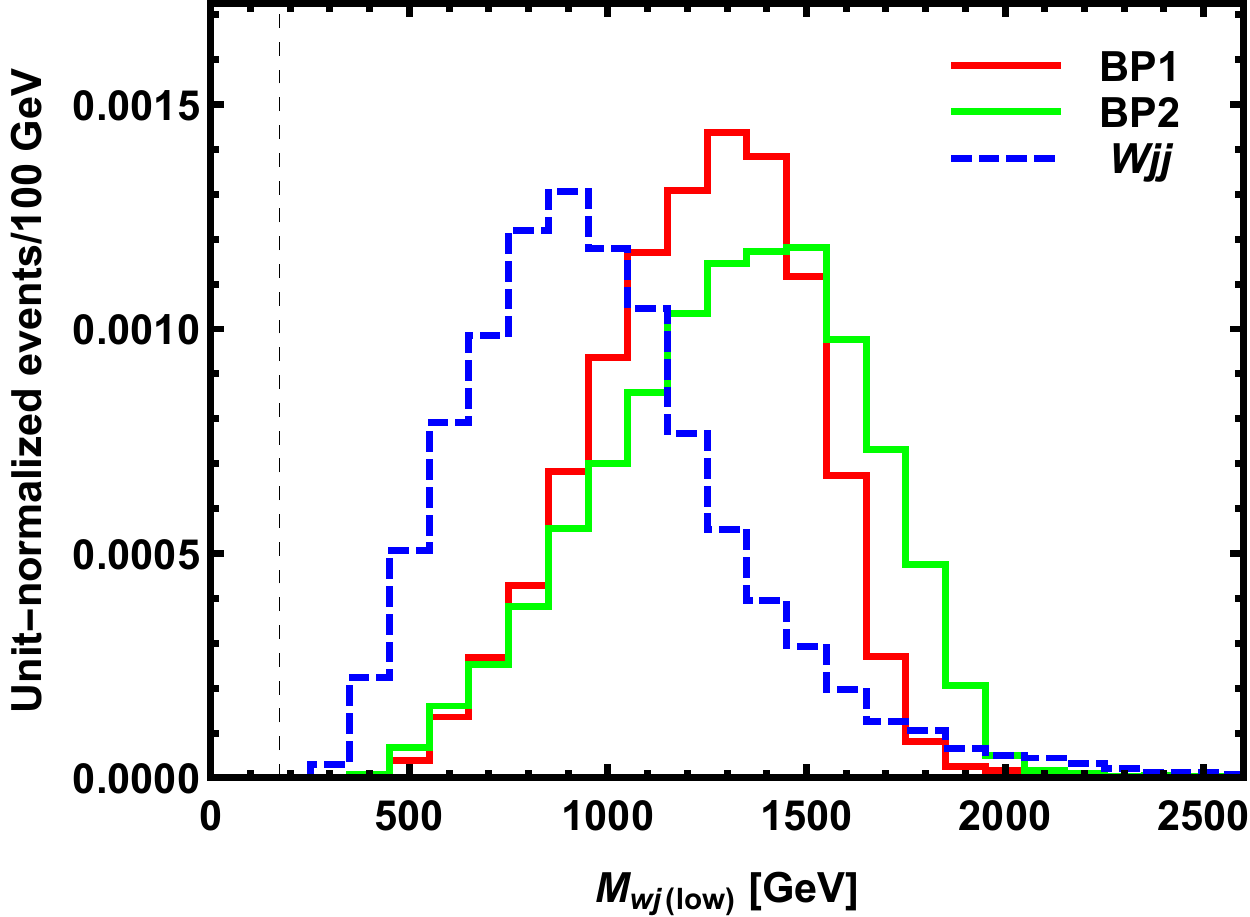}
\caption{\label{fig:mgls} $M_{j\ell(\textrm{low})}$ (left panel) and $M_{wj(\textrm{low})}$ distributions (right panel) for $W$-$Wgg$-BP1 (red solid histogram), $W$-$Wgg$-BP2 (green solid histogram) and $Wjj$ (blue dashed histogram) with events passing the selection criteria listed in-between~\eqref{eq:Nl} and~\eqref{eq:cutdelta} and the $W$ reconstruction procedure.
The black dashed lines mark $M_{bl}^{\max}$ value in the top quark decay and the top quark mass, respectively. }
\end{figure}

On the other hand, getting a background event from the $t\bar{t}$ process depends on its decay mode.
Due to the existence of an isolated lepton in the final state, dileptonic and semileptonic channels are relevant.
\bea
\hbox{Dileptonic: }&&t_1\bar{t}_2 \rightarrow b_1\ell^+_1 \bar{b}_2\ell^-_2+\nu\bar{\nu}  \label{eq:dilepton}\\
\hbox{Semileptonic: }&&t_1\bar{t}_2 \rightarrow b_1\ell^+_1 \bar{b}_2q_2\bar{q}_2+\nu \label{eq:semilepton}
\eea   
Here we assign the same number to all visible particles belonging to the same decay leg just for convenience of the later argument. 
For the fully leptonic $t\bar{t}$ in~\eqref{eq:dilepton}, an event appear as a background with one of the two leptons (say, $\ell_2$) missed or unaccepted.
In contrast, any semi-leptonic $t\bar{t}$ events can be recorded as background events since we require $N_j\geq2$. 

Obviously, we can achieve an $\mathcal{O}(1)$ suppression of the background events stemming from $t\bar{t}$, vetoing the events with at least one $b$-tagged jet.
However, this is not enough to make $t\bar{t}$ negligible due to its large production cross section. 
It turns out that $t\bar{t}$ background is subdominant, compared to $Wjj$~\cite{Aad:2014qxa,CMS:2016srs}.
We note that the cuts in Refs.~\cite{Aad:2014qxa,CMS:2016srs} are softer than the corresponding ones in our analysis, so that $t\bar{t}$ would come back as a comparable background in the phase space resulting from the set of cuts that we require.
Indeed, we find that various invariant mass variables play crucial roles in reducing the $t\bar{t}$ background as well. 
Considering first the dileptonic $t\bar{t}$ in~\eqref{eq:dilepton} in which $\ell_2$ is not recorded, we find that the following criteria available:
\begin{itemize}
\item[D-1.] Since $b_1$-$\ell_1$ invariant mass $M_{b_1\ell_1}$ is bounded by $M_{b_1\ell_1}^{\max}\,(=153 \hbox{ GeV})$, for any dileptonic $t\bar{t}$ events $M_{j\ell(\textrm{low})}\equiv \min[M_{j_h\ell}, M_{j_s\ell}]$ should be smaller than $M_{b\ell}^{\max}$. 
The left panel in Fig.~\ref{fig:mgls} exhibits $M_{j\ell(\textrm{low})}$ distributions for $W$-$Wgg$-BP1 (red solid histogram), $W$-$Wgg$-BP2 (green solid histogram) and $Wjj$ (dashed blue histogram) at the detector level with events passing the cuts from~\eqref{eq:Nl} through~\eqref{eq:cutdelta} and the $W$ reconstruction procedure described earlier. 
The black dashed line marks the position of $M_{b\ell}^{\max}$, from which we observe that more than 99.9\% of signal events have a $M_{j\ell(\textrm{low})}$ value exceeding $M_{b\ell}^{\max}$.
\item[D-2.] Since the invisible momentum comes from a neutrino and a missing $W$, we often fail in reconstructing $W$ with $\ell_1$. 
\end{itemize} 
When it comes to semileptonic $t\bar{t}$, two cases are possible:
\begin{itemize}
\item[S-1.] If one of the two hardest jets and $\ell_1$ belong to the same decay side (e.g., $\ell_1 b_1 \bar{b}_2$, $\ell_1 b_1 q_2$, or $\ell_1 b_1 \bar{q}_2$), the invariant mass between $b_1$ and the reconstructed $W$ should be the same as the top quark mass. 
Therefore, $M_{wj(\textrm{low})}\equiv \min[M_{wj_h},M_{wj_s}]$ does not exceed the top quark mass. 
The right panel in Fig.~\ref{fig:mgls} shows $M_{wj(\textrm{low})}$ distributions for $W$-$Wgg$-BP1 (red solid histogram), $W$-$Wgg$-BP2 (green solid histogram) and $Wjj$ (blue dashed histogram) at the detector level with events satisfying the same criteria described in D-1. The black dashed line represent the location of the top quark mass. 
We observe that every single signal event has a $M_{wj(\textrm{low})}$ value greater than 300 GeV. 
\item[S-2.] If the two hardest jets belong to the second decay side (e.g., $\ell_1 \bar{b}_2 q_2$, $\ell_1 \bar{b}_2 \bar{q}_2$, or $\ell_1 q_2 \bar{q}_2$), the dijet invariant mass should be either the same as the $W$ mass or smaller than $M_{bq}^{\max}\,(=153 \hbox{ GeV})$. 
Since the dijet invariant mass window cut to be used later is much larger than those values, we do not expect any background contribution from this case. 
\end{itemize}
From all these considerations thus far, we expect that $t\bar{t}$ is negligible as well, so we henceforth consider $Wjj$ as the main background to signal events. 

\begin{figure}[t]
\centering
\includegraphics[width=7.2cm]{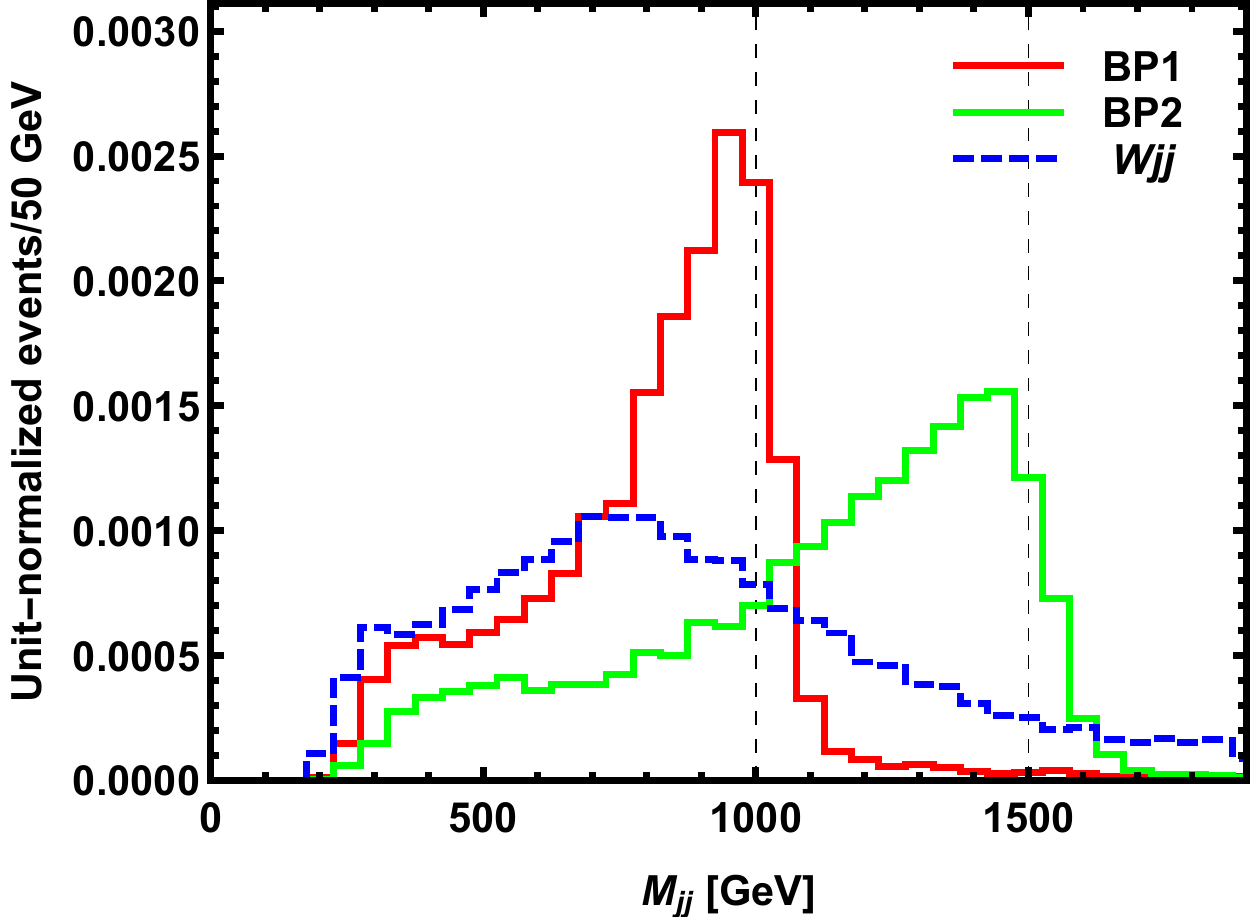} \hspace{0.2cm}
\includegraphics[width=7.2cm]{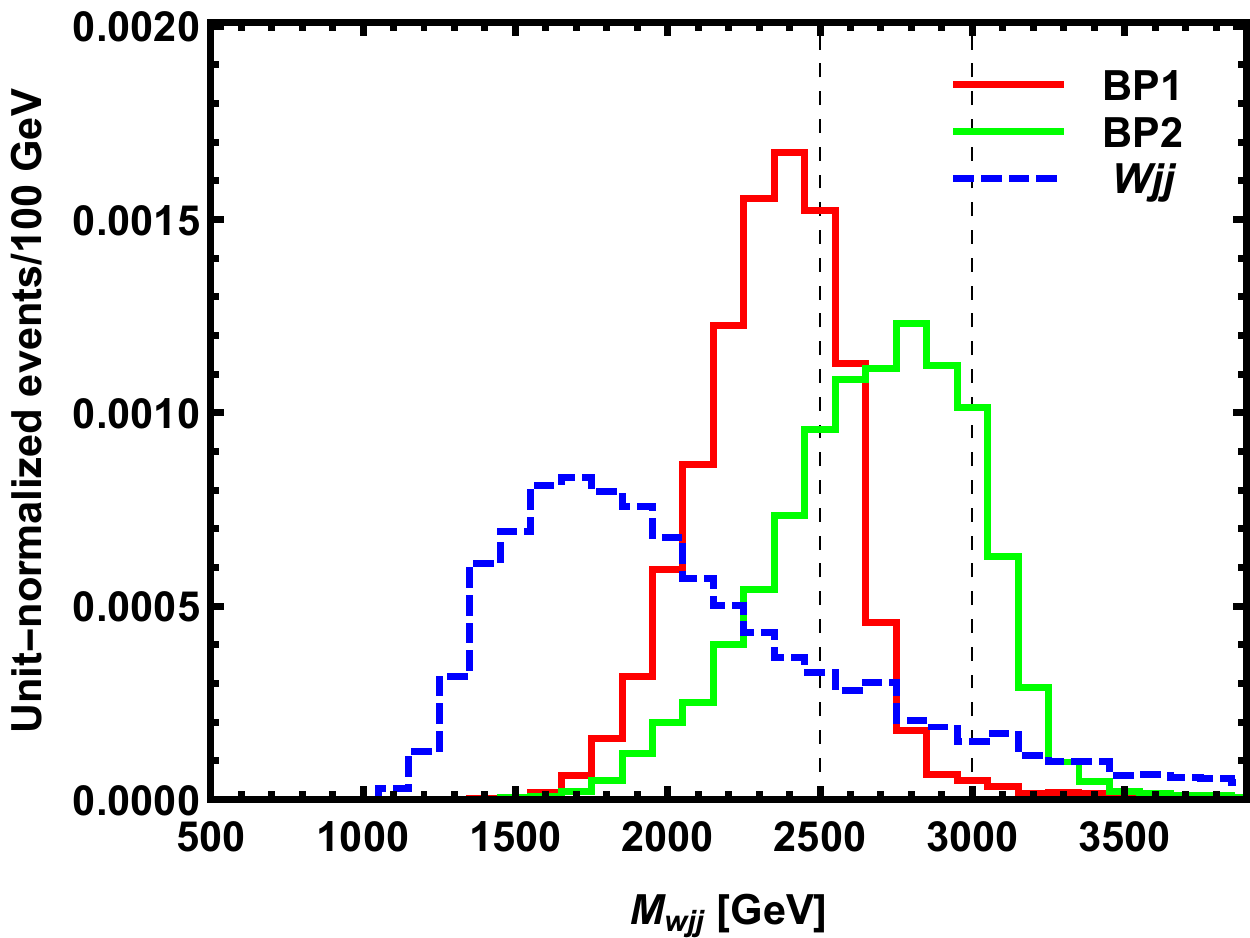} \vspace{0.2cm}
\includegraphics[width=7.2cm]{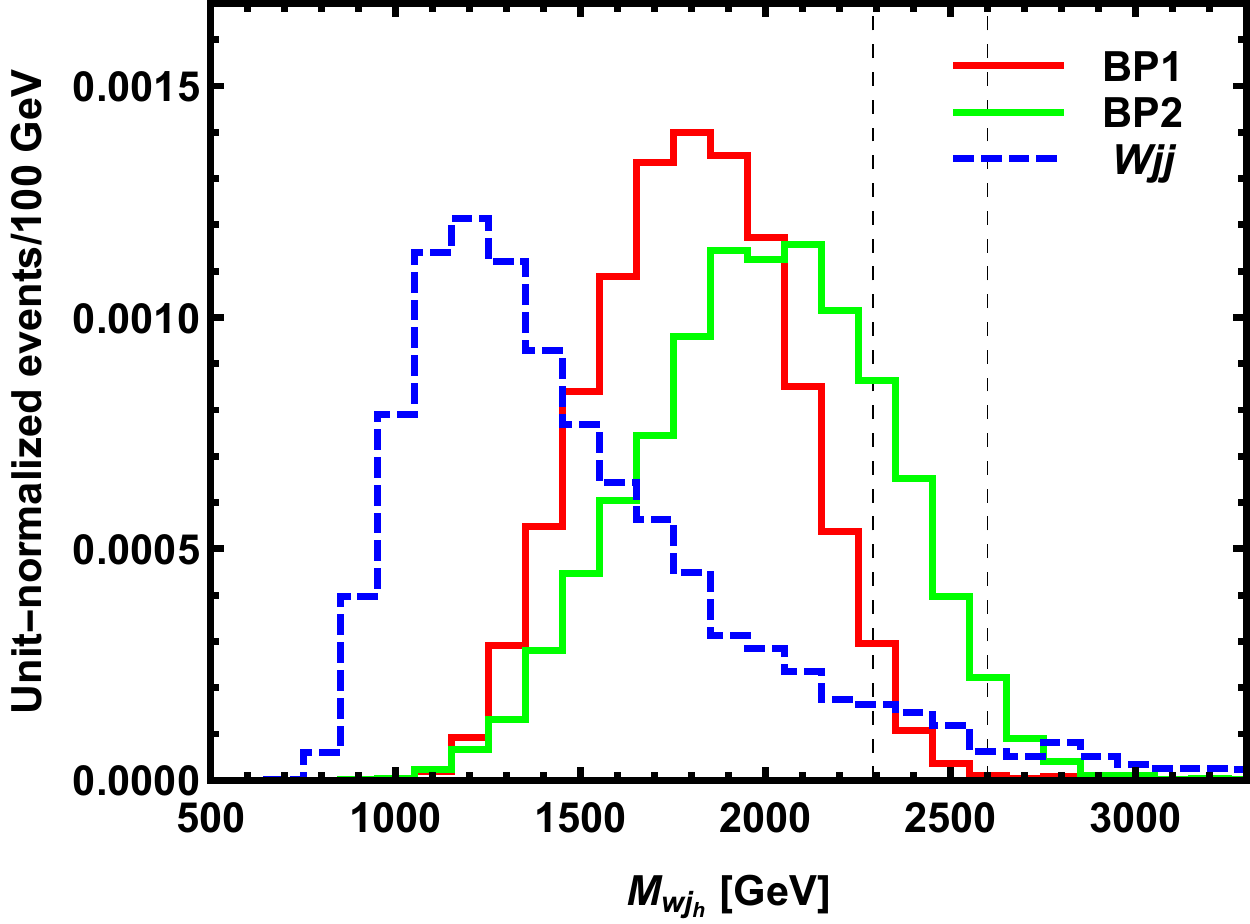}
\caption{\label{fig:threeinv} $M_{jj}$ (upper left panel), $M_{wjj}$ (upper right panel), and $M_{wj_h}$ (bottom panel) distributions for $W$-$Wgg$-BP1 (red solid histogram), $W$-$Wgg$-BP2 (green solid histogram) and $Wjj$ (blue dashed histogram) with events passing the selection criteria listed in-between~\eqref{eq:Nl} and~\eqref{eq:cutdelta} and the $W$ reconstruction procedure. 
The black dashed lines mark the input radion mass, the input KK $W$ mass, and the theoretical $M_{wj}^{\max}$ value, respectively. }
\end{figure}

\begin{table}[t]
\centering
\begin{tabular}{|c|c|c|c|}
\hline 
Cuts & $W$-$Wgg$-BP1 & $W$-$Wgg$-BP2 & $Wjj$  \\
\hline \hline
No cuts & 0.740  & 0.506 & (43,600) \\
$N_{\ell}=1$, $N_j \geq2$ with basic cuts & 0.478 & 0.339 & 14,500  \\
$\met >200$ GeV & 0.385 & 0.253 & 1,840 \\
$p_{T,\ell}> 150$ GeV, $|\eta_{\ell}|<2.5$ & 0.263 & 0.178 &275 \\
$p_{T,j_{h(s)}} >400 \,(200)$ GeV, $|\eta_j|<2.5$ &0.169 & 0.151 & 67.5 \\
$\Delta \phi(\met,j_{h(s)})>2.0\,(0.8)$ & 0.159 & 0.125  & 52.0 \\
$M_{w} \in [60,\,100]$ GeV & 0.159 & 0.125  & 51.9  \\	
\hline
$M_{wj_s} >300$ GeV & -- & 0.125 & 51.8 \\
$M_{jj} \in [1150,\,1550]$ GeV & --  & 0.0632 & 7.03  \\
$M_{wjj} \in [2625,\,3175]$ GeV & -- &  0.0515 & 1.18  \\  
$M_{wj_h} \in [1500,\,2700]$ GeV & -- & 0.0496 & 0.903  \\
\hline
$M_{wj_s} >300$ GeV & 0.159 & -- & 51.8 \\
$M_{jj} \in [675,\,1025]$ GeV & 0.104  & --  &  16.8  \\
$M_{wjj} \in [2175,\,2625]$ GeV & 0.0816 & -- & 2.17 \\  
$M_{wj_h} \in [1375,\,2250]$ GeV & 0.0781 & -- & 1.82  \\
\hline
$S/B$ & 0.043 & 0.055 & --\\
$S/\sqrt{B}$ ($\mathcal{L}=300$ fb$^{-1}$) & 1.0 & 0.90 & -- \\
$S/\sqrt{B}$ ($\mathcal{L}=3000$ fb$^{-1}$) & 3.2 & 2.9 &-- \\
\hline
\end{tabular}
\caption{Cut flows for signal and major background events in terms their cross sections. The cross sections are in fb. The number in the parentheses for $Wjj$ is obtained with basic cuts ($p_{T,j} >100$ GeV, $|\eta_j|<4$, $\Delta R_{jj}>0.4$, $\met >60$ GeV) at the generation level to avoid divergence. In the second row, the same basic cuts are imposed to both signal and background events along with jet and lepton multiplicity requirements. 
\label{tab:leptonicchannel} }
\end{table} 

In this channel, invariant mass window cuts defined by the masses of KK $W$ and radion are useful in separating signal events from background ones. 
The corresponding variables are $M_{wjj}$ and $M_{jj}$. 
In addition, since the signal process is characterized by a two-step cascade decay of a heavy resonance, the invariant mass formed by the reconstructed $W$ and a jet is also useful as pointed out in Sec.~\ref{sec:massvariables}. 
Two signal jets are not distinguishable here, motivating us to consider the prescriptions proposed in eqs.~\eqref{eq:setI1} and~\eqref{eq:setI2}.
In the channel of current interest, the two sets are translated as follows.
\bea
\hbox{Set 1: }&& M_{wj_h},\,\,\, M_{wj_s} \\
\hbox{Set 2: }&& M_{wj(\textrm{high})}= \max[M_{wj_h},M_{wj_s}],\,\,\,M_{wj(\textrm{low})}= \min[M_{wj_h},M_{wj_s}]
\eea
In this analysis, we choose Set 1 as it enables us to achieve slightly better signal sensitivity.
We further restrict ourselves to $M_{wj_h}$ along with $M_{wjj}$ and $M_{jj}$ as four variables are connected by the sum rule in eq.~\eqref{eq:sumrule} which is translated as 
\bea
M_{wjj}^2=M_{jj}^2+M_{wj_1}^2+M_{wj_2}^2=M_{jj}^2+M_{wj_h}^2+M_{wj_s}^2\,,
\eea 
where $W$ is assumed effectively massless. 
Fig.~\ref{fig:threeinv} demonstrates $M_{jj}$ (upper left panel), $M_{wjj}$ (upper right panel), and $M_{wj_h}$ (bottom panel) distributions for $W$-$Wgg$-BP1 (red solid histogram), $W$-$Wgg$-BP2 (green solid histogram) and $Wjj$ (blue dashed histogram) with events passing the selection criteria listed in-between~\eqref{eq:Nl} and~\eqref{eq:cutdelta} and the $W$ reconstruction procedure. 
The black dashed lines mark the input radion mass, the input KK $W$ mass, and the theoretical $M_{wj}^{\max}$ value, respectively.
We clearly see that they can be utilized in the posterior analysis in order to further separate the signal events from the background ones. 
Of course, detailed invariant mass windows depend on the mass spectrum that we aim to look for, and we provide the cut flows for $W$-$Wgg$-BP1, $W$-$Wgg$-BP2, and the major SM background (i.e., $Wjj$) in Table~\ref{tab:leptonicchannel}.
We observe that KK $W$ in both benchmark points may manifest its existence by $\sim 3\sigma$ with an integrated luminosity of 3000 fb$^{-1}$.


\section{Conclusion}
\label{sec:conclusion}

The standard warped extra dimensional model, with the entire SM arising as modes of fields propagating in the bulk between the UV and the IR branes, provides solutions to {\em both} the Planck-weak and flavor hierarchy problems of the SM.
However, without additional flavor structure, it requires the
IR brane scale, and therefore the mass of KK excitations corresponding to SM particles, to be $\gtrsim \mathcal{O}(10)$ TeV in order to be consistent with flavor/CP bounds. This leaves us with a little hierarchy problem.
This constraint also implies that the typical new particles of this model (i.e., the KK modes) are beyond LHC reach. Recent work~\cite{Agashe:2016rle} presented a variation, where the bulk is (mildly) extended, starting from the standard setup, in the IR direction down to {\em a few} TeV.
Moreover, only gauge and gravity fields have access to this additional space, while Higgs and matter fields are confined to the region of the bulk in-between
the UV and an intermediate brane corresponding to a scale of $\gtrsim \mathcal{O}(10)$ TeV. Note that the location of the intermediate brane in this setup is the same as the original IR brane in the standard setup.

It is possible to show (see Ref.~\cite{Agashe:2016rle}) that the solutions to the Planck-weak and flavor hierarchy, as well as consistency with 
electroweak (EW) and flavor precision tests remain unchanged (including the mild residual tuning) in this extension.
In fact, the constraint on the (ultimate) IR brane scale for this model originates instead from direct LHC searches (specifically for gauge KK modes), which is how the choice of a few TeV was made.
On the flip side, future LHC signals from gauge and gravity KK modes are thus possible (cf.~the standard scenario).

In this paper, for concreteness, we considered the simplest possibility within the above extended framework, where {\em all} gauge fields live down to a few TeV.
Such a geometry suppresses the usually dominant decay modes of lightest gauge/gravity KK particles into top/Higgs for all KK modes, thereby allowing other decay modes, thus far overwhelmed by top/Higgs, to make their case; in short, the LHC phenomenology can be significantly altered as compared to the standard setup. For example, gauge KK particles can decay into the corresponding SM gauge boson, in association with a radion, with the latter decaying into various pairs of SM gauge bosons. Interestingly, using AdS/CFT duality between this warped model and the idea of composite SM Higgs, it can be argued that the above decay channel for KK gauge bosons is roughly the 
``analogue'' of $\rho \rightarrow \pi \gamma$, followed by $\pi \rightarrow \gamma \gamma$ in QCD. 
%
%
Here, we studied in detail the LHC signals resulting from such cascade decays of gauge KK modes. It is clear that there is a plethora of final states possible from this decay (involving combinations of photons, ordinary jets, $W/Z$-jets, and leptons from $W/Z$ decay). In this work, we focussed on several among them with significant rates.

Overall, we found that the prospects for evidence of these KK particles via the new cascade decay channel look bright, with the KK gluon being the best shot (due mostly to largest production cross-section), whereas KK photon/$W/Z$ require higher luminosity (3000 fb$^{-1}$) for detection.

We would like to point out here that 
the {\em first} discovery 
for our benchmark points in this framework
will probably be of the {\em radion}, instead of gauge KK modes, for the following reason.
First of all, radion has to be {\em lighter} than KK gauge bosons (i.e., $\lesssim \mathcal{O}(1)$ TeV) in order for the above new cascade decay to be kinematically allowed. Moreover, in the above framework, the radion still has sizable couplings to gluons inside proton and to a lesser extent to photons, even though the couplings to top/Higgs are suppressed, just like for gauge KK.
Therefore, taking the mass and couplings of the radion into consideration, one then has to contend with stringent bounds from its {\em direct} production at the LHC from gluon fusion, in addition to a smaller contribution from the above decay of gauge KK, followed by its decay into the ``golden" diphoton final state.
Thus one has to make a compromise here between the above two considerations -- preference for light radion in order for new decay channel to be significant vs. LHC bounds disfavoring it.\footnote{Note that it is also difficult to {\em naturally} obtain a radion {\em much} lighter than gauge KK mode.} This implies that the radion has to live on the edge of current bounds, and will likely be discovered via direct production before gauge KK particles. 

{\bf Future work}: Beyond our specific study, the main motivation of our paper is that small plausible tweaks can change even the qualitative aspects of LHC signals of new physics in a BSM scenario. We considered the manifestations of this in the context of the warped framework, and showed how the LHC signals of the KK modes are dramatically impacted. Our philosophy in this regard is to explore all plausible scenarios that are not ruled out, and look for them in colliders. 

With this approach in mind, we would like to point out an interesting possiblity that is consistent with current bounds on different masses and interactions of the warped extra-dimensionial framework, and has a striking signal. Within the above brane configuration (i.e., UV, intermediate and IR), we can contemplate only {\em some} of the SM gauge fields to reside in the full bulk, as compared to \textit{all} of them. For example, suppose that only the EW gauge sector has this luxury of propagating all the way to the last brane. In other words, consider restricting the gluons to propagate only down to $\gtrsim \mathcal{O}(10)$ TeV.
In this scenario, we then lose the KK gluon (now with mass $\sim \mathcal{O}(10)$ TeV) signal, which was dominant one in our analysis. We thus seem to be headed in the ``wrong" direction. However, a more careful consideration suggests that this can rather be to our advantage. 
First note that the radion decouples from gluons in this new scenario, because it is peaked near the IR brane while the gluons do not reach there. Thus, in this scenario, the radion decays mostly to EW gauge bosons ($W/Z$ and photon, which {\em do} feel the IR extension).  This means that the BR to diphoton increases substantially as compared to above model. This is still not desirable, in terms of satisfying the relevant bound from direct production of radion. However, the production cross section of radion from gluon fusion becomes negligible due to the reduced radion-gluon coupling. In fact, we can show that the {\em net} radion rate into diphotons actually reduces, making it easily safe from current bounds (from direct production), even if it is lighter than EW gauge KK. As a corollary, radion discovery via this direct process can be {\em significantly} delayed.

Moving onto the effect on the EW gauge KK decaying into the new channel, we first note that the couplings of gauge KK modes to light quarks (involved in production of gauge KK modes) and to radion-SM gauge boson (relevant for this initial decay) are unchanged, as compared to the above model. This implies that, for the same masses, the net cross section for this cascade decay (i.e., into radion, plus SM gauge boson) is unchanged.
Remarkably, the associated radion now decays dominantly to a pair of EW gauge bosons, instead of the  more background-prone dijets in the previous model. This twist then gives rise to the clean {\em tri-EW boson} decay mode (i.e., combinations of $W/Z/\gamma$ final state) from EW gauge KK production and cascade decay (cf.~$W/Z/\gamma + {\rm dijet}$ having largest rate before).
It is this feature that ultimately makes the new decay channel for EW gauge KK more visible over background.

We plan to return to a detailed and systematic study of the above fascinating possibility.
For now, we simply comment on the signal for a related case, with only the hypercharge (approximately photon) occupying the extended bulk. The gluons {\em and} $SU(2)$ gauge bosons are allowed only in smaller bulk. As a consequence, KK photon and radion are the only players in the game here, with 
radion decaying mostly to diphoton, culminating in the potentially striking signal of KK photon production and decay to a {\em triphoton} final state, 
with negligible SM background!
%
%
%
%
%
Under these circumstances, 
it is possible that 
%
%
the signal appears imminently in the current run of the LHC. Of course, this would {\em simultaneously} constitute discovery for {\em radion}. In particular, the signal would feature both a {\em di}photon resonance (radion) and a {\em tri}photon resonance (KK photon).
%
%
Note that the KK $W/Z$ and KK gluon are much heavier in this case so that their signal is negligible, and possibly even beyond LHC kinematic reach. The radion mass is below KK photon, but still the direct production is small since its couplings to the gluons are suppressed in this model.
%
%
%
Thus, we see that simple variations, easily within the realm of possibility, can tip the scales in favor or against a given channel, switching the particle of
first discovery.

In summary, we feel that we might have uncovered just the tip of the iceberg in terms of LHC signals for the general framework of warped SM/composite Higgs: 
more channels for the above models can be studied and further extensions of the models are possible (for example, actually adding more branes corresponding to end of bulk region for
different gauge fields).
As the second phase of the LHC steps into full gear, our study 
%
%
underscores (using one of the leading candidates for beyond SM physics as illustration) why it is crucial to cast as wide a net as is possible in order to catch whatever guise new physics might take.

\section*{Acknowledgements}
We would like to thank Alberto Belloni, Kyle Cranmer, Sarah Eno, Andy Haas, Enrique Kajomovitz, Deepak Kar, Josh Kunkle, Greg Landsberg, Peter Maksimovic, and Daniel Whiteson for providing encouragement; Alberto Belloni, Suyong Choi, and Jae Hyeok Yoo for discussions on fake objects in detectors; Sung Hak Lim for discussions on the MDT, and Raman Sundrum for 
%
%
general
%
%
discussions. This work was supported in part by NSF Grant No.~PHY-1315155 and the Maryland Center for Fundamental Physics (MCFP). DK was supported in part by the U.S. Department of Energy under Grant DE-SC0010296 and is presently supported in part by the Korean Research Foundation (KRF) through the CERN-Korea Fellowship program. 
KA would like to thank the Aspen Center for Physics for hospitality during the initiation of this work.
%
%

\end{document}